\RequirePackage{lineno}
\documentclass[11pt,a4paper]{article}
\usepackage{epsf}
\usepackage{amsmath,amssymb}
\usepackage{url,cite,ifpdf,slashed,multirow,tabularx}
\usepackage{subfigure}  
\usepackage{ulem}
\usepackage{color}

\ifpdf      
\usepackage[dvipdfmx]{graphicx}                                 
\else                                      
  \usepackage[dvips]{graphicx}             
\fi

\def\slashchar#1{\setbox0=\hbox{$#1$} 
\dimen0=\wd0 
\setbox1=\hbox{/} \dimen1=\wd1 
\ifdim\dimen0>\dimen1 
\rlap{\hbox to \dimen0{\hfil/\hfil}} 
#1 
\else 
\rlap{\hbox to \dimen1{\hfil$#1$\hfil}} 
/ 
\fi}

  \makeatletter
    
    \@addtoreset{equation}{section}
  \makeatother
  
\begin{document}

\renewcommand{\thefootnote}{\fnsymbol{footnote}} 
\begin{titlepage}

\begin{center}

\hfill KEK--TH--1682\\
\hfill UT--13--38\\
\hfill IPMU--13--0209\\
\hfill November, 2013\\

\vskip .75in

{\LARGE \bf 
  Gauge invariant Barr-Zee type contributions \\[0.1em]
to  fermionic EDMs \\[0.5em]
in the two-Higgs doublet models
}

\vskip .6in

{\Large
  \textbf{Tomohiro Abe}$^{\rm (a)}$,
  \textbf{Junji Hisano}$^{\rm (b,c)}$, \\[0.4em]
  \textbf{Teppei Kitahara}$^{\rm (d)}$, and
  \textbf{Kohsaku Tobioka}$^{\rm (c,d)}$
}
\vskip 0.3in

$^{\rm (a)}${\it Theory Group, KEK, Tsukuba, 305-0801, Japan}
\vskip 0.1in
$^{\rm (b)}${\it Department of Physics, Nagoya University, Nagoya 464-8602, Japan}
\vskip 0.1in
$^{\rm (c)}${\it Kavli IPMU (WPI), University of Tokyo, Kashiwa, Chiba 277--8583, Japan}
\vskip 0.1in
$^{\rm (d)}${\it Department of Physics, University of Tokyo, Tokyo 113--0033, Japan}

\end{center}

\vskip .6in

\begin{abstract}

  We calculate all gauge invariant Barr-Zee type contributions to
  fermionic electric dipole moments (EDMs) in the two-Higgs doublet
  models (2HDM) with softly broken $Z_2$ symmetry.  We start by
  studying the tensor structure of $h \to VV'$ part in the Barr-Zee
  diagrams, and we calculate the effective couplings in a gauge
  invariant way by using the pinch technique. Then we calculate all
  Barr-Zee diagrams relevant for electron and neutron EDMs.  We make
  bounds on the parameter space in type-I, type-II, type-X, and type-Y
  2HDMs. The electron and neutron EDMs are complementary to each other
  in discrimination of the 2HDMs.  Type-II and type-X 2HDMs are
  strongly constrained by recent ACME experiment's result, and future
  experiments of electron and neutron EDMs may search ${\cal O}$(10)
  TeV physics.

\end{abstract}

\end{titlepage}

\setcounter{page}{1}
\renewcommand{\thefootnote}{\#\arabic{footnote}}
\setcounter{footnote}{0}


\section{Introduction}

The standard model (SM) has been worked very well for a long time, and
its last missing piece, the Higgs boson, was finally discovered by
the Large Hadron Collider (LHC) experiment at CERN~\cite{Aad:2012tfa,
  Chatrchyan:2012ufa}.  This is a triumph of the SM and a great step
to understand physics at the electroweak scale.  However, there are
many unsolved problems within the SM, for example, the observed 
dark matter particles and baryon asymmetry in the Universe. 
From theoretical viewpoint, the gauge hierarchy problem is still
in question. Hence, there
have been many attempts to solve such problems in frameworks beyond
the SM.

In a bottom-up approach towards new physics beyond the SM, an
attractive option is to study the two-Higgs doublet models
(2HDMs). They are simple and may be low-energy effective theories of
various new physics models. 
 Since 2HDMs generally have dangerous flavor changing neutral currents (FCNCs),
we particularly consider 2HDMs with softly broken $Z_2$ symmetry which
suppresses the FCNCs.   If two Higgs fields do not distinguish the generations of quarks and leptons, the models are classified, with respect to the
Yukawa interactions, into four types: type-I, type-II, type-X, and
type-Y. One of the important feature of 2HDMs is that there is a new CP
violation source in the Higgs potential.

In general, the powerful tool to seek new physics including 2HDMs is
of course the LHC which may directly probe physics up to a few
TeV. Another possibility is provided by low energy precision
measurements, such as in flavor physics. The remarkable feature is
that these measurements have a potential to investigate new physics
beyond the LHC reach by orders of magnitude.  In particular, the
electric dipole moments (EDMs) are interesting because the EDMs are
highly sensitive to CP violation in physics beyond the SM.  While the
SM predictions of EDMs are much lower than the current experimental
bounds, assuming the strong CP problem is solved by some mechanism,
such as the Peccei-Quinn symmetry \cite{Baron:2013eja, Baker:2006ts},
new physics around TeV scale would give large values within the reach of the future EDM
measurements \cite{EDMreview}.  In addition, the electroweak
baryogenesis (EWBG)~\cite{Kuzmin:1985mm, Cohen:1993nk, Rubakov:1996vz,
  Trodden:1998ym}, which needs a new CP violation source, may lead to larger values of EDMs than the SM predictions.
 
\begin{figure}[t]
\begin{minipage}{0.24\hsize}
\subfigure[]{\includegraphics[bb=0 0 200 130,
 width=0.9\hsize]{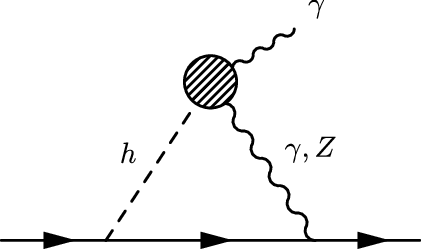} \label{fig:BZ1}}  
\end{minipage}
\begin{minipage}{0.24\hsize}
\subfigure[]{\includegraphics[bb=0 0 200 130,
 width=0.9\hsize]{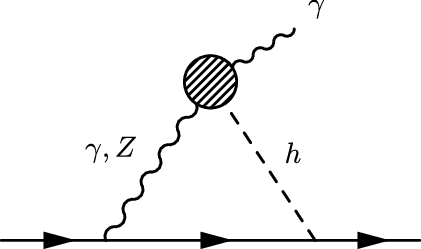} \label{fig:BZ2}}  
\end{minipage}
\begin{minipage}{0.24\hsize}
\subfigure[]{\includegraphics[bb=0 0 200 130,
 width=0.9\hsize]{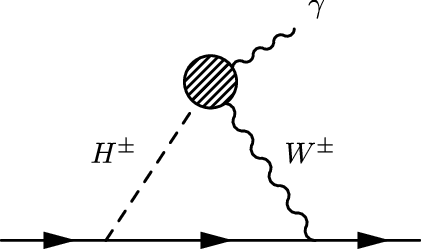} \label{fig:BZ3}}  
\end{minipage}
\begin{minipage}{0.24\hsize}
\subfigure[]{\includegraphics[bb=0 0 200 130,
 width=0.9\hsize]{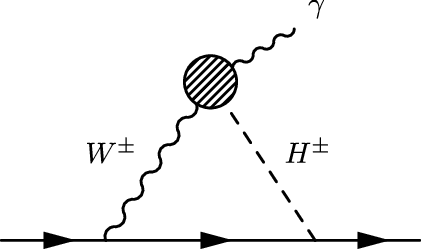} \label{fig:BZ4}}  
\end{minipage}
\caption{
Barr-Zee diagrams, which contribute to fermionic EDMs at two-loop level. 
}
\label{fig:Barr-Zee}
\end{figure}

The EDM measurements, therefore, are concrete tests on 2HDMs
containing a new CP phase.  In the models, the one-loop contributions
to the fermionic EDMs are too small to observed since those
contributions are proportional to the third power of small Yukawa
couplings. Some two-loop diagrams, called the Barr-Zee
diagrams~\cite{Barr:1990vd}, which we show in Fig.~\ref{fig:Barr-Zee},
may give sizable contributions to the EDMs, since they are suppressed
by only one power of small Yukawa couplings.  These diagrams contain
one-loop effective vertices, $h \gamma \gamma$, $h \gamma Z$, and
$H^{\mp} W^{\pm} \gamma$.  The Type-II case was evaluated in
Refs.~\cite{Leigh:1990kf, Chang:1990sf}, but the results in the
previous works are not gauge invariant. We improve this point by using
the pinch technique~\cite{Degrassi:1992ue, Degrassi:1992ff,
  Denner:1991kt} and make the Barr-Zee diagrams gauge invariant. We
also study EDMs in the other three types as well as the type-II.

We organize this paper as follows. 
In Sec.~\ref{sec:2HDM}, we briefly review the 2HDMs with softly-broken
$Z_2$ symmetry.
In Sec.~\ref{sec:subdiagram}, we
study the tensor structure of the effective vertices which are needed to
evaluate the Barr-Zee diagrams, and show the gauge invariant tensor
structure. After that,
%
we calculate the effective vertices
explicitly and show that the diagrams which include the gauge bosons
are not gauge invariant. This implies that we need some non-Barr-Zee diagrams
to make the effective vertices gauge invariant. We show it by using the
pinch technique.
The formulae of the gauge invariant Barr-Zee diagrams are given in
Sec.~\ref{sec:Barr-Zee}, and their numerical evaluation is presented
in Sec.~\ref{sec:neumerical}. There we discuss the complementarity
between the electron and neutron EDM measurements in discrimination of
2HDMs, and also prospects of future experiments.
Sec.~\ref{sec:summary} is devoted to conclusions and discussion.
Notations and details of the calculation are given in the Appendices.

\section{Models}
\label{sec:2HDM}

We briefly review the models discussed in this paper.
We have two Higgs doublets, $H_1$ and $H_2$, and they have the vacuum
expectation values (VEVs).  The Higgs doublets are parametrized as
follows,
\begin{align}
H_i
=&
\left(
\begin{matrix}
 \pi^{+}_i 
\\ 
 \frac{1}{\sqrt{2}}
\left(
v_i
+
\sigma_i
-
i \pi_i^{3}
\right)
\end{matrix}
\right)
,
\quad \quad
(i = 1, 2)
.
\label{eq:H-base}
\end{align}
In order to avoid the dangerous FCNC problems, we introduce the $Z_2$ symmetry.
The $Z_2$ symmetry is assumed to be softly broken so that the domain-wall formation in the early universe is suppressed.
Under this symmetry, the Higgs doublets are translated into 
$H_1 \to + H_1$ and $H_2 \to - H_2$, and the Higgs potential is given as 
\begin{align}
 V=&
m_1^2 H_1^{\dagger} H_1
+
m_2^2 H_2^{\dagger} H_2
-
\left(
\left(
\textrm{Re}m_3^2
+
i
\textrm{Im}m_3^2
\right)
H_1^{\dagger} H_2
+
(h.c.)
\right)
\nonumber \\
&
+ \frac{1}{2} \lambda_1 (H_1^{\dagger} H_1)^2
+ \frac{1}{2} \lambda_2 (H_2^{\dagger} H_2)^2
+ \lambda_3 (H_1^{\dagger} H_1) (H_2^{\dagger} H_2)
+ \lambda_4 (H_1^{\dagger} H_2) (H_2^{\dagger} H_1)
\nonumber \\
&
+
\left(
\lambda_5
e^{i2\phi}
 (H_1^{\dagger} H_2)^2 
+ 
(h.c.)
\right)
.
\end{align}
The third and last terms in this potential contain complex
parameters. While one of them can be eliminated by redefinition of
Higgs fields, another phase is physical so that CP symmetry is
broken. In this paper we take the Higgs VEVs, $v_1$ and $v_2$, real
using the gauge symmetry and also redefinition of a Higgs field. In
this basis, two phases in the potential are related to each others by
the stationary condition of the potential, $V' = 0$.  In this paper we
choose $\phi$ as an input parameter for CP violation.

We also use the following variables for convenience in this paper,
\begin{align}
&
 \cos\beta = \frac{v_1}{v},\ 
 \sin\beta = \frac{v_2}{v},
\\
&
 M^2
\equiv
 \frac{v_1^2 + v_2^2}{v_1 v_2}
 \text{Re} m_3^2
.
\label{eq:def_M}
\end{align}
and where
\begin{align}
 v =  \sqrt{v_1^2 + v_2^2} = (\sqrt{2} G_F)^{-1/2} \simeq 246\text{~GeV}.
\end{align}
$G_F$ is the Fermi constant.
It is easy to find the charged Higgs boson mass,
\begin{align}
 m_{H^{\pm}}^2
=&
M^2
-
\frac{1}{2}
v^2
(\lambda_4 + \lambda_5 \cos(2\phi))
.
\end{align}
On the other hand, since CP symmetry is broken in the Higgs potential,
we need to diagonalize a 3 by 3 matrix to find the neutral Higgs masses.

The Yukawa interaction in this model is given by
\begin{align}
\mathcal{L}_{\textrm{Yukawa}} = 
 -\overline{q}_L  \widetilde{H}_2 y_u u_R 
 -\overline{q}_L  H_i y_d d_R
-\overline{\ell}_L
 H_j
y_e
 e_R
+
h.c.
,
\end{align}
where 
$\widetilde{H}_2 = \epsilon H_2^{\ast}$, 
and $i, j = 1$ or 2, depending on
the type of 2HDMs. 
While up-type quarks couple to only to $H_2$, leptons and down-type quarks couple to either $H_1$ or $H_2$ due to the 
$Z_2$ symmetry. We summarize which Higgs fields couple to fermions in
Table~\ref{tab:type_summary}. 

The detail information of the models, such as mass eigenvalues, mixings, and interactions of the Higgs bosons, are given in Appendix~A.

\begin{table}[tb]
\begin{center}
\caption{Summary of the Higgs fields which couple to quarks and leptons
 in four types.}
 \label{tab:type_summary}
  \vskip .15in
\begin{tabular}{c|c c c c}
\hline \hline
 Type    & I     & II     & X     & Y     \\
\hline 
 $u$     & $H_2$ & $H_2$  & $H_2$ & $H_2$ \\ 
 $d$     & $H_2$ & $H_1$  & $H_2$ & $H_1$ \\ 
 $\ell$  & $H_2$ & $H_1$  & $H_1$ & $H_2$ \\ 
\hline \hline
\end{tabular}
\end{center}
\end{table}

\section{Effective vertices}
\label{sec:subdiagram}

\begin{figure}[tbp]
\begin{center}
 \includegraphics[bb=0 0 200 130,angle=0, width=0.3\hsize ]{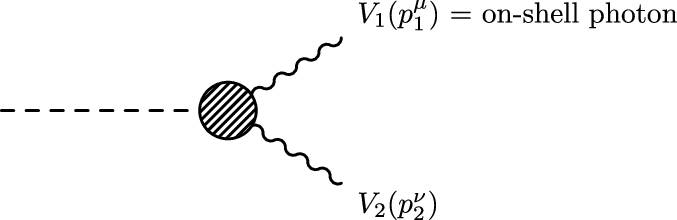}
\end{center}
 \caption{Effective Higgs boson-vector boson-vector boson vertices. 
}
 \label{fig:hvv}
 \end{figure}

In this section we calculate effective vertices relevant for the
Barr-Zee diagrams in a gauge invariant way. To make our point clear, we
start by exploring the relevant form of the effective vertices
shown in Fig.~\ref{fig:hvv}. Then we calculate  effective $h\gamma \gamma$, $h
Z\gamma$  and $H^{\mp}W^{\pm}\gamma$ vertices . We also calculate the pinch terms to make the vertices
gauge invariant. 

\subsection{Tensor structure of the effective vertices}\label{subsec:tensor}
We study the tensor structure of the effective vertices shown in
Fig.~\ref{fig:hvv}. 
This part has two Lorentz indices, and does not contain
$\gamma$-matrices. Then it is generally written as
\begin{align}
 \Gamma^{\mu \nu}
=&
  A_{0} g^{\mu \nu}
+ A_{1} p_1^{\mu} p_1^{\nu}
+ A_{2} p_2^{\mu} p_2^{\nu}
+ A_{12} p_1^{\mu} p_2^{\nu}
+ A_{21} p_2^{\mu} p_1^{\nu}
+ i\Gamma_{5} \epsilon^{\mu \nu \rho \sigma} p_{1 \rho} p_{2 \sigma}
,
\end{align}
where $p_1^{\mu}$ and $p_2^{\nu}$ are the momenta of $V_1$ and $V_2$,
respectively, and their direction is outgoing. We consider the case that 
$V_1$ is on-shell photon, and thus the terms proportional to
$p_1^{\mu}$ are dropped. 
In addition, the gauge symmetry of photon requires 
$  \Gamma^{\mu \nu} p_{1 \mu} = 0. $
Then, the effective vertex for $h$-$V_1$-$V_2$ in the
case that $V_1$ is on-shell photon is defined with only two form factors as 
\begin{align}
 \Gamma^{\mu \nu} (p_1, p_2)
=&
 \Gamma(p_1, p_2) 
\left(
 -(p_1 p_2) g^{\mu \nu} + p_2^{\mu} p_1^{\nu}
\right)
+
i \Gamma^{5} (p_1, p_2) 
\epsilon^{\mu \nu \rho \sigma} p_{1 \rho} p_{2 \sigma}
.
\label{eq:gauge_inv_coupling_form}
\end{align}
Note that this tensor structure is led from the gauge symmetry of
on-shell photon. Then all the effective vertices must be this form.
We emphasize this point because sometimes this point seems
overlooked, for example the tensor structure in Eq.~(9) in
Ref.~\cite{Leigh:1990kf} is different from
Eq.~(\ref{eq:gauge_inv_coupling_form}).

However, in the actual calculation, we would find terms proportional to
$p_{2}^{\mu} p_{2}^{\nu}$ and $g^{\mu \nu}$, which should vanish and
do not appear in Eq.~(\ref{eq:gauge_inv_coupling_form}), namely we would
find the effective vertices become
\begin{align}
 \widetilde{\Gamma}^{\mu \nu}(p_1, p_2)
=&
 \Gamma^{\mu \nu} (p_1, p_2)
+
 \Gamma^{P}(p_1, p_2) g^{\mu \nu}
+
 \Gamma^{D}(p_1, p_2) p_2^{\mu} p_2^{\nu}
,
\label{eq:gauge_variant_coupling_form}
\end{align}
where $\Gamma^{\mu \nu} (p_1, p_2)$ is defined 
in Eq.~(\ref{eq:gauge_inv_coupling_form}).
These extra terms, $\Gamma^{P}$ and $\Gamma^{D}$, are apparently against
the gauge invariance, but, 
nevertheless, they would appear. See, for example, Eq.~(9) in
Ref.~\cite{Leigh:1990kf}. As we will see the following sections, we find
they disappear {\it if} we take on-shell conditions for all the external legs.
However, we should keep them off-shell except for 
a single photon 
because we use the effective vertices
to calculate the Barr-Zee diagrams. Hence we need to consider how to
deal with these gauge variant terms.

Fortunately, it is found that the $p_{2}^{\mu} p_{2}^{\nu}$ term does
not contribute to the EDMs at two-loop level.
If $\Gamma^{\mu \nu}(p_1, p_2)$ contains terms proportional to $p_2^{\mu}
p_2^{\nu}$, the diagrams shown in Fig.~\ref{fig:Barr-Zee} contain the
following structures,
\begin{align}
&
\overline{u}(p+q)
\slashchar{\ell} 
\frac{1}{\slashchar{p} + \slashchar{q} - \slashchar{\ell} - m_{f}}
u(p)
\label{eq:p2p2_1}
, \\ 
&
\overline{u}(p+q)
\frac{1}{\slashchar{p} + \slashchar{\ell} - m_{f}}
\slashchar{\ell} 
u(p)
,
\label{eq:p2p2_2}
\end{align}
where Eq.~(\ref{eq:p2p2_1}) (Eq.~(\ref{eq:p2p2_2})) comes from 
Figs.~\ref{fig:BZ1} and \ref{fig:BZ3} (Figs.~\ref{fig:BZ2} and \ref{fig:BZ4}).
If we omit ${\cal O}(y_f^2)$ terms, we can ignore the mass term in the
fermion propagator and the mass of the external fermions. Then, by
using the equation of motion of the external fermions,
\begin{align}
&
\overline{u}(p+q)
(\slashchar{\ell} - \slashchar{p} - \slashchar{q})
\frac{1}{\slashchar{p} + \slashchar{q} - \slashchar{\ell}}
u(p)
, \\ 
&
\overline{u}(p+q)
\frac{1}{\slashchar{p} + \slashchar{\ell}}
(\slashchar{\ell} + \slashchar{p})
u(p)
.
\end{align}
Now it is apparent that these terms do not contain $\sigma^{\mu \nu}
\gamma^5$ structure because all the $\gamma$-matrices are canceled out.
Therefore the terms which are proportional to $p_2^{\mu}
p_2^{\nu}$ in the effective vertices do not contribute to the EDMs.
Then we can safely drop the $\Gamma^D$ term from
Eq.~(\ref{eq:gauge_variant_coupling_form}).

On the other hand, the $\Gamma^P$ term in
Eq.~(\ref{eq:gauge_variant_coupling_form}) remains as long as we take
off-shell conditions. This is nothing strange because the gauge
invariance is promised for $S$-matrix, not for effective
coupling. Then the gauge invariance will recover once we calculate
non-Barr-Zee diagrams as well as the Barr-Zee diagrams, namely a full
two-loop order calculation manifestly gives the gauge invariant
results. However, it is very tough work to accomplish it. Instead of
the full two-loop order calculation, we make the effective vertex
gauge invariant by borrowing some terms from non-Barr-Zee
diagrams. This technique is known as the pinch technique, and the
borrowed terms are called pinch terms \cite{Degrassi:1992ue,
  Degrassi:1992ff, Denner:1991kt}.  As we will see in the fallowing
section, we find that $\Gamma^P$ term in
Eq.~(\ref{eq:gauge_variant_coupling_form}) is completely compensate
with the pinch terms.

Hereafter we calculate both $ -(p_1 p_2) g^{\mu \nu} + p_2^{\mu}
p_1^{\nu}$ 
and $g_{\mu \nu}$ terms, and demonstrate the latter term completely
vanishes thanks to the pinch terms.

\subsection{Effective $h\gamma\gamma$ and $hZ\gamma$  vertices --- $W$ boson loop --- }
\label{sec:barr-zee_2}

Now we move on to calculate the effective vertices for $h\gamma
\gamma$ and $h Z \gamma$, which appear in Figs.~\ref{fig:BZ1} and \ref{fig:BZ2}.
In the following, $p_1$ is the momentum of the external (on-shell) photon where  $p_1^2=0$, and
$p_2$ is the momentum of the virtual gauge boson in the Barr-Zee
diagram. 
Note that the diagrams which contain both $W$ and $H^{\pm}$ in the loop
are absent in the 2HDM because $g_{\gamma W^{\pm} H^{\mp}} = g_{Z W^{\pm} 
H^{\mp}} = 0$, where $H^{\pm}$ is a physical charged scalar not a NG boson.

In this subsection,  we focus on $W$ boson loops of  the $h\gamma\gamma$ and $hZ\gamma$ effective vertices because we find 
these  are not gauge invariant as long as we keep off-shell conditions.
We work in 't Hooft-Feynman gauge 
and find the $h\gamma\gamma$ and $hZ\gamma$ effective vertices are given by
\begin{align}
  \Gamma^{\mu \nu}_{h G \gamma}(p_1, p_2)
=&
+ \frac{e}{(4\pi)^2}
\frac{1}{m_W^2}
g_{WWh} g_{WWG}
\nonumber\\ 
& \quad \times \Biggl[
\Gamma^{A}_{hG\gamma}
(p_{2}^{\mu} p_1^{\nu} - p_2 p_1 g^{\mu \nu})
+
\Gamma^{P}_{hG\gamma} (p_2^2  - m_G^2) g^{\mu \nu}
+
\Gamma^{B}_{hG\gamma} p_{2}^{\mu} p_2^{\nu}
\Biggr.
\nonumber\\ 
\Biggl. & \quad \quad \quad  +
\Gamma^{C}_{hG\gamma} [(p_1 + p_2)^2  - m_h^2] g^{\mu \nu}
\Biggr] 
.
\label{eq:hGA_effective}
\end{align}
where
\begin{align}
\Gamma^{A}_{hG\gamma}
=&
4 
\left(
-4 J_1(m_W^2) + 6J_2(m_W^2) 
+
\frac{m_G^2}{m_W^2}
( J_1(m_W^2) - J_2(m_W^2))
+
\left(
1 - \frac{1}{2} \frac{m_G^2}{m_W^2}
\right)
\frac{m_h^2}{m_W^2}
J_2(m_W^2)
\right)
\label{eq:hVA_eff_dam}
, \\ 
\Gamma^{P}_{hG\gamma}
=&
+ 3 J_1(m_W^2)   \label{GammaPhGg}
, \\ 
\Gamma^{B}_{hG\gamma}
=&
- 3 J_1(m_W^2)
+
\frac{m_G^2}{m_W^2}
( J_1(m_W^2) - J_2(m_W^2))
+
\frac{1}{2}
\frac{m_G^2}{p_2^2}
(1 - 2 J_1(m_W^2))
\nonumber\\
&+
\frac{m_G^2}{m_W^2}
\frac{(p_1 + p_2)^2}{p_2^2}
J_2(m_W^2)
, 
\end{align}
\begin{align}
\Gamma^{C}_{hG\gamma}
=&
- 
\left(
1
-
\frac{m_G^2}{m_W^2}
\right)
J_1(m_W^2)
,\qquad \qquad \qquad \qquad \qquad \qquad \qquad \qquad \qquad \qquad \qquad \qquad \quad
\end{align}
where $G$ stands for $Z$ or $\gamma$, 
 and where 
\begin{align}
 J_1(m^2)
=&
 \int_0^{1} dx
 \int_0^{1-x} dy
\frac{1}{1 - \frac{p_2^2}{m^2} x (1-x) - \frac{(p_1+p_2)^2 -
 p_2^2}{m^2} xy} 
\label{eq:def_J1}
, \\ 
 J_2(m^2)
=&
 \int_0^{1} dx
 \int_0^{1-x} dy
\frac{xy}{1 - \frac{p_2^2}{m^2} x (1-x) - \frac{(p_1+p_2)^2 -
 p_2^2}{m^2} xy} 
\label{eq:def_J2}
.
\end{align}
The explicit forms of couplings, such as $g_{WWh}$ and $g_{WWG}$,  are given in Appendix~\ref{sec:app_2HDM}.  This result is consistent with previous works,
for example in Eq.~(9) in Ref.~\cite{Leigh:1990kf}.

Although the gauge invariance requires $\Gamma^{P} = \Gamma^{B} =
\Gamma^C =0$ as we discussed in Sec.~\ref{sec:subdiagram}, it is not
satisfied in Eq.~(\ref{eq:hGA_effective}).  So we should consider the
gauge invariance for the EDM calculation carefully. As discussed in
Ref.~\cite{Leigh:1990kf}, the $\Gamma^{C}$ term does not contribute to
the EDMs.  Because this term is proportional to inverse of neutral
Higgs propagator, it can reduce neutral Higgs propagator in
Barr-Zee diagram. Then we can apply the vertex relation\footnote{We
  show this vertex relation in Appendix~\ref{WWh}.}  $\sum_h g_{\ell
    \ell h}^{A} g_{WWh} = 0$, where $g_{\ell \ell h}^{A}$ is axial-scalar
  coupling of external fermion $\ell$ with neutral Higgs bosons $h$ and 
  $\sum_h$ is summation for three neutral Higgs bosons.  The
$\Gamma^{B}$ terms do not contribute to the EDMs neither, because
these terms do not keep $\sigma^{\mu \nu} \gamma^{5}$ structure as we
discussed in Sec.~\ref{subsec:tensor}.  Then only the $\Gamma^P$
terms are problematic. Actually the $\Gamma^P$ terms vanish once we
consider the pinch contributions as will be shown.

There are many two-loop diagrams which contribute to the EDMs, as well
as the Barr-Zee diagrams. Once we calculate all the diagrams, the
result must be gauge invariant. Therefore the gauge variant terms we
discussed above should be canceled out by contributions from
non-Barr-Zee diagrams. In order to see this cancellation, we do not
need to calculate all the diagrams, but only \textit{the pinch
  contributions}.  The gauge invariance of
Eq.~(\ref{eq:hGA_effective}) would be recovered by borrowing some
terms from non-Barr-Zee diagrams.

\begin{figure}[tbp]
\begin{minipage}{0.24\hsize}
\subfigure[]{
\includegraphics[bb=0 0 200 130,
width=0.9\hsize]{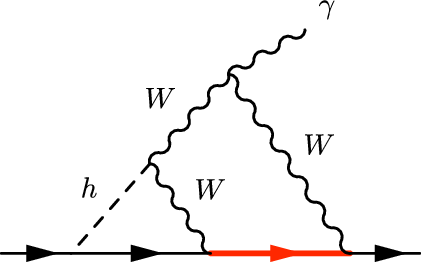}
\label{fig:pinch_a}
}  
\end{minipage}
\begin{minipage}{0.24\hsize}
\subfigure[]{
\includegraphics[bb=0 0 200 130,
width=0.9\hsize]{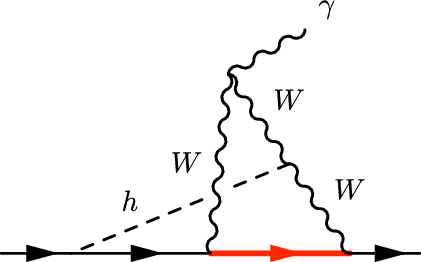}
\label{fig:pinch_b}
}  
\end{minipage}
\begin{minipage}{0.24\hsize}
\subfigure[]{
\includegraphics[bb=0 0 200 130,
width=0.9\hsize]{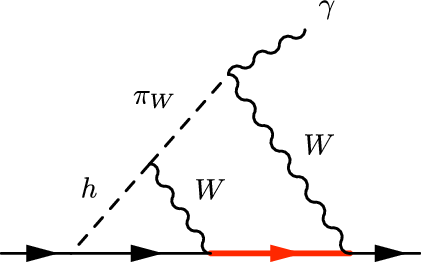}
\label{fig:pinch_c}
}  
\end{minipage}
\begin{minipage}{0.234\hsize}
\subfigure[]{
\includegraphics[bb=0 0 200 130,
width=0.9\hsize]{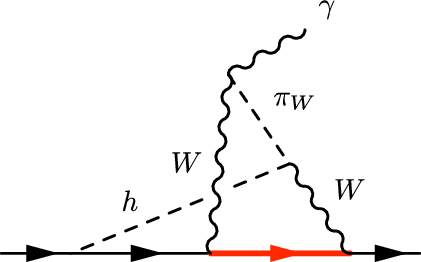}
\label{fig:pinch_d}
}  
\end{minipage}
\caption{Diagrams containing the pinch terms for the effective
  $h\gamma \gamma$ and $h Z \gamma$ vertices. We pinch the fermion
  lines shown with red color. The dashed lines attached to the fermion
  lines are the physical scalars, and those not attached are would-be
  NG bosons.  }
 \label{fig:pinch_1}
%
%
%
%
\begin{center}
\begin{minipage}{0.23\hsize}
\subfigure[]{
\includegraphics[bb=0 0 200 130,
width=0.9\hsize]{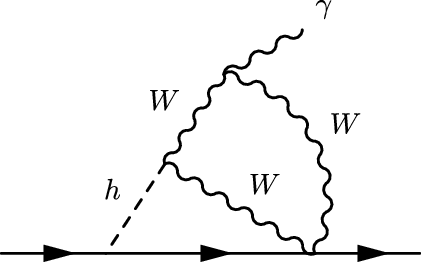}
\label{fig:pinched_b1}
}  
\end{minipage}
\quad
\begin{minipage}{0.23\hsize}
\subfigure[]{
\includegraphics[bb=0 0 200 130,
width=0.9\hsize]{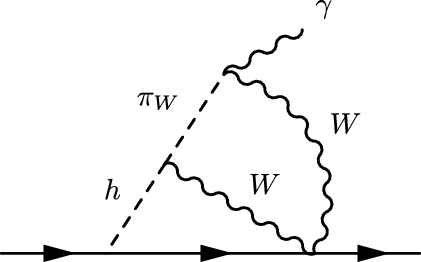}
\label{fig:pinched_b2}
}  
\end{minipage}
\caption{ Diagrams (a) and (b) are the diagrams after pinched away the
  red lines.  Figs.~\ref{fig:pinch_a} and \ref{fig:pinch_b} and
  Figs.~\ref{fig:pinch_c} and \ref{fig:pinch_d} become diagrams (a)
  and (b), respectively.  }
 \label{fig:pinch_2}
 \end{center}
 \end{figure}

For this purpose, we calculate the diagrams shown in
Fig.~\ref{fig:pinch_1}.  These diagrams contain derivative couplings
which are contracted with the gamma matrices by the Lorentz
index.  Then, these terms cancel out internal fermion
propagators.  We pick up the terms in which the fermion lines with red
color in Fig.~\ref{fig:pinch_1} are canceled out, and they are just
the pinch contributions which make Barr-Zee contributions gauge
invariant.  These terms are schematically shown in
Fig.~\ref{fig:pinch_2}.
 In 't Hooft-Feynman gauge\footnote{In other gauge, we would need
   other diagrams as well as shown in Fig.~\ref{fig:pinch_1}.}, we
 find
\begin{align}
\left. 
\text{Fig.~\ref{fig:pinch_1}}
\right|_{\textrm{pinch}}
=&
\sum_h
\int_{\ell}
i\tilde{\Gamma}^{\mu \nu}_{hG\gamma} (-q,\ell)
\frac{i}{(q-\ell)^2 - m_h^2}
\frac{-i g_{\nu \rho }}{\ell^2 - m_G^2}
(-i \gamma^{\rho} g_{G \ell \ell })
\frac{i}{\slashchar{p} + \slashchar{q} - \slashchar{\ell} - m_f}
(-i g_{\ell \ell h})
,
\label{eq:pinch_pre}
\end{align}
where
\begin{align}
\tilde{\Gamma}^{\mu \nu}_{h G \gamma} (p_1,p_2)
=&
-
g^{\mu \nu}
 3 \frac{e}{(4\pi)^2} 
\frac{g_{WWh}}{m_W^2}
g_{WWG}
(p_2^2 - m_G^2)
J_1(m_W^2)
\label{eq:pinch_G}
.
\end{align}
Here, $J_1$ is given in Eq.~(\ref{eq:def_J1}), $g_{G \ell \ell}$ and 
$g_{\ell \ell h}$ are couplings of external fermion $\ell$ with gauge and Higgs bosons,
respectively,
 and $\int_{\ell} = \int d^4 \ell /( 2\pi)^4$.
Since Fig.~\ref{fig:BZ1} with effective
$h G \gamma$ vertices is calculated as
\begin{align}
\textrm{Fig.~\ref{fig:BZ1}}
=&
\sum_h
 \int_{\ell}
i \Gamma^{\mu \nu}_{hG\gamma} (-q, \ell) 
\frac{-i g_{\nu \rho}}{\ell ^2 - m_G^2}
\frac{i}{(\ell - q)^2 -m_h^2}
\left(
- i \gamma^{\rho} g_{G \ell \ell}
\right)
\frac{i}{\slashchar{p} + \slashchar{q} - \slashchar{\ell} - m_f}
(-ig_{\ell \ell h})
,
\label{eq:to_find_pinch}
\end{align}
we find
Eq.~(\ref{eq:pinch_G}) is nothing but
parts of the effective vertices
by comparing Eq.~(\ref{eq:to_find_pinch}) to Eq.~(\ref{eq:pinch_pre}), 
and cancels the second term in Eq.~(\ref{eq:hGA_effective}) ($\Gamma^P$) which is gauge
variant term.
In other words, the pinch term certainly cancels the gauge variant
term and make the effective coupling gauge invariant.

After adding the pinch terms, we finally find the gauge invariant
$W$ loop contributions to the effective $h\gamma\gamma$ and $hZ\gamma$
vertices for the Barr-Zee diagrams,  
\begin{align}
 \Gamma^{\mu \nu}_{h G \gamma} (p_1, p_2)
=&
\frac{e}{(4\pi)^2}
\frac{1}{m_W^2}
g_{WWh} g_{WWG}
\Gamma^{A}_{hG \gamma}
\left(
 -(p_1 p_2) g^{\mu \nu} + p_2^{\mu} p_1^{\nu}
\right),
\end{align}
where
$\Gamma^{A}_{h G \gamma}$ is given in Eq.~(\ref{eq:hVA_eff_dam}).

\subsection{Effective $h\gamma\gamma$ and $hZ\gamma$  vertices --- fermion, $H^{\pm}$ loop --- }
For the Barr-Zee diagram calculation, we need other contributions to
effective $h\gamma\gamma$ and $hZ\gamma$ vertices. 
We calculate the fermion loop contribution to the effective
$h\gamma\gamma$ and $hZ\gamma$ vertices.
We denote the fermion as $f$. Note
that they are independent from the gauge
fixing terms. Hence $\Gamma^{P}$ and $\Gamma^{D}$ in
Eq.~(\ref{eq:gauge_variant_coupling_form}) are zero. We find
$\Gamma$ and $\Gamma_5$ defined in
Eq.~(\ref{eq:gauge_inv_coupling_form}) are
\begin{align}
 \Gamma_{hG\gamma}(p_1,p_2)
=&
+
\frac{N_c}{(4 \pi)^2}
2 e Q_f
g_{ffh}^{V}
(g_{Gff}^{L} + g_{Gff}^{R})
\frac{2}{m_f}
\left(
J_1(m_f^2)
-
4
J_2(m_f^2)
\right)
\label{eq:fermion_sub-diagram1}
, \\ 
 \Gamma^5_{hG\gamma}(p_1,p_2)
=&
+
\frac{N_c}{(4 \pi)^2}
2 e Q_f
(i g_{ffh}^{A})
(g_{Gff}^{L} + g_{Gff}^{R})
\frac{2}{m_f}
J_1(m_f^2)
,
\label{eq:fermion_sub-diagram2}
\end{align}
where $N_c$ is the color factor, for example $N_c = 3$ for the top quark
loop, $Q_f$ is the QED charge of the fermion in the loop, for example $Q_f =
2/3$ for the top quark loop.

The diagrams with the charged Higgs boson loop are also independent from the gauge
fixing terms. Thus $\Gamma^{P}$ and $\Gamma^{D}$ in
Eq.~(\ref{eq:gauge_variant_coupling_form}) are zero. We find
$\Gamma$ and $\Gamma_5$ defined in
Eq.~(\ref{eq:gauge_inv_coupling_form}) are
\begin{align}
 \Gamma_{hG\gamma}(p_1, p_2)
=&
 -4 
\frac{1}{(4\pi)^2}
e
g_{H^{+}H^{-} h}
g_{GH^{+}H^{-}}
\frac{2}{m_{H^{\pm}}^2}
J_2(m_{H^{\pm}}^2)
\label{eq:scalar_eff_vertex}
, \\ 
\Gamma^5_{hG\gamma} (p_1, p_2) =& 0
.
\end{align}

\subsection{Effective $H^{\mp} W^{\pm}\gamma$  vertices --- $W$, $H^{\pm}$ loop --- }

The effective vertices for $H^{\mp} W^{\pm} \gamma$, shown in
Figs.~\ref{fig:BZ3} and \ref{fig:BZ4}, are also necessary to calculate
the all the Barr-Zee diagrams. Note that these Barr-Zee contributions
have not been studied in the literature yet, and we calculate 
for the first time them. To find a gauge invariant set for the Barr-Zee diagrams, we need
to take into account for the pinch contributions. Calculations are
tedious and long, so the details are given in
Appendix~\ref{sec:HWA_coupling_appendix}. After summing up all terms
which are relevant for the EDM calculations, we find the following
gauge invariant effective vertex:
\begin{align}
\Gamma^{\mu \nu}_{H^{-} W^{+} \gamma}(p_1,p_2)
&=
+
\frac{1}{(4\pi)^{2}}
\left(
p_{2 \mu} p_{1 \nu} - p_2 p_1 g_{\mu \nu}
\right)
\nonumber\\ 
&
\times
\Biggl(
+
\sum_h
e g_{W^{+} H^{-} h} g_{WWh}
 \int_{0}^{1}\!\!\!dz \int_{0}^{1-z}\!\!\!\!\!\!\!\!dy \
\frac{
-2 yz - 4z + 4 - \frac{m_{H^{\pm}}^2 - m_h^2}{m_W^2} 2 y z
}{
m_W^2 (1-z) + m_h^2 z - p_2^2 z (1-z) -2 p_1 p_2 yz
}
\nonumber \\ 
&
\qquad
-
\sum_h
e g_{W^{+} H^{-} h} g_{H^{+}H^{-}h}
 \int_{0}^{1}\!\!\!dz \int_{0}^{1-z}\!\!\!\!\!\!\!\!dy \
\frac{
4 y z
}{
m_{H^{\pm}}^2 (1-z) + m_h^2 z - p_2^2 z (1-z) -2 p_1 p_2 yz
}
\Biggr)
,
\end{align}
\begin{align}
\Gamma^{\mu \nu}_{H^{+} W^{-} \gamma}(p_1,p_2)
=&
\left(
\Gamma^{\mu \nu}_{H^{-} W^{+} \gamma}(p_1,p_2)
\right)^{*}
.
\end{align}
Here we have already omitted the terms which do not contribute to the EDM
calculations.\footnote{These terms do not contribute to the on-shell
$H^{-} \to W^{-} \gamma$ process neither.}

There might also be fermion loops in the effective $H^{\mp}
W^{\pm}\gamma$ vertices. It is found that the fermion loops in the
effective $H^{\mp} W^{\pm}\gamma$ vertices do not contribute to the
EDMs if we consider only the CP phase in the Higgs potential in
2HDMs. While another CP phase is present in the Cabibbo-Kobayashi-Maskawa (CKM)  matrix, the contributions to
the EDMs should be much suppressed due to the GIM mechanism. Then, we
do not calculate the fermion loop contributions to the effective
$H^{\mp} W^{\pm}\gamma$ vertices in this paper.

\section{EDM from Barr-Zee diagram}
\label{sec:Barr-Zee}

In this section we calculate diagrams in Fig.~\ref{fig:Barr-Zee}.
The EDM, $d_{\ell}$, for fermion $\ell$ is defined through
\begin{align}
 {\cal H}_{\text{eff}}
=&
i
\frac{d_{\ell}}{2}
\overline{\psi}_{\ell}
\sigma_{\mu \nu} \gamma_5
\psi_{\ell}
F^{\mu \nu}
\label{eq:def_of_EDM}
,
\end{align}
where
\begin{align}
 \sigma_{\mu \nu}
=&
 \frac{i}{2}
[\gamma_{\mu}, \gamma_{\nu}]
.
\end{align}
Once we get the gauge invariant effective vertex whose tensor structure
is given in Eq.~(\ref{eq:gauge_inv_coupling_form}), we
find the neutral Higgs boson contributions to 
$d_{\ell}$ as
\begin{align}
\left(
 d_{\ell}
\right)^{
 \text{Fig.~\ref{fig:BZ1}}}_{+
 \text{Fig.~\ref{fig:BZ2}}
}
=&
\frac{1}{2}
\sum_{G = Z, \gamma}
\sum_{h}
\left(
g_{G \ell \ell}^{L} + g_{G \ell \ell}^{R}
\right)
\int_{\ell}
\left(
i g_{\ell \ell h}^{A} \Gamma_{h G \gamma} (0, \ell) 
+
g_{\ell \ell h}^{V} \Gamma^5_{h G \gamma} (0, \ell) 
\right)
\frac{1}{\ell^2 - m_G^2}
\frac{1}{\ell^2 - m_h^2}
.
\label{eq:df_formula}
\end{align}
where $g^{L(R)}_{G \ell \ell}$ is for couplings of left(right)-handed
fermion $\ell$ with gauge boson $G$, and $g^{V(A)}_{\ell \ell h}$ is for
(axial) scalar couplings with scalar boson $h$.  Here we keep only the
leading term for $p$ and $q$, and ignore mass term in the fermion
propagator, and we have used a relation, $ \epsilon^{\mu \nu \alpha
  \beta} \gamma_{\alpha} \gamma_{\beta} = -i \gamma^{5} [
\gamma^{\mu}, \gamma^{\nu}] .  $

Note that we work in 't Hooft-Feynman gauge in
Eq.~(\ref{eq:df_formula}). If we work in other gauge, gauge boson
propagators contain the terms that proportional to $\ell_{\nu}$ and
contract with the effective vertices. Since $\Gamma^{\mu \nu}(-q,
\ell) \ell_{\nu} = 0$, the terms proportional to $\ell_{\nu}$ in the
gauge boson propagators always vanish.  Therefore the Barr-Zee
diagrams are gauge invariant {\it as long as} the effective vertices
are gauge invariant.

In the similar manner, we find the charged Higgs boson contribution to the leptonic EDMs as
\begin{align}
\left(
d_{\ell}
\right)^{
 \text{Fig.~\ref{fig:BZ3}}
}_{+
 \text{Fig.~\ref{fig:BZ4}}}
=&
\frac{1}{2 \sqrt{2}}
\frac{e}{s}
\int_{\ell}
\frac{1}{\ell^2 - m_W^2}
\frac{1}{\ell^2 - m_{H^{\pm}}^2} i
{\rm Im}
\left(
g_{ \bar{\nu} e H^{+}}^{R}
\Gamma_{H^{-} W^{+} \gamma}(0, \ell)
\right)
.
\label{chargedhiggs_edm}
\end{align}
Here we have used the following relations,
\begin{align}
 g_{\bar{e} \nu H^{-} }^{L} 
=&
\left(
 g_{\bar{\nu} e H^{+} }^{R}  
\right)^{*}
,
\\
\Gamma_{H^{+} W^{-} \gamma}(0, \ell)
=&
\left(
\Gamma_{H^{-} W^{+} \gamma}(0, \ell)
\right)^{*}
.
\end{align}
The charged Higgs contributions to the up-type and down-type
quark EDMs are derived by replacing
$g_{\bar{\nu}e H^{+}}^R \Gamma_{H^{-}W^{+}\gamma}$ in Eq.~(\ref{chargedhiggs_edm})
by 
$g_{ \bar{d} uH^{-} }^R \Gamma_{H^{+}W^{-}\gamma}$ 
and 
$g_{ \bar{u}d H^{+}}^R \Gamma_{H^{-}W^{+}\gamma}$,
respectively.
We denote $s$ and $c$ as sine and cosine of the Weinberg angle, respectively,  in the following.

The chromo-EDMs (cEDMs) also contribute to the neutron EDM.  Its
definition is similar to Eq.~(\ref{eq:def_of_EDM}), replace $F_{\mu
  \nu}$ by $g_s G_{\mu \nu}$,
\begin{align}
{\cal H}_{\text{eff}}
=&
i
\frac{d_{q}^c}{2}
\overline{q}
g_s \sigma_{\mu \nu} \gamma_5
G^{\mu \nu}
q
,\label{eq:cEDM}
\end{align}
where $g_s$ and $G_{\mu \nu}$ are the QCD coupling and the field
strength of the gluon, respectively.

The formulae of EDMs include complicated functions. 
Here, we show the approximated expressions in the {\it
decoupling limit} for qualitative discussion, while
all plots are drawn by using the exact formulae.
The exact formula are given in Appendix~\ref{sec:EDMdetails}.
In the decoupling limit all the non-SM particles are degenerated,
heavier than the electroweak scale, and decoupled from the SM sector. We
can take such a limit by $M \to \infty$ where $M$ is defined in
Eq.~(\ref{eq:def_M}). 

Since the results depend on the Yukawa
structure, we introduce the following notation to simplify our expressions:
\begin{align}
 {\cal G}^A_{x}
=&
\bordermatrix{
     & \text{Type-I} & \text{Type-II} & \text{Type-X} & \text{Type-Y} \cr
u/c/t    & 1 & 1 & 1 & 1 \cr
d/s/b    & -1 & \tan^2\beta & -1 & \tan^2\beta \cr
e/\mu/\tau & -1 & \tan^2\beta & \tan^2\beta & -1  \cr
}
, \\
 {\cal S}_{x}
=&
\bordermatrix{
     &    \cr
u/c/t    & -1  \cr
d/s/b    &  1  \cr
e/\mu/\tau &  1  \cr
}
,
\end{align}
where index $A$ represents type of the model, and index $x$ is for flavor.

It is found that the EDMs for fermion $\ell$ in the decoupling limit are approximated to be
\begin{align}
\left(
 \frac{d_{\ell}}{e}
\right)_{W}
\simeq&
-
X
{\cal G}^{A}_{\ell}
\times
\Biggl(
e 
\left(
15
+
2 
\ln\left( \frac{M}{{\rm TeV}} \right)
\right)
\left(
g_{\gamma \ell \ell}^L
+
g_{\gamma \ell \ell}^R
\right)
\nonumber\\
& \quad \quad \quad \quad+
g_{WWZ}
\left(
6.5
+
0.71
\ln \left(\frac{M}{{\rm TeV}} \right)
\right)
\left(
g_{Z \ell \ell}^L
+
g_{Z \ell \ell}^R
\right)
\Biggr)
,
\\
\left(
 \frac{d_{\ell}}{e}
\right)_{\textrm{top}}
\simeq&
+ 
X
\times
\Biggl(
e\left(
5.3 {\cal G}^{A}_{\ell}
+
7.6
\right)
\left(
g_{\gamma \ell \ell}^L
+
g_{\gamma \ell \ell}^R
\right)
+
e\left(
1.4 {\cal G}^{A}_{\ell}
+
2.0
\right)
\left(
g_{Z \ell \ell}^L
+
g_{Z \ell \ell}^R
\right)
\Biggr)
,
\\
\left(
 \frac{d_{\ell}}{e}
\right)_{\textrm{bottom}}
\simeq&
+ 
X
\times
\Biggl(
e\left(
0.018 {\cal G}^{A}_{\ell}
+
0.022 {\cal G}^{A}_{b}
\right)
\left(
g_{\gamma \ell \ell}^L
+
g_{\gamma \ell \ell}^R
\right) \Biggr.
\nonumber\\
&
\qquad \quad
\Biggl.+
e\left(
0.0075 {\cal G}^{A}_{\ell}
+
0.0087 {\cal G}^{A}_{b}
\right)
\left(
g_{Z \ell \ell}^L
+
g_{Z \ell \ell}^R
\right)
\Biggr)
,
\\
\left(
 \frac{d_{\ell}}{e}
\right)_{\textrm{tau}}
\simeq&
+ 
X
\times
\Biggl(
e\left(
0.024 {\cal G}^{A}_{\ell}
+
0.029 {\cal G}^{A}_{\tau}
\right)
\left(
g_{\gamma \ell \ell}^L
+
g_{\gamma \ell \ell}^R
\right) \Biggr.
\nonumber\\
&
\qquad \quad
\Biggl.+
e\left(
0.00034 {\cal G}^{A}_{\ell}
+
0.00038 {\cal G}^{A}_{\tau}
\right)
\left(
g_{Z \ell \ell}^L
+
g_{Z \ell \ell}^R
\right)
\Biggr)
,
\\
\left(
 \frac{d_{\ell}}{e}
\right)_{H^{\pm}}
\simeq&
+
X
{\cal G}^{A}_{\ell}
\times
\Biggl(
0.34 e
\left(
g_{\gamma \ell \ell}^L
+
g_{\gamma \ell \ell}^R
\right)
+
0.34
g_{ZH^{+}H^{-}}
\left(
g_{Z \ell \ell}^L
+
g_{Z \ell \ell}^R
\right)
\Biggr)
,
\\
\left(
 \frac{d_{\ell}}{e}
\right)_{HW\gamma}
\simeq&
 - X
{\cal G}^{A}_{\ell}
{\cal S}_{\ell}
\times
\left(
0.23
+0.20
\ln\left( \frac{M}{{\rm TeV}} \right)
\right)
,\\
 (d^c_q)_{\textrm{top}}
\simeq&
+ 
X
\times
g_s^2
\left(
4.0
{\cal G}^{A}_{q}
+
5.7
\right)
, \\
 (d^c_q)_{\textrm{bottom}}
\simeq&
+ 
X
\times
g_s^2
\left(
0.053
{\cal G}^{A}_{q}
+
0.065
{\cal G}^{A}_{b}
\right)
,
\end{align}
where
\begin{align}
 X
=&
\frac{1}{(4\pi)^{4}} 
\frac{m_{\ell}}{M^2}
\cos^2\beta
\lambda_5
\sin 2\phi
,
\label{eq:EDM_factor}
\end{align}
and we use $\overline{MS}$ mass of $M_Z$ scale, 
$m_e = 0.511$ MeV, $m_{\tau} = 1.75$ GeV, $m_u = 1.40$ MeV, $m_t = 170.9$ GeV, $m_d = 2.92$ MeV and $m_b = 2.94$ GeV. 
Notice that the EDMs and cEDMs are proportional to $\lambda_5 \sin
2\phi = \textrm{Im}[\lambda_5 \exp(i2\phi)]$, namely the imaginary part of the
coupling which is needed for CP violation.

It is found that the $W$ loop contributions are dominant in large
parameter region. Among the contributions from fermion loops, only the
top quark contributions are relevant in the decoupling limit as long
as $\tan\beta \lesssim 10$.  In the similar manner, we can make
approximation of cEDMs.  The diagrams with charged Higgs boson in
$h\gamma\gamma$, $hZ\gamma$, and $H^{\mp} W^{\pm} \gamma$ couplings
are smaller than the other contributions.  Note that the contributions
from $Z$ boson exchange diagrams are proportional to $\left( g_{Z \ell
    \ell}^L + g_{Z \ell \ell}^R \right)$.  Although this factor is
numerically small at electron EDM case, one must not ignore at quark
EDM case.  Actually, $Z$ boson exchange diagrams occupy $30$--$50 \%$
of all contribution at down quark EDM case.

In the decoupling limit the bottom quark and
  tau lepton contributions are small because of their small Yukawa couplings. In
  the non-decoupling region, however, these are not necessarily valid.
  Their leading contributions are given by diagrams in which heavy
  Higgs propagate, and their values are approximately ${\cal O} ( X
  {\cal G}^{A}_{\ell} {\cal G}^{A}_{b / \tau} (m_{h_3}^2 - m_{h_2}^2)
  / M^2 )$, where $h_3$ and $h_2$ is the heaviest and the
  next heaviest Higgs bosons, respectively.  These contributions are enhanced by $\tan
  ^2 \beta$ when $\tan\beta \gg1$. Thus, when $\tan\beta$ is large,
  the contribution may be sizable in the non decoupling region.

\section{Numerical results}
\label{sec:neumerical}

\begin{figure}[tb]
\begin{center}
\includegraphics[width=0.5\hsize]{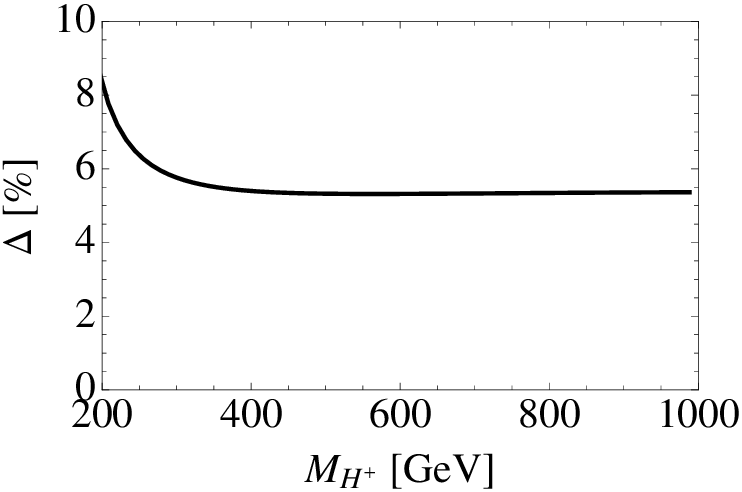}
\caption{Numerical improvement of electron EDM by the pinch contributions in the  type-II 2HDM.
We take  $\tan\beta = 10$, $\lambda_1 = \lambda_3 = \lambda_4 = \lambda_5 \sin 2 \phi =
0.5$ and require the 126~GeV Higgs mass.
}
\label{fig:diff}
\end{center}
\end{figure}

Now we evaluate the EDMs numerically. At first, in
Fig.~\ref{fig:diff}, we show the numerical improvement by the pinch
contributions. Here we consider the electron EDM in the type-II 2HDM.  The
vertical axis in the Fig.~\ref{fig:diff} is difference of the gauge
invariant EDM contribution and non-invariant one, $\Delta$, defined as
\begin{align}
 \Delta
=&
\frac{
 (d_e)_{\text{gauge non-inv.}}
-
 (d_e)_{\text{gauge inv.}}
 }{
 (d_e)_{\text{gauge inv.}}
}
,
\end{align}
where the gauge non-invariant EDM contribution $ (d_e)_{\text{gauge non-inv.}}$  is gotten by calculating only Barr-Zee diagrams \cite{Leigh:1990kf, Chang:1990sf}.
The horizontal axis is the mass of charged Higgs boson.  We take
$\tan\beta = 10$, $\lambda_1 = \lambda_3 = \lambda_4 = \lambda_5 \sin
2 \phi = 0.5$ and require the mass of lightest neutral scalar to be
126~GeV, then $\lambda_2$ is uniquely determined.  We find that the
pinch contributions are $5$--$8$\%. This is not big improvement from
the numerical point of view. However, we would like to emphasize that
our result is now gauge invariant, which must be satisfied when we
discuss observables.

\begin{figure}[tb]
\begin{minipage}{0.49\hsize}
\includegraphics[width=0.9\hsize]{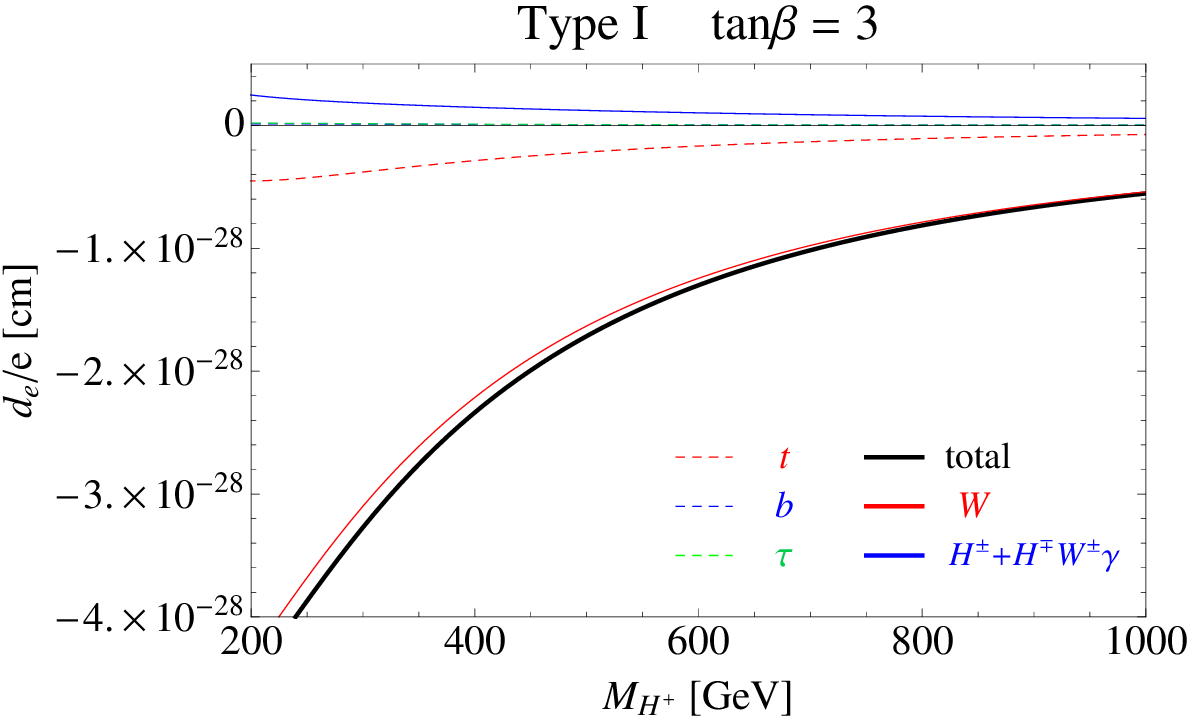} 
\end{minipage}
\begin{minipage}{0.49\hsize}
\includegraphics[width=0.9\hsize]{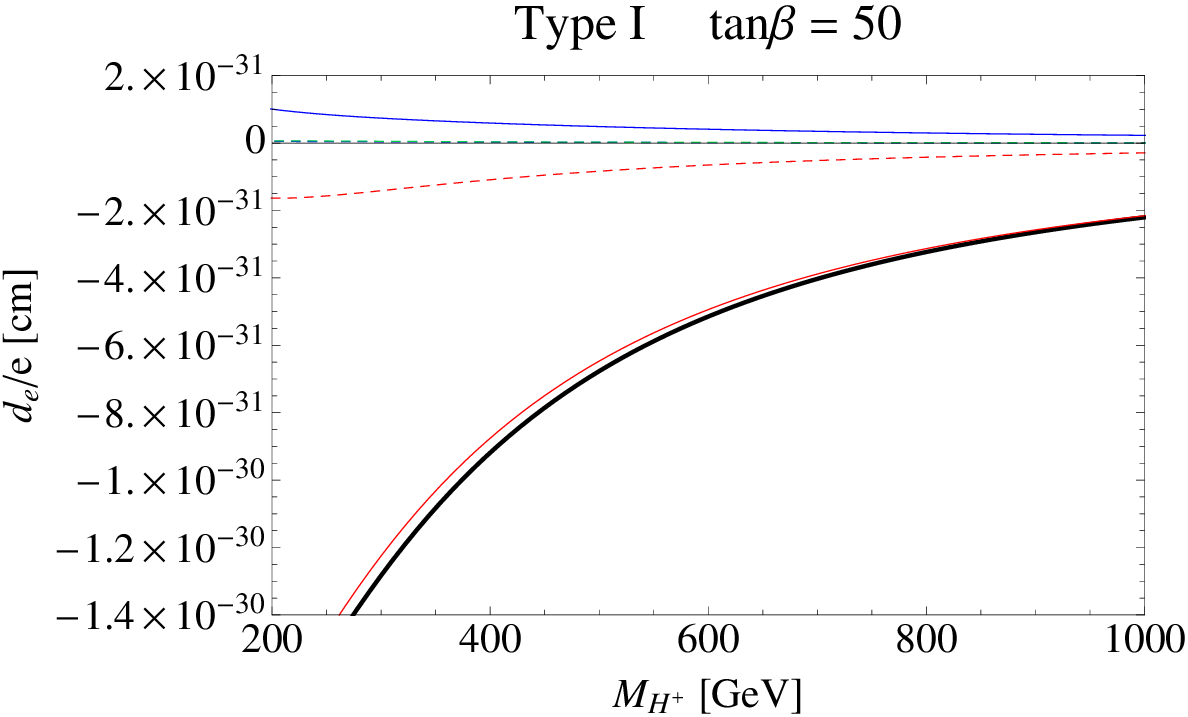} 
\end{minipage}
\caption{ Anatomy of the type-I electron EDM. Various Barr-Zee contributions to the electron EDM are shown as functions of charged Higgs mass $M_H^+$. We take $\tan \beta = 3$
  or $50$, and $\lambda_1 = \lambda_3 = \lambda_4 = \lambda_5 \sin 2
  \phi = 0.5$. The mass of lightest
  neutral Higgs is 126~GeV.  We see that $W$ loop is the dominant
  contribution.  The qualitative feature are independent from
  $\tan\beta$.  }
\label{fig:eEDM_anatomy_type1}\vskip .3in
\begin{minipage}{0.49\hsize}
\includegraphics[width=0.9\hsize]{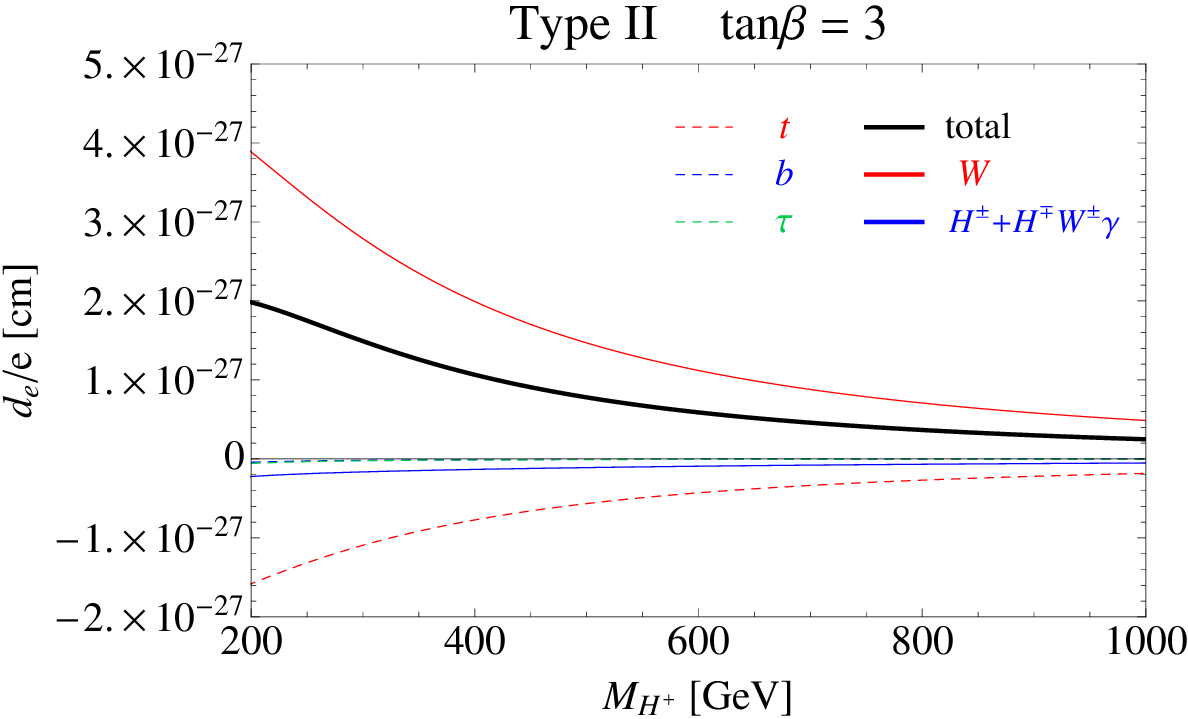} 
\end{minipage}
\begin{minipage}{0.49\hsize}
\includegraphics[width=0.9\hsize]{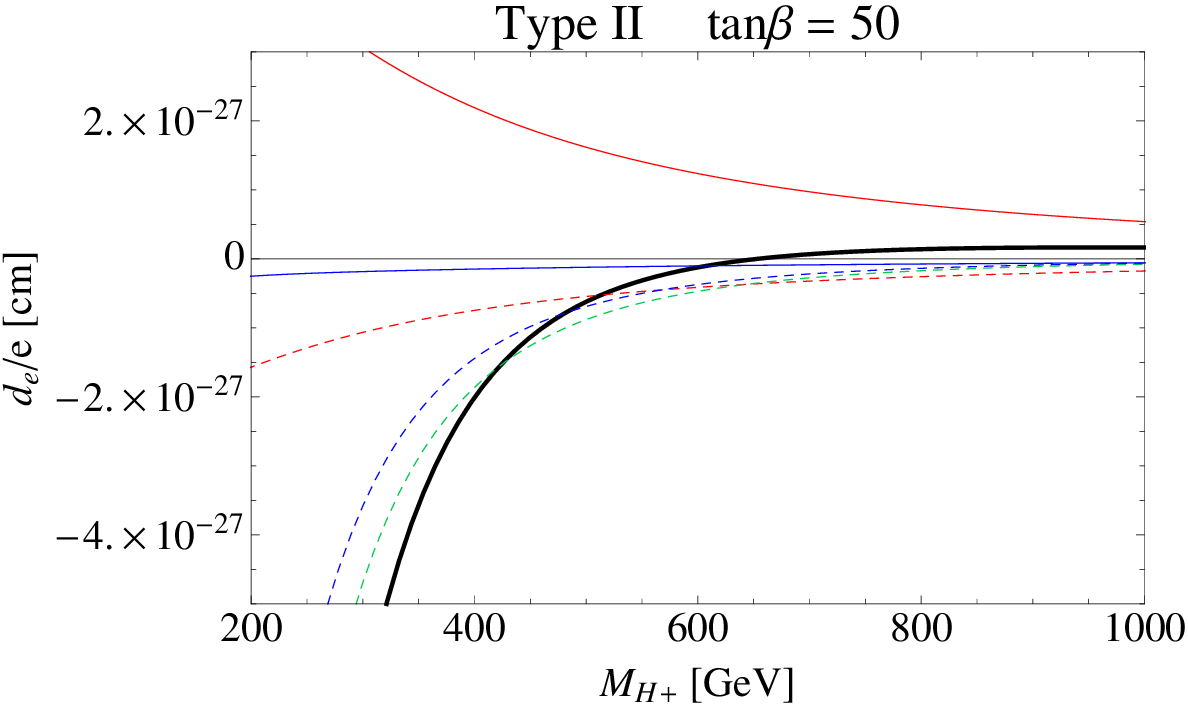}
\end{minipage}
\caption{ Anatomy of the type-II electron EDM. The input parameters
  are the same as in Fig.~\ref{fig:eEDM_anatomy_type1}.  In contrast
  of the type-I case, the qualitative feature depends on
  $\tan\beta$. For large $\tan\beta$, bottom quark and tau lepton contributions are
  sizable due to the $\tan\beta$ enhancement of their Yukawa
  couplings.  }
\label{fig:eEDM_anatomy_type2}
\end{figure}

Next, we discuss dependence of the electron EDM on the types of 2HDMs.
The contributions from each types of diagrams to the electron EDM for
type-I and II cases in Figs.~\ref{fig:eEDM_anatomy_type1} and
\ref{fig:eEDM_anatomy_type2}, respectively. Here we take $\tan \beta =
3$ or $50$, and $\lambda_1 = \lambda_3 = \lambda_4 = \lambda_5 \sin 2
\phi = 0.5$ as a benchmark. We also require the mass of lightest
neutral Higgs to be 126~GeV.

It is found that in the type-I case the $W$ boson contribution to
$h\to\gamma \gamma$ is dominant and that all contributions to the
  electron EDM are proportional to $1/\tan^2\beta$ for $\tan\beta
\gtrsim 1$.  On the other hand, the electron EDM in the type-II case
is qualitatively different from the type-I case.  Even when
$\tan\beta$ is large, the $W$ boson and top quark contributions are
not suppressed and the bottom quark and tau lepton contributions also
become dominant due to the non-decoupling effect. Since the
signs of the bottom quark and tau lepton contributions are opposite to
that of the $W$ boson, the accidental cancellation occurs in some
parameter region. Thus, the $\tan\beta$ dependence is non-trivial in
the type-II case.

\begin{figure}[tbp]
\begin{minipage}{0.24\hsize}
\subfigure[]{\includegraphics[
 width=2.0\hsize]{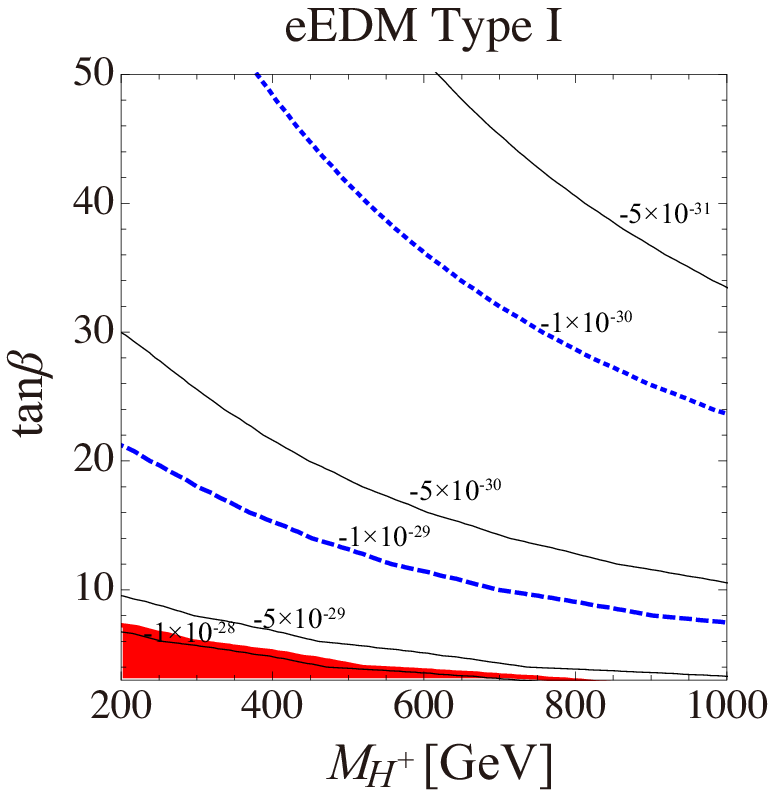} \label{fig:eEDM1}
}  
\end{minipage}\quad \quad \quad \quad \quad \quad \quad \quad \quad \quad \quad 
\begin{minipage}{0.24\hsize}
\subfigure[]{\includegraphics[
 width=2.0\hsize]{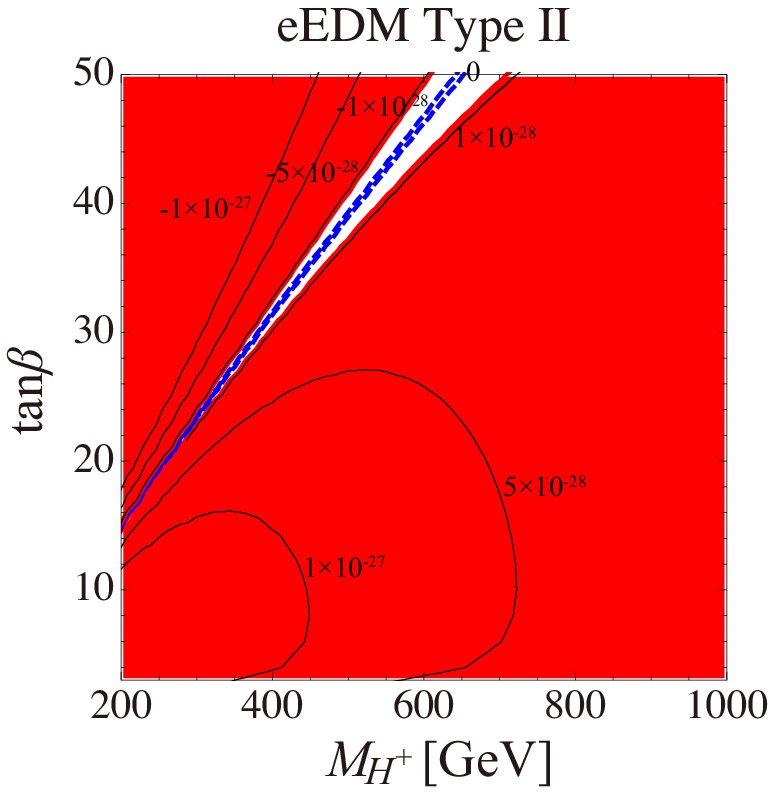} \label{fig:eEDM2}
} 
\end{minipage}\\
 \quad 
\begin{minipage}{0.24\hsize}
\subfigure[]{\includegraphics[
 width=2.0\hsize]{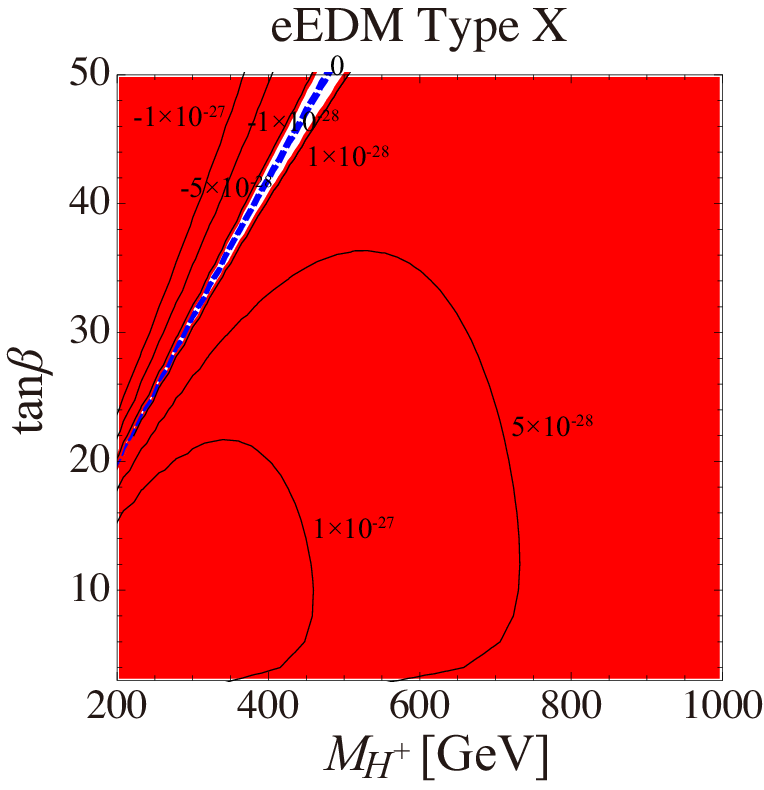} }
\end{minipage}\quad \quad \quad \quad \quad \quad \quad \quad \quad \quad \quad 
\begin{minipage}{0.24\hsize}
\subfigure[]{\includegraphics[
 width=2.0\hsize]{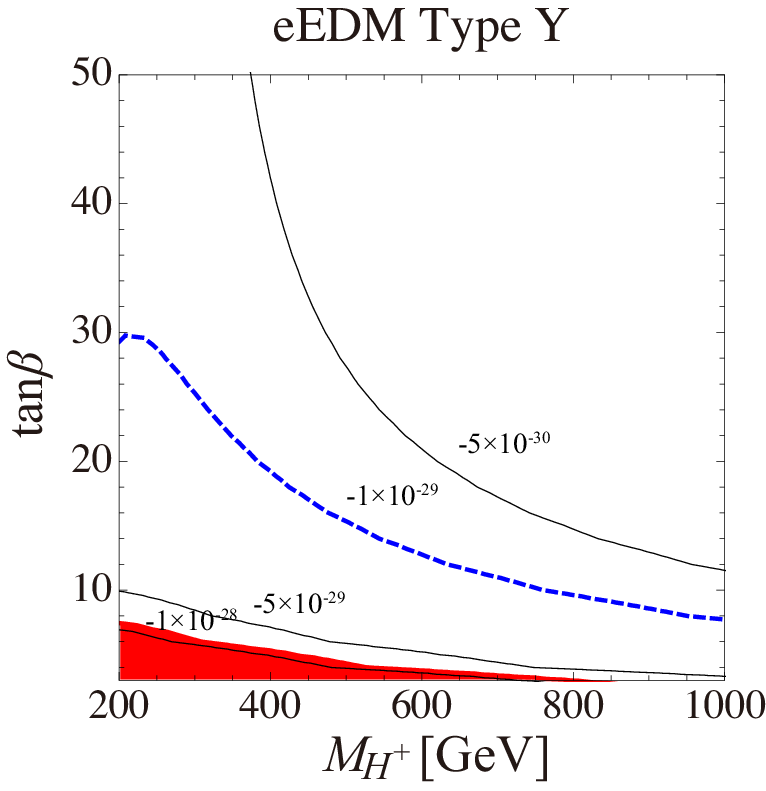} }
\end{minipage}
\caption{Electron EDM on charged Higgs boson mass and $\tan\beta$
  plane in four types of 2HDMs. We take $\lambda_1 = \lambda_3 =
  \lambda_4 = \lambda_5 \sin 2 \phi = 0.5$ and $\lambda_2 = 0.25$. The
  regions filled with red color show the current
  bound~\cite{Baron:2013eja}. The blue dashed lines are the future
  prospects given in Table~\ref{tab:EDM_future}.  }
\label{fig:eEDM}
\end{figure}

In Figs.~\ref{fig:eEDM}, the electron EDM is shown in four types of
2HDMs as functions of $\tan\beta$ and charged Higgs boson mass. 
We take $\lambda_1 = \lambda_3 = \lambda_4 = \lambda_5 \sin 2 \phi = 0.5$
and $\lambda_2 = 0.25$.  The regions filled with red color in
the figures show the excluded regions by the latest  
upper bound on electron EDM, which is derived by the ACME
experiment, 
\begin{align}
 |d_e| <& 8.7 \times 10^{-29} e \text{ cm } (90\%~\text{CL
\cite{Baron:2013eja}})
.
\end{align}
The blue dashed lines are the future prospects given in
Table~\ref{tab:EDM_future}.
\begin{table}[h]
\centering 
\begin{tabular}{|c|c|}
\hline
 experiments
&  sensitivities on $d_e$ 
\\ 
\hline
 Fr \cite{Sakemi:2011zz}
&  $1 \times 10^{-29} e$ cm
\\ 
\hline
 YbF molecule \cite{Kara:2012ay}
&  $1 \times 10^{-30} e$  cm 
\\ 
\hline
 WN ion \cite{Kawall:2011zz}
&  $1 \times 10^{-30} e$  cm
\\
\hline
\end{tabular}
\caption{Future prospects on electron EDM.}
\label{tab:EDM_future}
\end{table}

The electron EDM in the type-X and Y models has similar behavior
  to the type-II and I ones, respectively, because leptons couple to
  $H_2$ in type-I and Y models, and to $H_1$ in type-II and X models.
  We find that type-II and type-X 2HDMs are strongly constrained by
  the recent ACME experimental result, except for regions where the
  cancellation among diagrams occurs, as shown in
  Fig.~\ref{fig:eEDM}. Furthermore, the future experiments could cover
  wide parameter regions with charged Higgs mass smaller than 1~TeV
  even in type-I and Y cases.

Next let us consider the neutron EDM. Even when the Peccei-Quinn mechanism
\cite{Peccei:1977hh}
is operative, the neutron EDM is generated by higher-dimensional
CP-violating operators in QCD, such as quark EDMs and also cEDMs with
mass dimension up to 5. The neutron EDM is evaluated from the up and
down quarks EDM and cEDM with the QCD sum rules
\cite{Pospelov:2000bw, Hisano:2012sc, Fuyuto:2013gla}. The evaluation
still ${\cal O}(1)$ uncertainties from the excited state contribution to the
correlation function \cite{Pospelov:2000bw}, and also from input
parameters \cite{Hisano:2012sc}. In this paper we use the result in Ref.~\cite{Fuyuto:2013gla} since it gives more conservative prediction for the neutron EDM,
\begin{align}
 d_n
=&
0.79 d_d
-0.20 d_u
+e (0.59 d_d^{c} + 0.30 d_u^{c})
.
\end{align}
Here, the Peccei-Quinn mechanism is assumed. 

\begin{figure}[tbp]
\begin{minipage}{0.49\hsize}
\includegraphics[width=0.9\hsize]{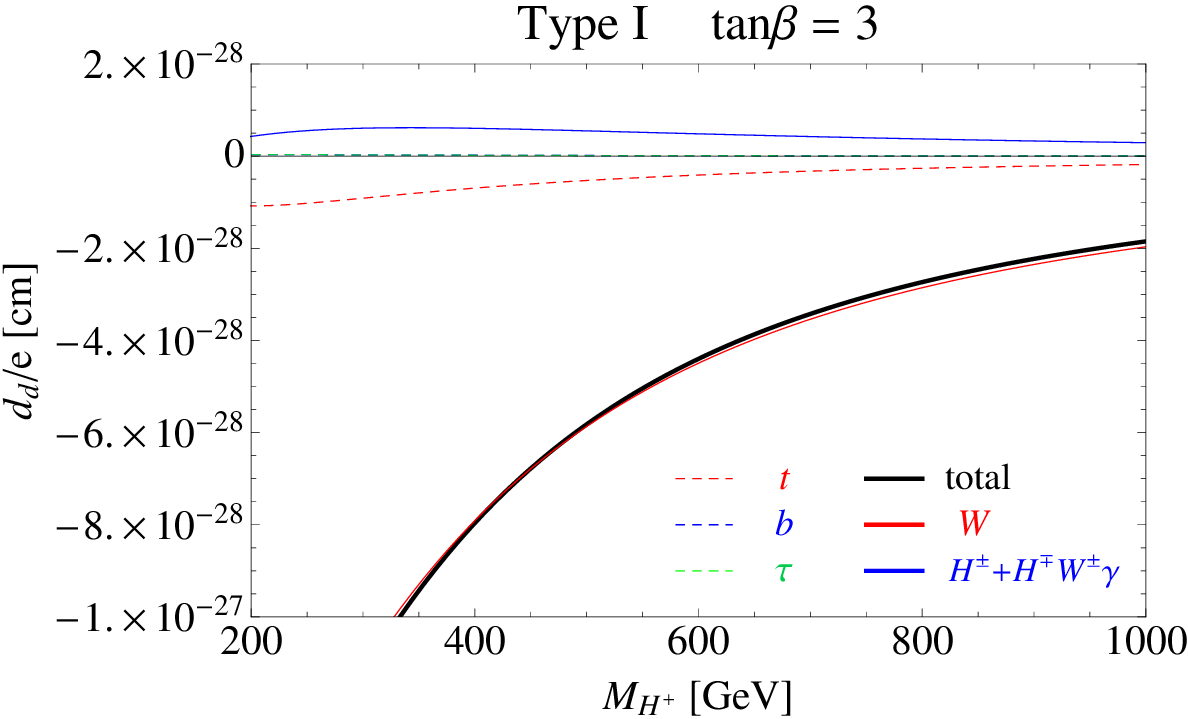} 
\end{minipage}
\begin{minipage}{0.49\hsize}
\includegraphics[width=0.9\hsize]{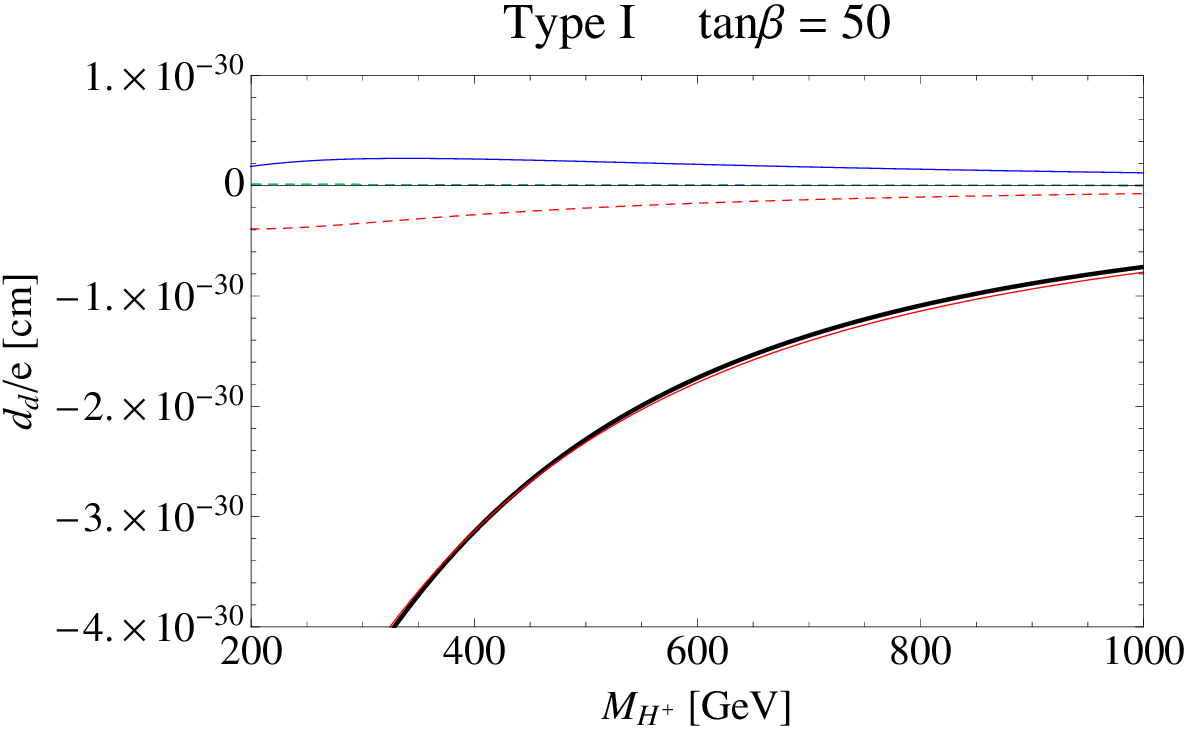} 
\end{minipage}
\\
\begin{minipage}{0.49\hsize}
\includegraphics[width=0.9\hsize]{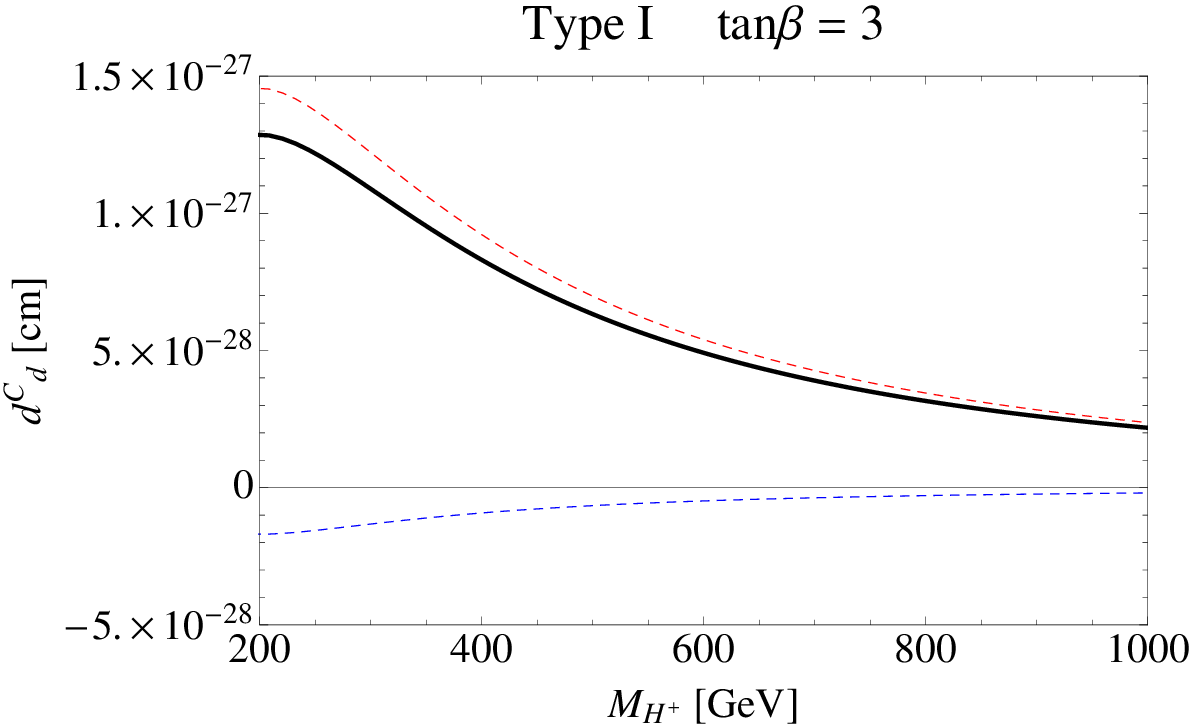} 
\end{minipage}
\begin{minipage}{0.49\hsize}
\includegraphics[width=0.9\hsize]{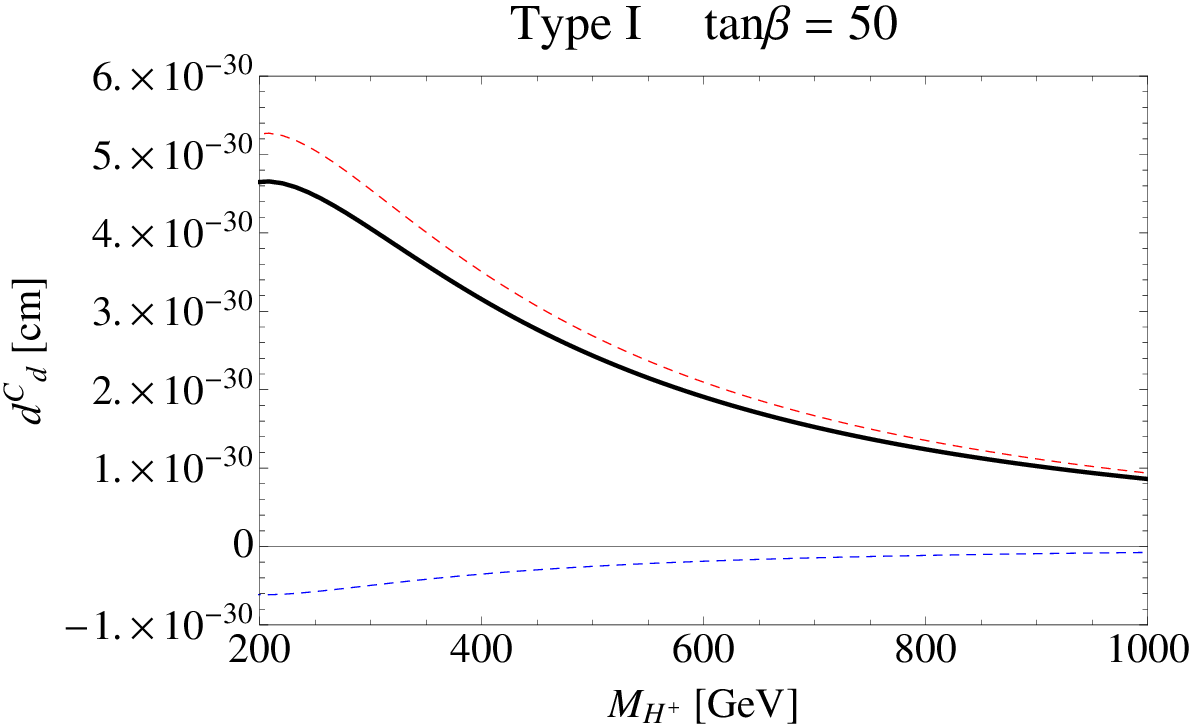}
\end{minipage}
\caption{ Anatomy of the type-I down quark EDM and cEDM.  
Various Barr-Zee contributions to the EDM and cEDM are shown as functions of charged Higgs mass $M_H^+$. We take
  $\tan\beta=3$ and 50. Other input parameters are the same as in
  Fig~\ref{fig:eEDM_anatomy_type1}.  We see that $W$ and top give
  dominant contributions to EDM and cEDM, respectively.  }
\label{fig:dEDM_anatomy_type1}\vskip .3in
\begin{minipage}{0.49\hsize}
\includegraphics[width=0.9\hsize]{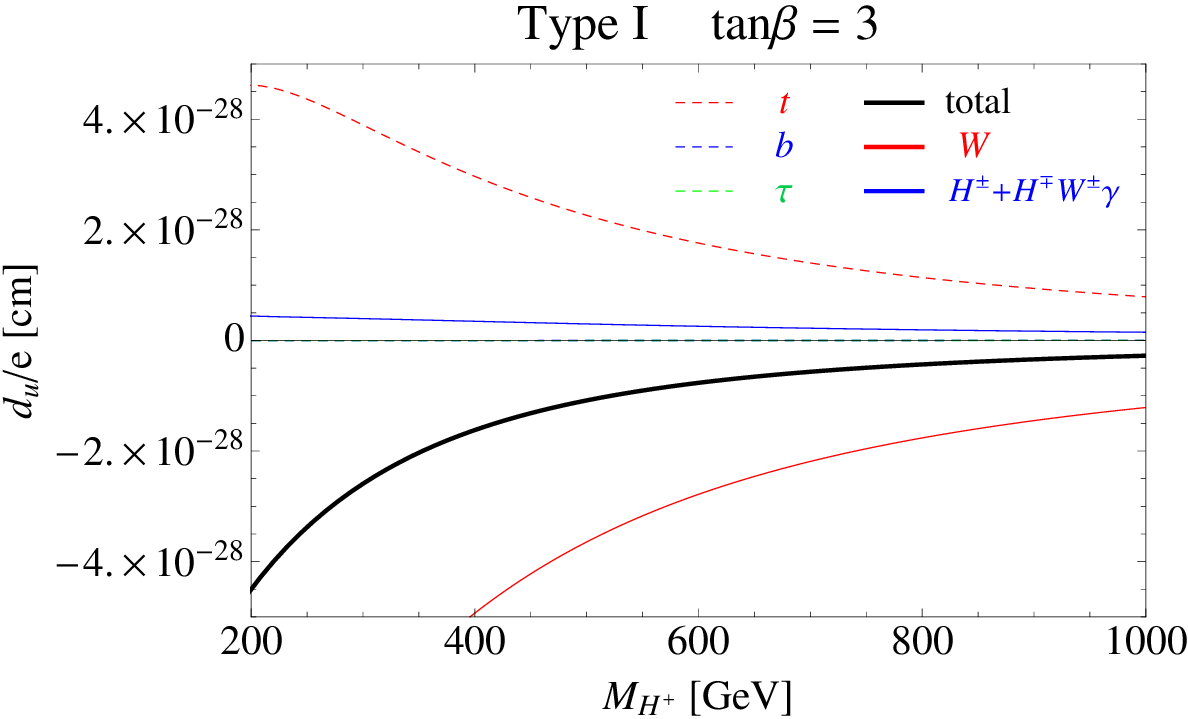}
\end{minipage}
\begin{minipage}{0.49\hsize}
\includegraphics[width=0.9\hsize]{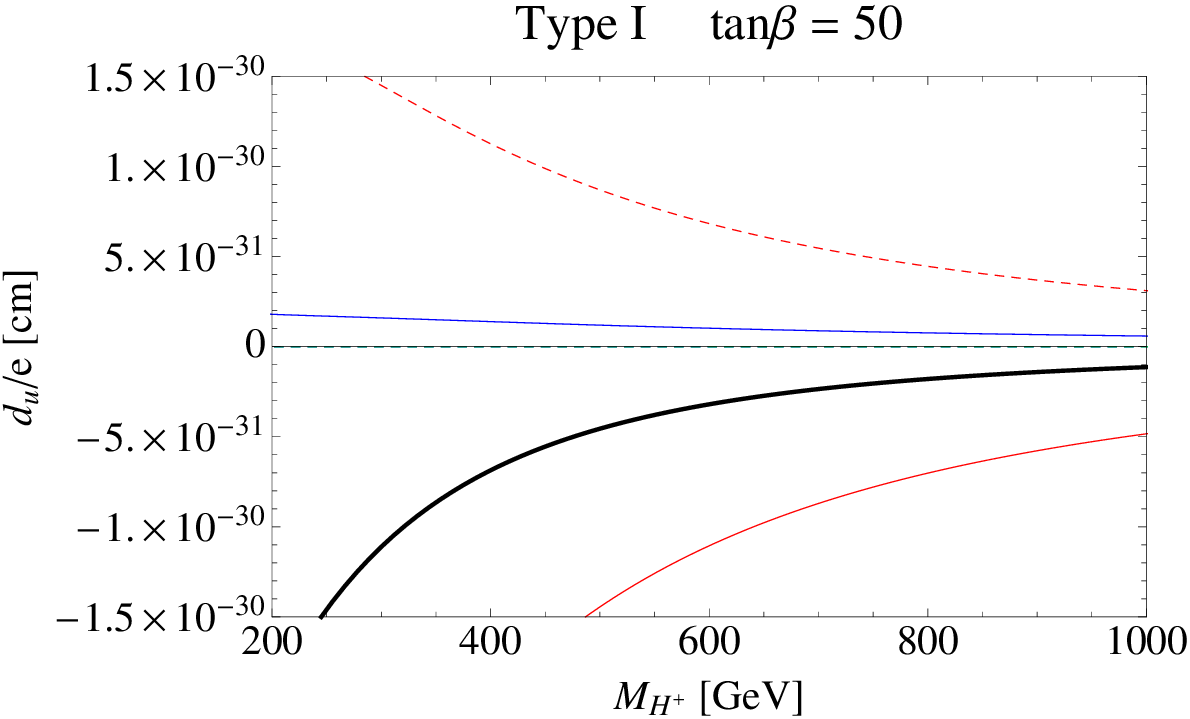} 
\end{minipage}
\\
\begin{minipage}{0.49\hsize}
\includegraphics[
 width=0.9\hsize]{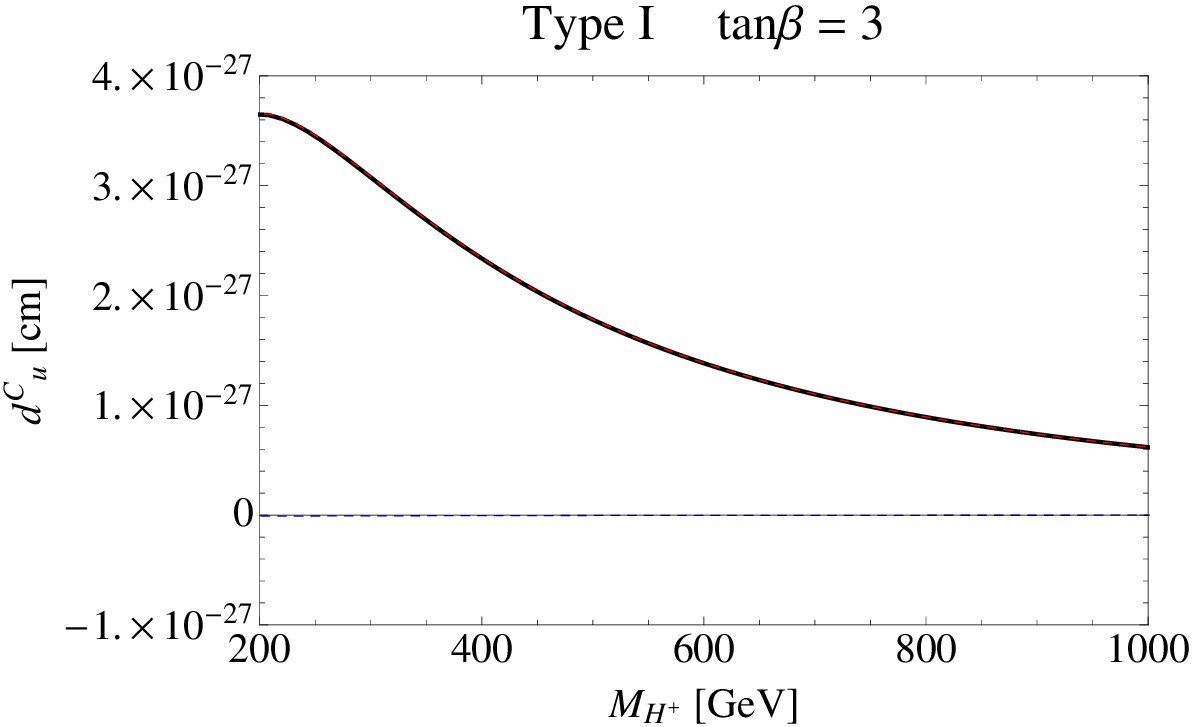}
\end{minipage}
\begin{minipage}{0.49\hsize}
\includegraphics[width=0.9\hsize]{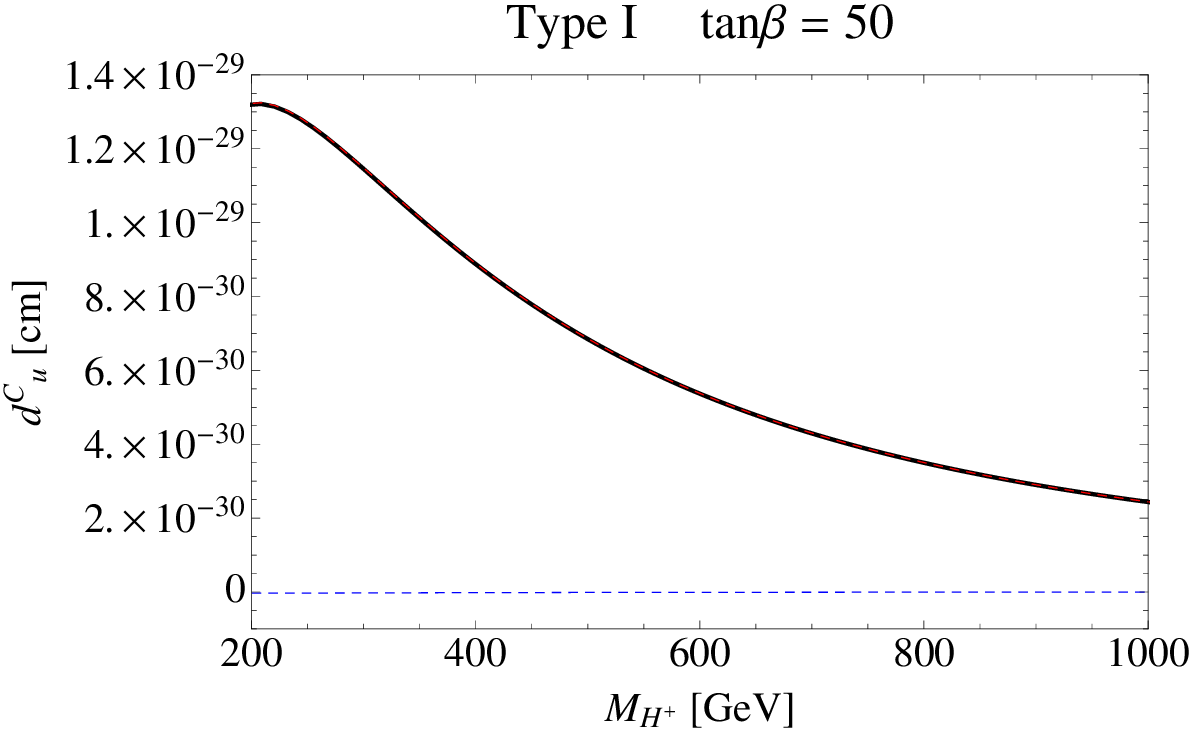} 
\end{minipage}
\caption{ Anatomy of the type-I up quark EDM and cEDM.  We
  take$\tan\beta=3$ and 50.  Other input parameters are the same as in
  Fig~\ref{fig:eEDM_anatomy_type1}.  We see that $W$ and top give
  dominant contributions.  }
\label{fig:uEDM_anatomy_type1}
\end{figure}

Before going to evaluate the neutron EDM, we discuss behaviors of the quark EDMs
and cEDMs in the 2HDMs. We plot the contributions from each types of
diagrams to the down and up quark EDMs and cEDMs in the type-I case in
Figs.~\ref{fig:dEDM_anatomy_type1} and
\ref{fig:uEDM_anatomy_type1}. The input parameters are the same as in
Fig~\ref{fig:eEDM_anatomy_type1}. We see that the $W$ boson and top
quark contributions give the dominant contributions to the EDMs and
cEDMs, respectively, and the $\tan\beta$ dependence is
$1/\tan^2\beta$, as expected from Eq.~(\ref{eq:EDM_factor}). It is
found that the sizes of cEDMs and EDMs are comparable to each others
so that both contributions have to be included in evaluation of the
neutron EDM.

\begin{figure}[tbp]
\begin{minipage}{0.49\hsize}
\includegraphics[width=0.9\hsize]{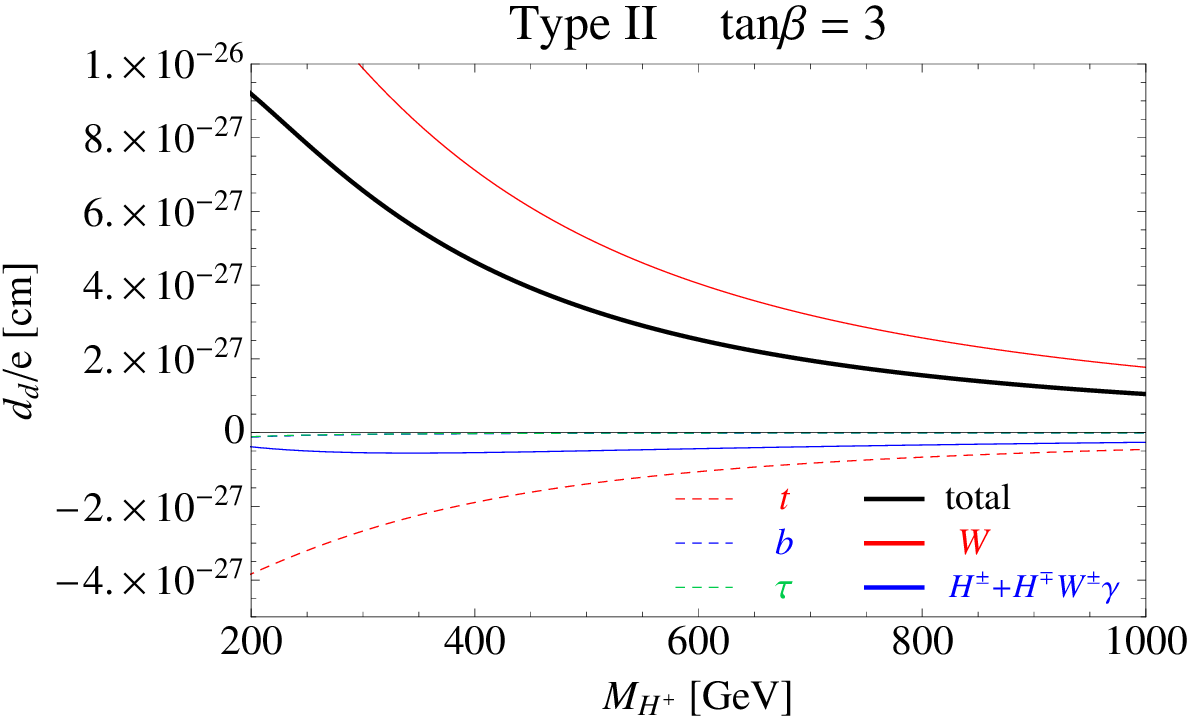} 
\end{minipage}
\begin{minipage}{0.49\hsize}
\includegraphics[width=0.9\hsize]{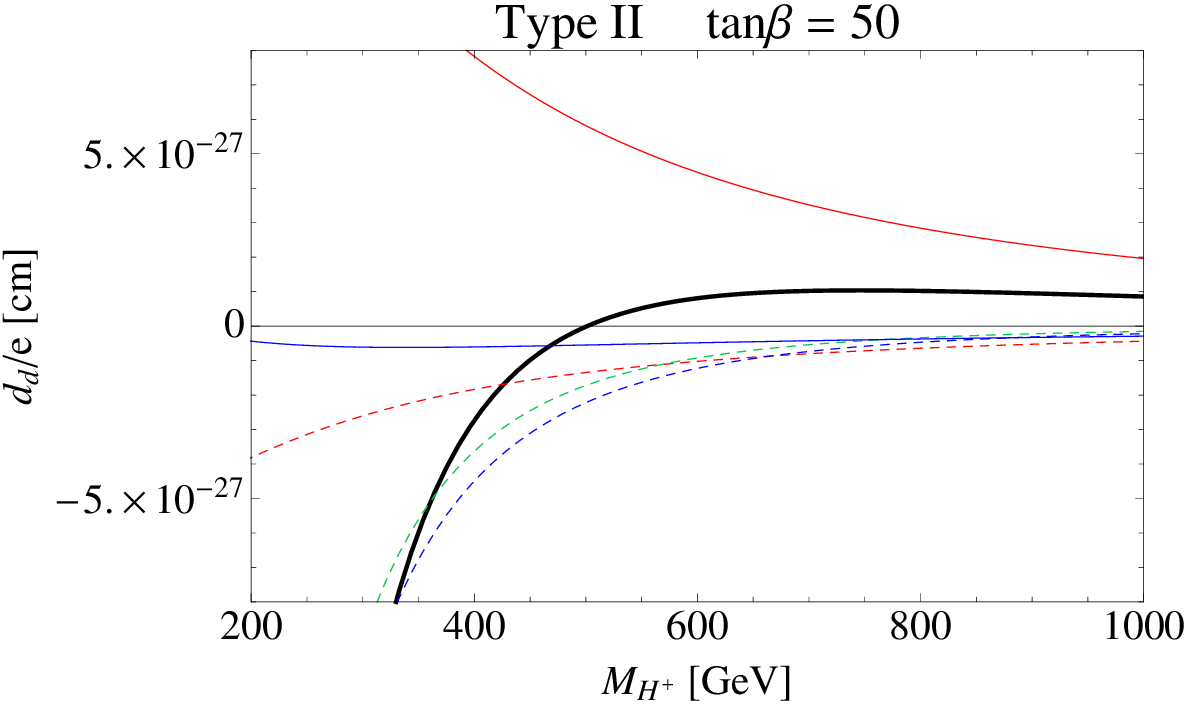} 
\end{minipage}
\\
\begin{minipage}{0.49\hsize}
\includegraphics[width=0.9\hsize]{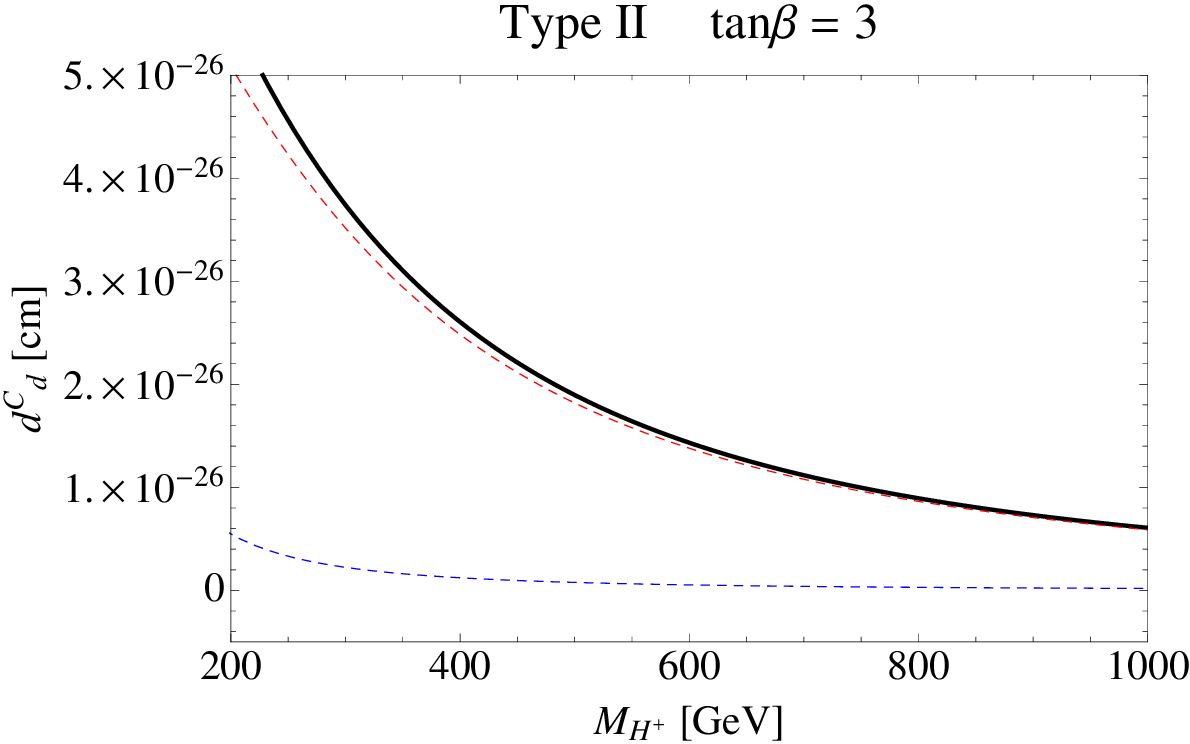} 
\end{minipage}
\begin{minipage}{0.49\hsize}
\includegraphics[width=0.9\hsize]{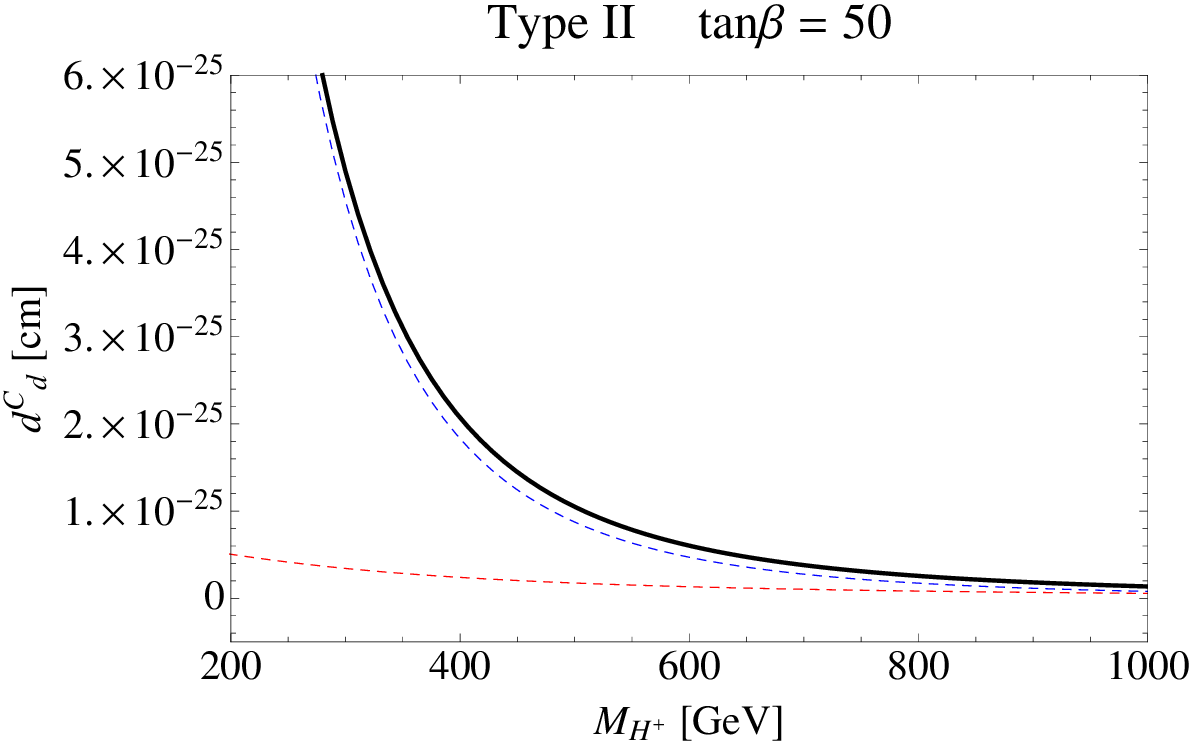} 
\end{minipage}
\caption{ Anatomy of the type-II down quark EDM and cEDM.  We take
  $\tan\beta=3$ and 50. Other input parameters are the same as in
  Fig~\ref{fig:eEDM_anatomy_type1}.  In contrast of the type-I case,
  the qualitative feature depends on $\tan\beta$. For large
  $\tan\beta$, the bottom quark and  tau lepton contributions are sizable
  due to the $\tan\beta$ enhancement of their Yukawa couplings.  }
\label{fig:dEDM_anatomy_type2}\vskip .1in

\begin{minipage}{0.49\hsize}
\includegraphics[width=0.9\hsize]{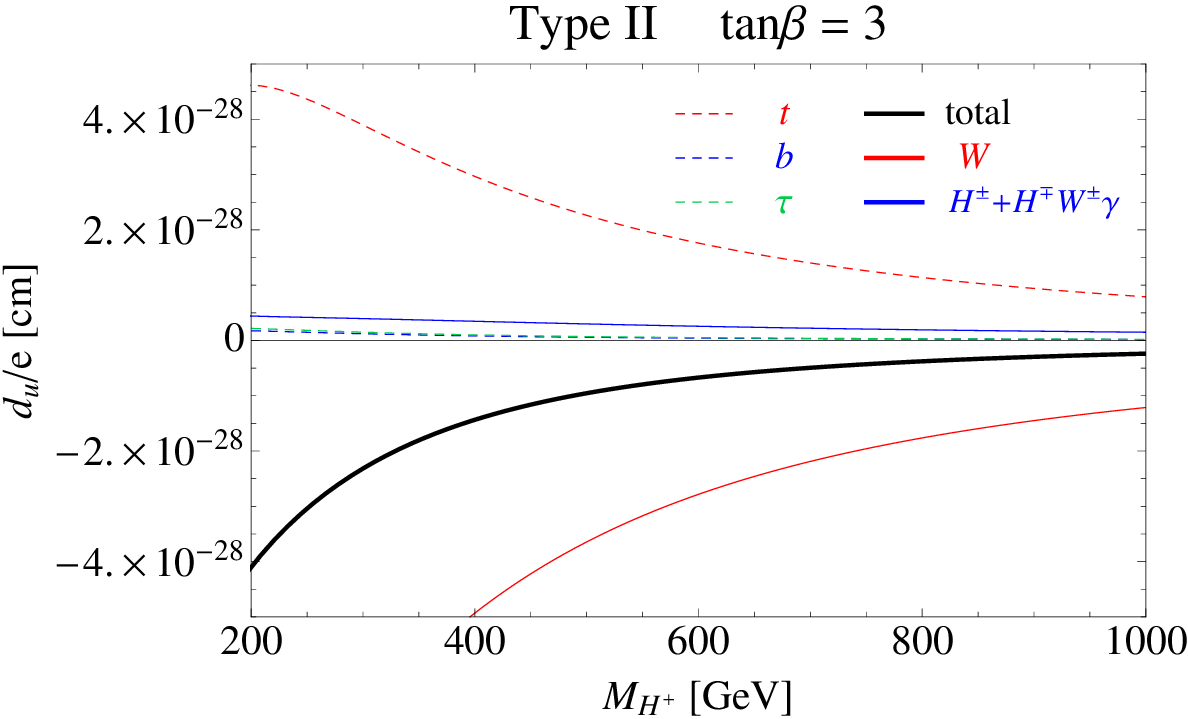} 
\end{minipage}
\begin{minipage}{0.49\hsize}
\includegraphics[width=0.9\hsize]{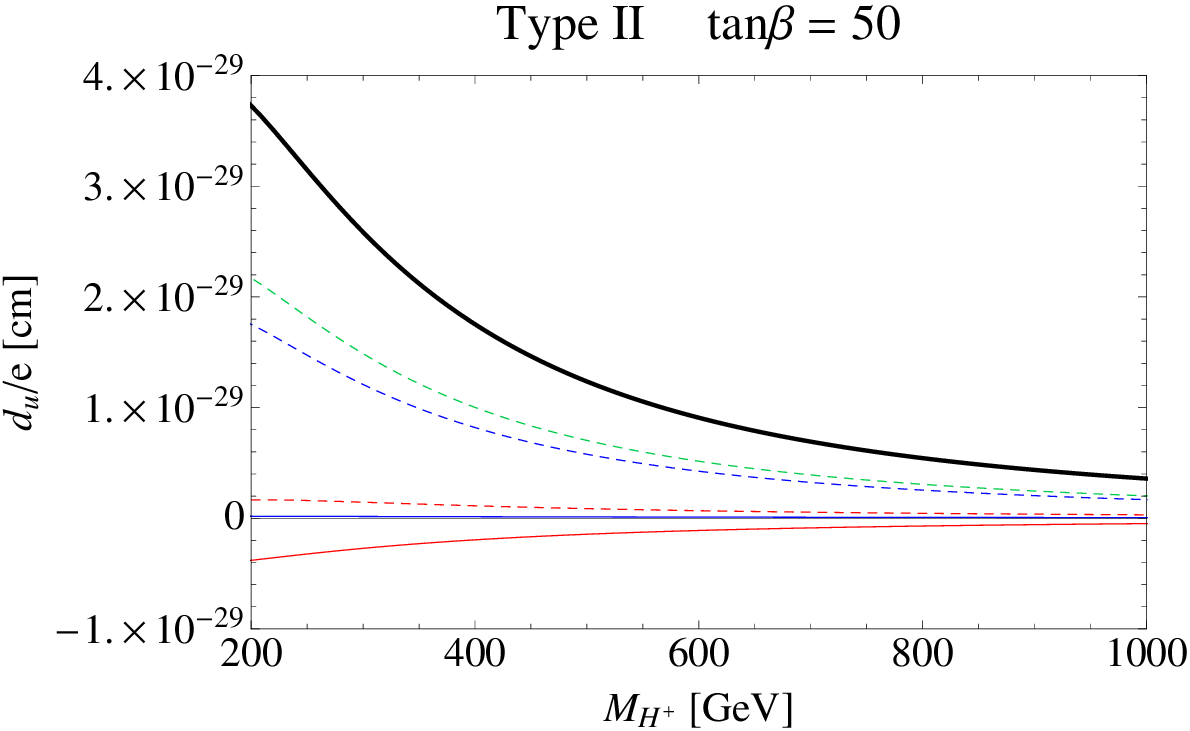} 
\end{minipage}
\\
\begin{minipage}{0.49\hsize}
\includegraphics[width=0.9\hsize]{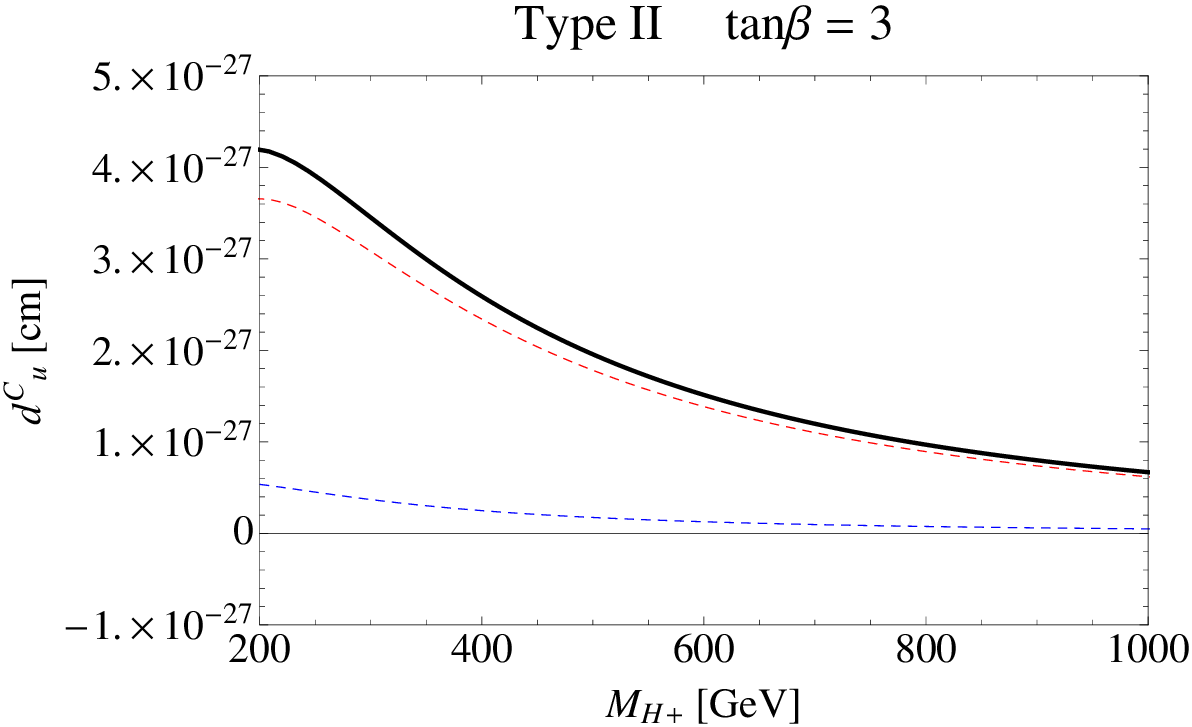} 
\end{minipage}
\begin{minipage}{0.49\hsize}
\includegraphics[width=0.9\hsize]{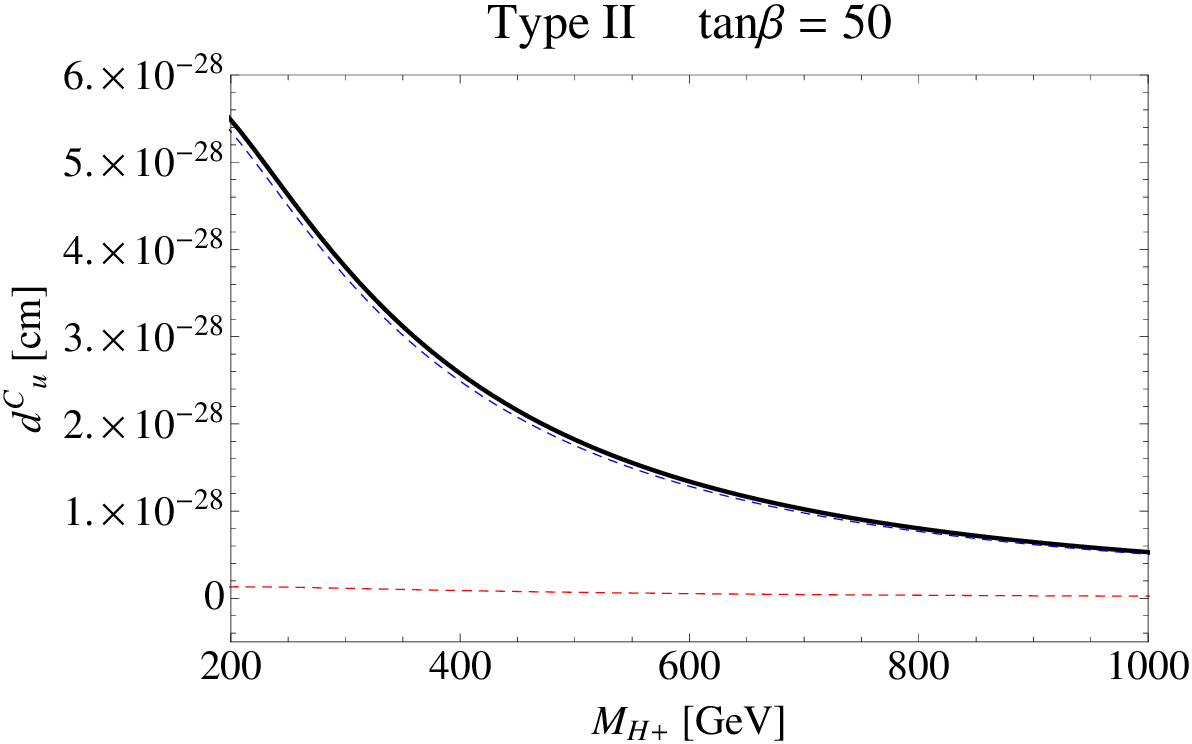} 
\end{minipage}
\caption{ Anatomy of the type-II up quark EDM and cEDM.  We take
  $\tan\beta=3$ and 50. Other input parameters are the same as in
  Fig~\ref{fig:eEDM_anatomy_type1}.  In contrast of the type-I case,
  the qualitative feature depends on $\tan\beta$. For large
  $\tan\beta$, bottom and tau contributions are sizable due to the
  $\tan\beta$ enhancement of their Yukawa couplings.  }
\label{fig:uEDM_anatomy_type2}
\end{figure}

In Figs.~\ref{fig:dEDM_anatomy_type2} and
\ref{fig:uEDM_anatomy_type2}, the contributions from each types of
diagrams to the down and up quark EDMs and cEDMs in the type-II
case are also shown. The EDMs and cEDMs have qualitatively different
behaviors from the type-I case. We find that the
largest contribution to the neutron EDM comes from down quark
cEDM. The top quark loop dominates in the down quark cEDM (and also
the up quark cEDM) for small $\tan\beta$, while the bottom quark one
quickly dominates it when $\tan\beta$ is large. The later comes from
the non-decoupling effect.  Thus, the neutron EDM would be enhanced
when $\tan\beta$ is large.  It is also found that the down quark EDM
has similar behavior to the electron EDM in the type-II case, though
it is smaller than the down quark cEDM in the neutron EDM.


Here, we ignore the QCD corrections to the quark EDMs and cEDMs. The
QCD corrections may change them up to ${\cal O}(10)$\%
\cite{Degrassi:2005zd,Hisano:2012cc}, while the neutron EDM evaluation
from the quark EDMs and cEDMs may have larger uncertainties.  See
Ref.~\cite{Hisano:2012cc} for evaluation for the QCD corrections to the
Barr-Zee diagrams.

\begin{figure}[tbp]
\begin{minipage}{0.24\hsize}
\subfigure[]{\includegraphics[
 width=2.\hsize]{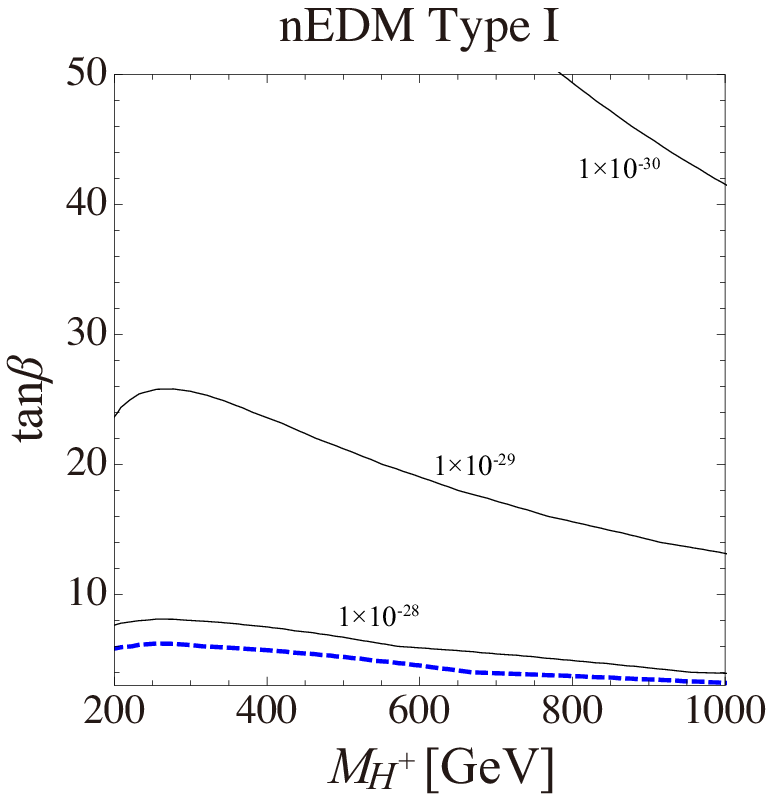} \label{fig:nEDM1}
}  
\end{minipage}\quad \quad \quad \quad \quad \quad \quad \quad \quad \quad \quad 
\begin{minipage}{0.24\hsize}
\subfigure[]{\includegraphics[
 width=2.\hsize]{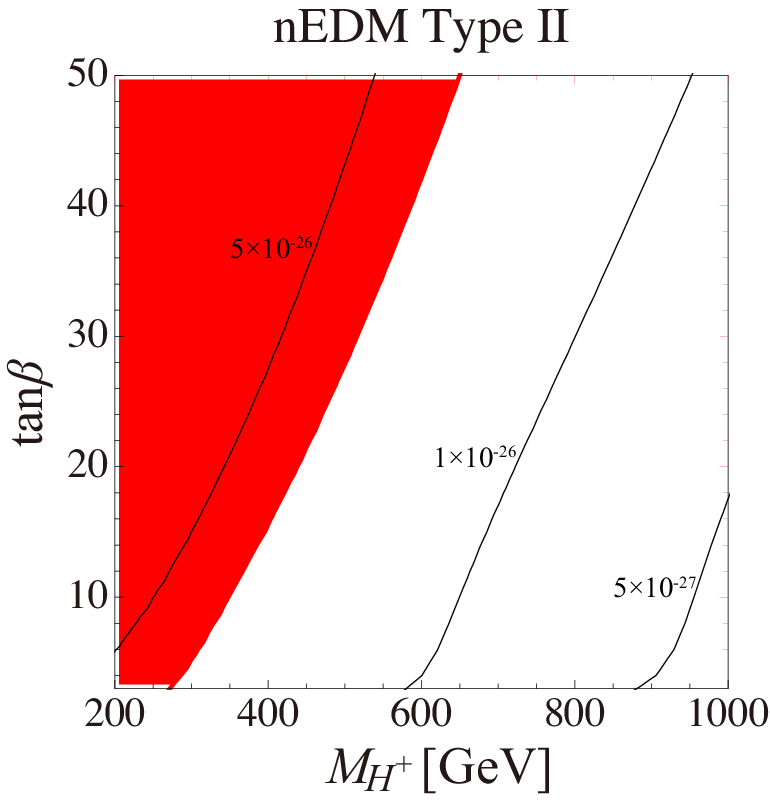}  \label{fig:nEDM2}}
\end{minipage} \\
\quad
\begin{minipage}{0.24\hsize}
\subfigure[]{\includegraphics[
 width=2.\hsize]{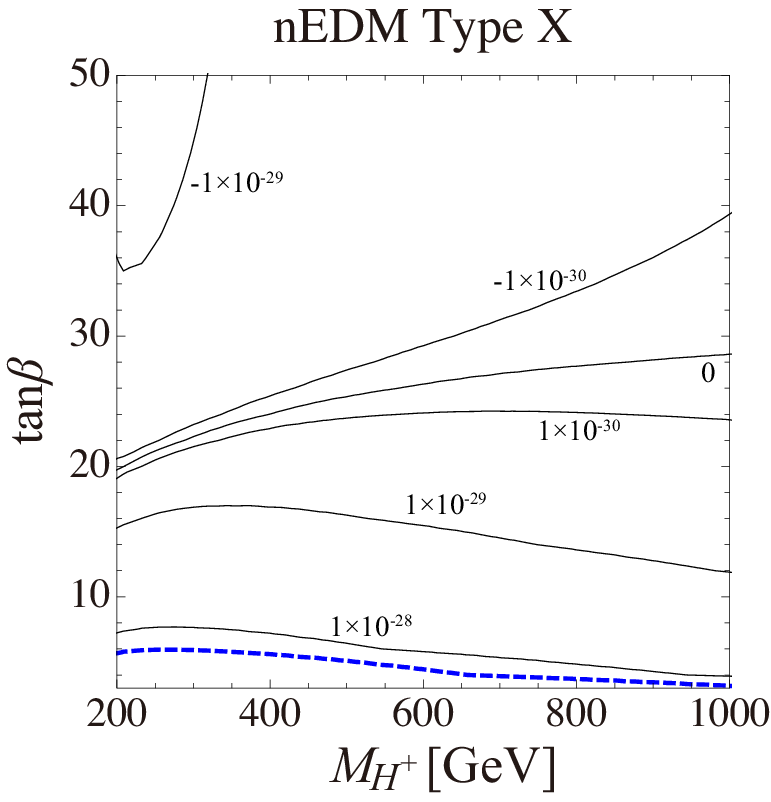}  \label{fig:nEDMX}}
\end{minipage}\quad \quad \quad \quad \quad \quad \quad \quad \quad \quad \quad 
\begin{minipage}{0.24\hsize}
\subfigure[]{\includegraphics[
 width=2.\hsize]{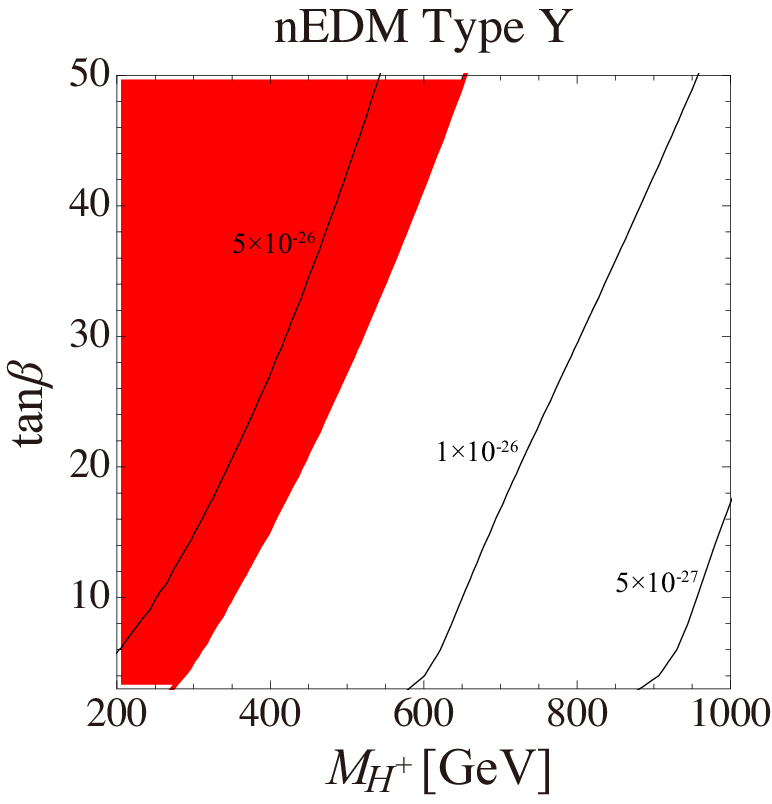} }
\end{minipage}
\caption{Neutron EDM on charged Higgs boson mass and $\tan\beta$
  plane. The input parameters are the same as in Fig.~{fig:eEDM}.  The
  region filled with red color show the current
  bound~\cite{Baker:2006ts}. The blue dashed lines are the future
  prospects given in Table~\ref{tab:nEDM_future}.  }
\label{fig:nEDM}
\end{figure}

Now we show the neutron EDM in four types of 2HDMs in
Fig.~\ref{fig:nEDM}. The regions filled with red color in Fig.~\ref{fig:nEDM} show the
excluded region by the current neutron EDM data,
\begin{align}
  |d_n| <& 2.9 \times 10^{-26} e \text{ cm } (90\%~\text{CL
    \cite{Baker:2006ts}}) .
\end{align}
The blue dashed lines are the future prospects given in Table~\ref{tab:nEDM_future}.
\begin{table}[h]
\centering 
\caption{Future prospects for neutron EDM}
\begin{tabular}{|c|c|}
\hline
 experiments
&  sensitivities on $|d_n|$ 
\\ 
\hline
 cyro EDM \cite{Balashov:2007zj}
&  $1.7 \times 10^{-28} e $ cm
\\ 
\hline
 PSI (Phase II) \cite{PSI}
&  $5 \times 10^{-28} e $ cm 
\\
\hline
\end{tabular}
\label{tab:nEDM_future}
\end{table}

It is found that the neutron EDM in the type-X case has similar
behavior to the type-I in low $\tan\beta$ region because the down
quark Yukawa couplings in these two types are the same. The difference
in high $\tan\beta$ region between Figs.~\ref{fig:nEDM1} and
\ref{fig:nEDMX} is due to the large $\tan\beta$ enhancement of the tau
lepton Yukawa coupling.  The behavior of the neutron EDM in the type-Y
case is quite similar to the type-II case. This is because the cEDM
contribution is dominant in both cases.

It is found in comparison of  Fig.~\ref{fig:eEDM} with
Fig.~\ref{fig:nEDM} that both measurements of the electron and
neutron EDMs are complementary to each others in order to
discriminate the 2HDMs. We may choose one from the four models in
future.

Before closing this section, we would like to give a comment on the
constraints on the parameter space.  We have shown that some parameter
regions are constrained by EDMs in Figs.~\ref{fig:eEDM} and
\ref{fig:nEDM}.  The constrained regions have an overlap with other
constraints, such as flavor physics \cite{Haisch:2008ar,
  Mahmoudi:2009zx} or direct search of heavy Higgs bosons
\cite{CMS:gya}.  Note that it is known that the custodial $SU(2)$
symmetry is broken in the Higgs potential in 2HDMs with the CP
violation, and $\rho$ parameter might deviate from one at the one loop
level \cite{Pomarol:1993mu}.  However, if heavy Higgs boson mass scale
$M$ is large or if coupling $\lambda_1 - \lambda_5$ are not large,
this contribution is small.  We checked that this contribution does
not conflict with the current bound in all figure of this paper.

\section{Conclusions and discussion}
\label{sec:summary}

\begin{figure}[tb]
\begin{minipage}{0.32\hsize}
\includegraphics[ width=1.5\hsize]{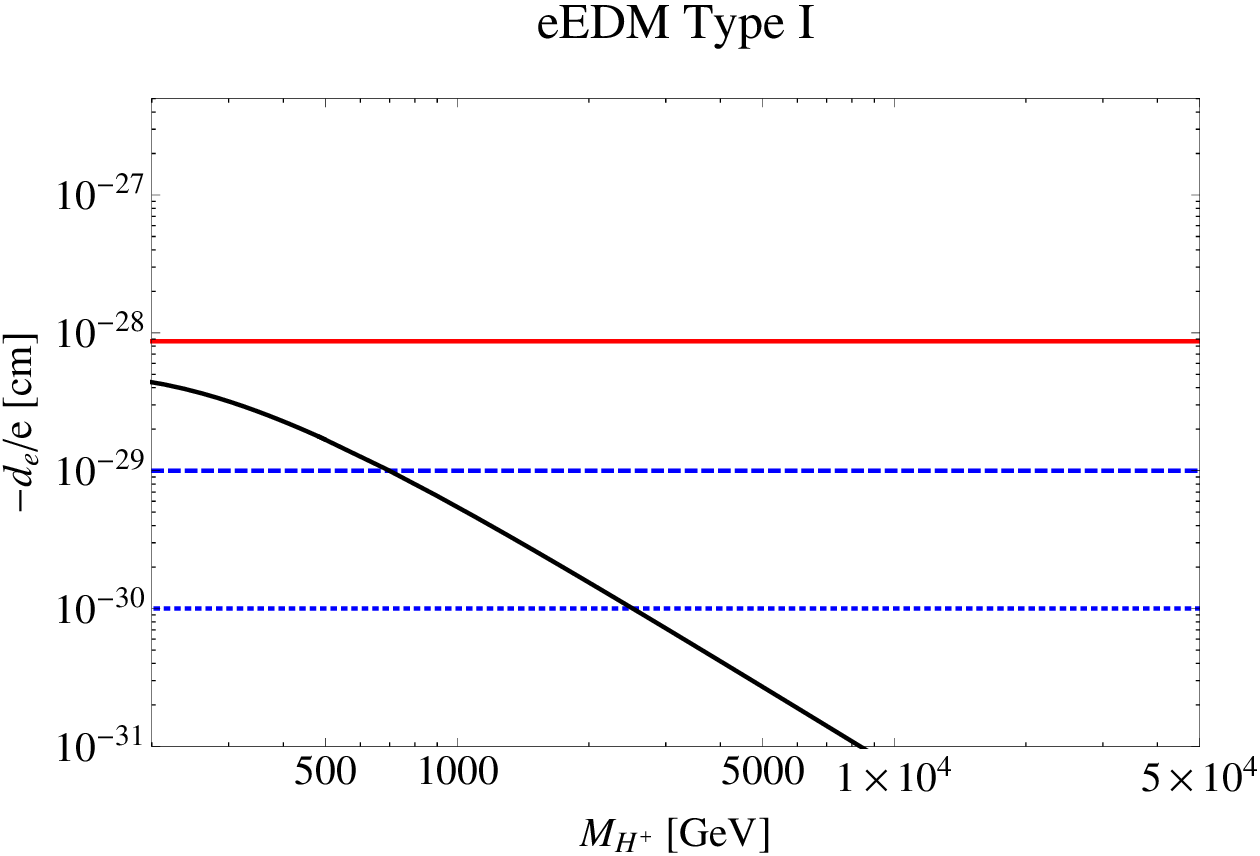} 
\end{minipage}\quad \quad \quad \quad \quad \quad \quad
\begin{minipage}{0.32\hsize}
\includegraphics[width=1.5\hsize]{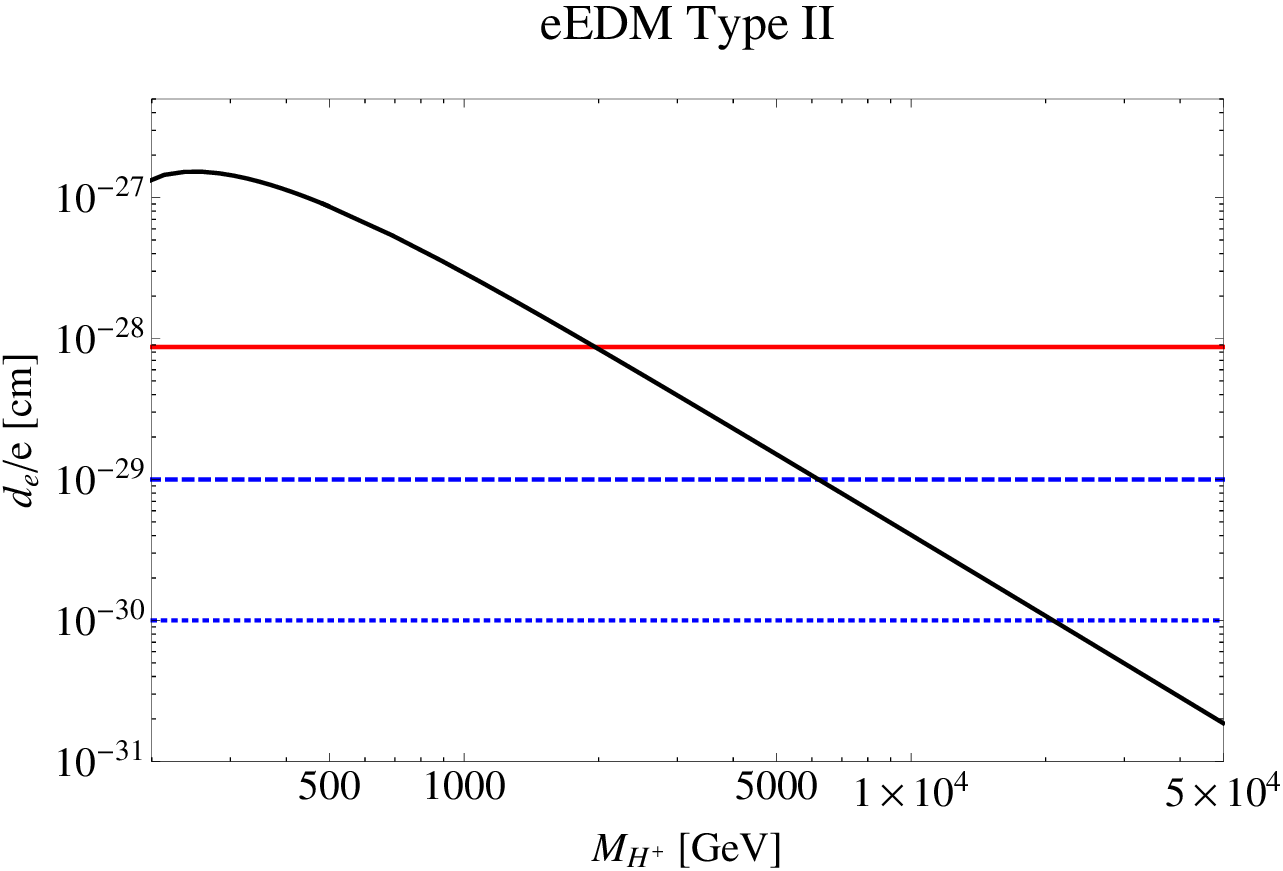} 
\end{minipage} \\
\begin{center}
\begin{minipage}{0.32\hsize}
\includegraphics[width=1.5\hsize]{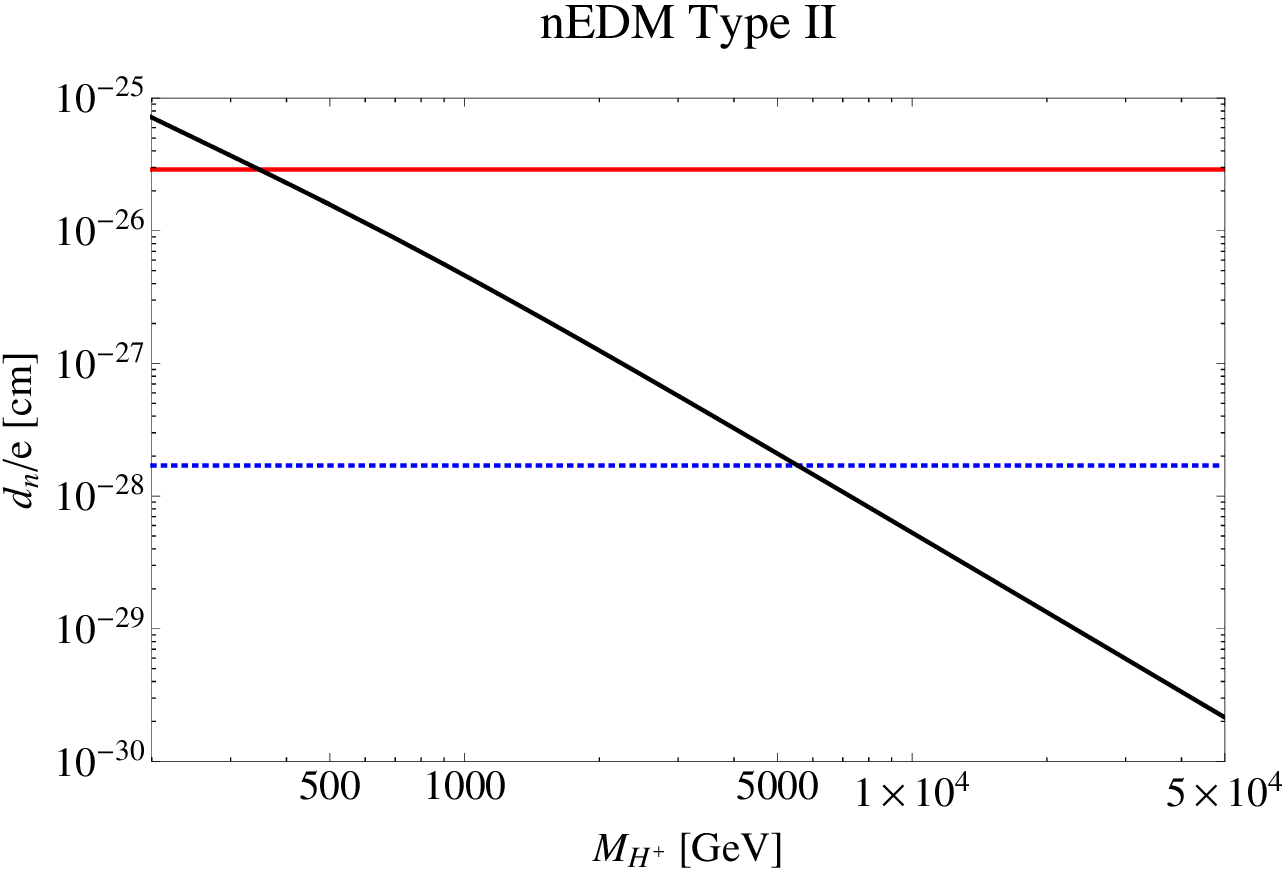} 
\end{minipage} \hspace{30mm}
\caption{Electron and Neutron EDMs at large $m_{H^{\pm}}$ region in the type-II case. We take  $\tan\beta = 10$, $\lambda_1 = \lambda_3 = \lambda_4 = \lambda_5 \sin 2 \phi =0.5$ and require the 126~GeV Higgs boson mass. 
The red and blue lines are current bounds~\cite{Baron:2013eja, Baker:2006ts}
and future prospects given in Tables~\ref{tab:EDM_future} and
\ref{tab:nEDM_future}, respectively.
} 
\end{center}
\label{fig:EDM_in_large_MH}
\end{figure}

In this paper, we evaluated fermionic EDMs in 2HDMs with softly broken
$Z_2$ symmetry.  We started by calculating the Barr-Zee diagrams in a
gauge invariant way by using the pinch technique. The modification by
the gauge invariant calculation is $5$--$8$\% numerically. This does not
change the previous result drastically, but important because physical
quantities must be calculated in a gauge invariant way.  We evaluated
the electron and neutron EDMs in all four types in the 2HDMs. We find
that type-II and type-X 2HDMs are strongly constrained by the latest
ACME experiment bound on the electron EDM. The electron and neutron
EDM measurements will  improve in the  future experiments.
They are possible to seek physics at ${\cal O}$(10)~TeV
scale (Fig.~\ref{fig:EDM_in_large_MH}).  The electron and neutron EDMs
have different sensitivities on the 2HDMs, and they are complementary
to each other in discrimination of the type of 2HDMs.

 We have not addressed that the contributions from non-Barr-Zee type
diagrams in this paper. Although they are naively expected to be smaller than the
contributions from the Barr-Zee diagrams, they would become important once
experiments find the EDMs and start precise measurements. To evaluate them, we
need to calculate all diagrams at two-loop level. This issue may be
discussed elsewhere.

It is worth referring to relation between EWBG and EDMs.  In the
2HDMs, it is known that EWBG may occur through a strongly first order
electroweak phase transition \cite{Turok:1990zg, Cohen:1991iu,
  Cline:1996mga, Fromme:2006cm, Dorsch:2013wja,Cline:2011mm}.  For
example, Ref.~\cite{Dorsch:2013wja} numerically showed that the 2HDMs
with softly-broken $Z_2$ symmetry may accommodate a strongly first
order phase transition when the lightest neutral Higgs boson is around 125 GeV. 
In order to achieve the EWBG, one
needs some CP violation phases in Higgs potential.  The EDM searches
could indirectly constrain parameter space which achieve the EWBG.
In this paper, we find that low $\tan\beta$ regions in 2HDMs are disfavored by electron EDM. 
On the other hand, in fact, a strongly first order phase transition,
which is needed for EWBG, prefers low $\tan \beta$ region
\cite{Dorsch:2013wja}.  Therefore there is a tension between EWBG
and current bound on the EDM.

\section*{Acknowledgments}
The authors thank the Yukawa Institute for Theoretical Physics at
Kyoto University, where this work was initiated during the YITP
workshop on ``LHC vs. Beyond the Standard Model
(YITP-W-12-21)'', March 19-25, 2013, and also acknowledge the
participants of the workshop for very active discussions.
They would also like to thank Koji Tsumura, Eibun Senaha, Ryosuke Sato, Yasuhiro Yamamoto, and Motoi Endo for useful discussions and comments.
The work of J.H. is supported by Grant-in-Aid for Scientific research
from the Ministry of Education, Science, Sports, and Culture (MEXT),
Japan, No. 24340047, No. 23104011 and No. 22244021, and also by World
Premier International Research Center Initiative (WPI Initiative),
MEXT, Japan.  The figures for this paper were drawn using
Feynmf~\cite{Ohl:1995kr}.

\appendix
\section{2HDMs}
\label{sec:app_2HDM}
In this appendix, we present mass spectrum and also interactions in
2HDMs, which are used in text.

\subsection{Relations between mass and gauge eigenstates}

While eight scalar fields are
present in 2HDMs,
\begin{align}
 \sigma_{1,2},\
 \pi_{1,2}^{\pm},\
 \pi_{1,2}^{3}
,
\nonumber
\end{align}
as in Eq.~(\ref{eq:H-base}), %
those states are not mass eigenstates, namely their mass
matrices are not diagonalized.  We call them the gauge
eigenstates. Corresponding to them, there
are eight mass eigenstates, which  we denote them as
\begin{align}
\begin{matrix}
 h_{1,2,3}  & \text{(neutral Higgs bosons)}, \\
 H^{\pm}    & \text{(charged Higgs bosons)}, \\
 \pi_{Z, W^{\pm}}    & \text{(would-be NG bosons)}.
\end{matrix}
\nonumber
\end{align}
These two-types of states are related with orthogonal or unitary matrices which diagonalize the
mass matrices.
For the fields which include would-be NG bosons the matrices are given as 
\begin{align}
\left(
\begin{matrix}
 \pi_{Z} \\
 \pi_{A} \\
\end{matrix}
\right)
=&
\left(
\begin{matrix}
 \cos\beta     &    \sin\beta \\
 -\sin\beta     &   \cos\beta 
\end{matrix}
\right)
\left(
\begin{matrix}
 \pi_1^{3} \\
 \pi_2^{3}
\end{matrix}
\right)
,
\nonumber\\
\left(
\begin{matrix}
 \pi_{W^{\pm}} \\
 H^{\pm} \\
\end{matrix}
\right)
=&
\left(
\begin{matrix}
 \cos\beta     &    \sin\beta \\
 -\sin\beta     &   \cos\beta 
\end{matrix}
\right)
\left(
\begin{matrix}
 \pi_1^{\pm} \\
 \pi_2^{\pm}
\end{matrix}
\right)
.
\end{align}
The matrix $U$ for physical neutral Higgs bosons is given by a  3 by 3 matrix as 
\begin{align}
\left(
\begin{matrix}
 h_1 \\
 h_2 \\
 h_3 \\
\end{matrix}
\right) 
=& U^{T} \left(\begin{matrix}
 \sigma_1 \\
 \sigma_2 \\
 \pi_A \\
\end{matrix}
\right) 
=
\left(
\begin{matrix}
  \omega_{h1}^{\sigma_1} 
& \omega_{h1}^{\sigma_2} 
& \omega_{h1}^{\pi_A} 
\\ 
  \omega_{h2}^{\sigma_1} 
& \omega_{h2}^{\sigma_2} 
& \omega_{h2}^{\pi_A} 
\\ 
  \omega_{h3}^{\sigma_1} 
& \omega_{h3}^{\sigma_2} 
& \omega_{h3}^{\pi_A}
\end{matrix}
\right)
\left(
\begin{matrix}
 \sigma_1 \\
 \sigma_2 \\
 \pi_A
\end{matrix}
\right)
\end{align}
where
\begin{align}
& \sum_X \omega_{i}^{X} \omega_{j}^{X} = \delta_{ij}, 
\,
\quad
\sum_i \omega_{i}^{X} \omega_{i}^{Y} = \delta^{XY}.
\label{eq:orthonormal}
\end{align}
These relations are useful to find relations among some couplings.

\subsection{Higgs masses in 2HDMs}

The mass terms for the neutral physical Higgs bosons are given by
\begin{align}
{\cal L}
\supset
-
\frac{1}{2}
\left(
\begin{matrix}
 \sigma_1 &
 \sigma_2 &
 \pi_A
\end{matrix}
\right)
{\cal \widetilde{M}}^2_{\text{N}}
\left(
\begin{matrix}
 \sigma_1 \\
 \sigma_2 \\
 \pi_A
\end{matrix}
\right)
,
\end{align}
where
\begin{align}
\left(
{\cal \widetilde{M}}^2_{\text{N}}
\right)_{11}
=&
v_1^2 \lambda_1
+
M^2 \sin^2 \beta
, \nonumber\\ 
\left(
{\cal \widetilde{M}}^2_{\text{N}}
\right)_{22}
=&
v_2^2 \lambda_2
+
M^2 \cos^2 \beta
, \nonumber\\ 
\left(
{\cal \widetilde{M}}^2_{\text{N}}
\right)_{33}
=&
M^2
-
v^2
\lambda_5
\cos(2\phi)
, \nonumber\\ 
\left(
{\cal \widetilde{M}}^2_{\text{N}}
\right)_{21}
=
\left(
{\cal \widetilde{M}}^2_{\text{N}}
\right)_{12}
=&
\left(
v^2  \lambda_{345}
- 
M^2 
\right)
\sin\beta \cos\beta
, \nonumber\\ 
\left(
{\cal \widetilde{M}}^2_{\text{N}}
\right)_{31}
=
\left(
{\cal \widetilde{M}}^2_{\text{N}}
\right)_{13}
=&
\frac{1}{2} 
v^2
\lambda_5
\sin(2\phi)
\sin\beta
, \nonumber\\ 
\left(
{\cal \widetilde{M}}^2_{\text{N}}
\right)_{32}
=
\left(
{\cal \widetilde{M}}^2_{\text{N}}
\right)_{23}
=&
\frac{1}{2}
v^2
\lambda_5
\sin(2\phi)
\cos\beta
,
\end{align}
where
\begin{align}
 \lambda_{345}
=&
 \lambda_3 + \lambda_4 + \lambda_5 \cos(2\phi)
,
\end{align}
and $M^2$ is defined in Eq.~(\ref{eq:def_M}).
This mass matrix satisfies
\begin{align}
{\cal \widetilde{M}}^2_{\text{N}}
=&
U
\left(
\begin{matrix}
 m_{h_1}^2 & &  \\
 & m_{h_2}^2  &  \\
 & & m_{h_3}^2   
\end{matrix}
\right)
U^{T}.
\label{eq:UdiagU}
\end{align}
In large $M$ limit, we find the following expressions for mass and
mixing angles.
\begin{align}
 m_{h_1}^2
=&
\frac{
   v_1^4 \lambda_1 
 + v_2^4 \lambda_2
 + 2 v_1^2 v_2^2 \lambda_{345}
}
{v^2}
+
{\cal O}(M^{-2})
,
\nonumber\\ 
 m_{h_2}^2
=&
M^2
\left(
1
+
{\cal O}(M^{-2})
\right)
,
\nonumber\\ 
 m_{h_3}^2
=&
M^2
\left(
1
+
{\cal O}(M^{-2})
\right)
\nonumber.
\end{align}
\begin{align}
\left(
\begin{matrix}
 \omega^{\sigma_1}_{h_1} \\
 \omega^{\sigma_2}_{h_1} \\
 \omega^{\pi_A}_{h_1}
\end{matrix}
\right)
=&
\left(
\begin{matrix}
 \cos \beta
\left(
1
-
X \sin^2 \beta 
\right) \\
\sin \beta
\left(
1
+
X \cos^2 \beta 
\right)\\
-\frac{v_1 v_2 \lambda_5 \sin(2\phi) }{M^2}
\end{matrix}
\right)
+
{\cal O}(M^{-4})
,\nonumber\\
\left(
\begin{matrix}
 \omega^{\sigma_1}_{h_2} \\
 \omega^{\sigma_2}_{h_2} \\
 \omega^{\pi_A}_{h_2}
\end{matrix}
\right)
=&
\left(
\begin{matrix}
- \sin\beta \sin\theta \\
  \cos\beta \sin\theta \\
 \cos\theta
\end{matrix}
\right)
+
{\cal O}(M^{-2})
,
\nonumber\\ 
\left(
\begin{matrix}
 \omega^{\sigma_1}_{h_3} \\
 \omega^{\sigma_2}_{h_3} \\
 \omega^{\pi_A}_{h_3}
\end{matrix}
\right)
=&
\left(
\begin{matrix}
- \sin\beta \cos\theta \\
  \cos\beta \cos\theta \\
-\sin\theta
\end{matrix}
\right)
+
{\cal O}(M^{-2})
,
\end{align}
where
\begin{align}
\tan(2\theta)
=&
\frac{
- (\cos^2\beta - \sin^2\beta) 
}{
\sin^2\beta \cos^2\beta
\left( 
\lambda_1
+
\lambda_2
-
2 \lambda_{345}
\right)
+
\lambda_5 \cos 2\phi
}
 \lambda_5 \sin(2\phi)
,
\nonumber\\ 
X
=&
 \frac{
 v_1^2 \lambda_1 
- v_2^2 \lambda_2
- (v_1^2 -v_2^2) \lambda_{345}
}{
M^2
}.
\end{align}

\subsection{Interactions in 2HDMs}

Couplings which are relevant to calculation for the gauge invariant
Barr-Zee contributions are written in this subsection.  Our convention
of the sign in covariant derivative is
\begin{align}
 D_{\mu}
=&
 \partial_{\mu}
+ i g V_{\mu}
.
\end{align}

\subsubsection{$V$-$\bar{f}$-$f$ couplings}
These couplings are the same as the SM case, but we show them here to
establish our conventions.
For neutral gauge bosons,
\begin{align}
 {\cal L}
\supset&
 - \sum_{G = \gamma, Z} \overline{f} \gamma^{\mu} g_{Gff} f G_{\mu}
,
\end{align}
where $g_{Gff}$ contains chirality structure,
\begin{align}
g_{Gff}
=
g_{G ff}^{L} P_L   +  g_{G ff}^{R} P_R
,
\end{align}
where
\begin{align}
g_{\gamma ff}^{L}
=&
e Q 
,\nonumber \\
g_{\gamma ff}^{R}
=&
e Q 
,\nonumber \\
g_{Zff}^{L} 
=&
  \frac{e}{sc}\left( T^3 - s^2 Q \right) 
, \nonumber\\
g_{Zff}^{R} 
=&
  \frac{e}{sc}\left( - s^2 Q \right) 
.
\end{align}
For $W$ boson,
\begin{align}
 {\cal L}
\supset&
 - \frac{1}{\sqrt{2}} 
 \overline{u} \gamma^{\mu} g_{W u d} d W^{+}_{\mu}
+
h.c.
,
\end{align}
where
\begin{align}
 g_{W u d}
=&
 V_{\text{CKM}} \frac{e}{s} P_L
,
\end{align}
where $V_{\text{CKM}}$ is for the CKM matrix.

\subsubsection{Yukawa couplings }

The Yukawa interaction terms are described as 
\begin{align}
&
-
\left(
\begin{matrix}
 \overline{u} &  \overline{d}
\end{matrix}
\right)
\left(
\begin{matrix}
m_u^{\text{diag.}}
+
\sum_{s} g_{uus} s
& 
\sum_{s} g_{\overline{u}ds^{+}} s^{+}
\\ 
\sum_{s} g_{\overline{d}us^{-}} s^{-}
& 
m_d^{\text{diag.}}
+
\sum_s g_{dds} s
\end{matrix}
\right)
\left(
\begin{matrix}
 u \\
 d
\end{matrix}
\right)
,
\end{align}
where $s = { h_1, h_2, h_3, \pi_Z}$, and $s^{\pm} = { H^{\pm},
\pi_{W^{\pm}}}$.
We define $g^{V}$ and $g^{A}$ as 
\begin{align}
 g = & g^{V} + i \gamma^5 g^{A}.
\end{align}
Finally we find explicit expressions of the couplings. For the neutral Higgs bosons,
\begin{align}
 g_{uuh}^{V}
=&
 \frac{m_u^{\text{diag.}}}{v}
\frac{1}{\sin\beta}
\omega_h^{\sigma_2}
,\nonumber \\ 
 g_{uuh}^{A}
=&
\frac{m_u^{\text{diag.}}}{v}
\frac{1}{\tan\beta}
\omega_h^{\pi_A}
,\nonumber \\ 
 g_{ddh}^{V}
=&
\begin{cases}
 \frac{m_d^{\text{diag.}}}{v}
\frac{1}{\cos\beta}
\omega_h^{\sigma_1}
 & ( i = 1)
\\
 \frac{m_d^{\text{diag.}}}{v}
\frac{1}{\sin\beta}
\omega_h^{\sigma_2}
 & ( i = 2)
\end{cases}
, \nonumber\\ 
 g_{ddh}^{A}
=&
\begin{cases}
\frac{m_d^{\text{diag.}}}{v}
\tan\beta \omega_h^{\pi_A}
 & ( i = 1)
\\
- \frac{m_d^{\text{diag.}}}{v}
\frac{1}{\tan\beta}
\omega_h^{\pi_A}
 & ( i = 2)
\end{cases}
.
\end{align}
Here, $i$ corresponds to the same suffix of $H_i$ which couples to
down-type quarks.

For the physical charged Higgs boson,
\begin{align}
g_{\overline{u} d H^{+}}^{V}
&=
\frac{1}{\sqrt{2}}
\left(
V_{\text{CKM}}
\frac{m_d^{\text{diag.}}}{v_i}
\left(
- \delta_{1 i} \sin\beta
+ \delta_{2 i} \cos\beta
\right)
-
\frac{m_u^{\text{diag.}}}{v_2}
V_{\text{CKM}}
\cos\beta
\right)
, \nonumber\\ 
g_{\overline{u} d H^{+}}^{A}
&=
-
\frac{i}{\sqrt{2}}
\left(
V_{\text{CKM}}
\frac{m_d^{\text{diag.}}}{v_i}
\left(
- \delta_{1 i} \sin\beta
+ \delta_{2 i} \cos\beta
\right)
+
\frac{m_u^{\text{diag.}}}{v_2}
V_{\text{CKM}}
\cos\beta
\right)
, \nonumber\\ 
g_{\overline{d} u H^{-}}^{V}
&=
-
\frac{1}{\sqrt{2}}
\left(
V_{\text{CKM}}^{\dagger}
\frac{m_u^{\text{diag.}}}{v_2}
\cos\beta
-
\frac{m_d^{\text{diag.}}}{v_i}
V_{\text{CKM}}^{\dagger}
\left(
- \delta_{1 i} \sin\beta
+ \delta_{2 i} \cos\beta
\right)
\right)
, \nonumber\\ 
g_{\overline{d} u H^{-}}^{A}
&=
\frac{i}{\sqrt{2}}
\left(
V_{\text{CKM}}^{\dagger}
\frac{m_u^{\text{diag.}}}{v_2}
\cos\beta
+
\frac{m_d^{\text{diag.}}}{v_i}
V_{\text{CKM}}^{\dagger}
\left(
- \delta_{1 i} \sin\beta
+ \delta_{2 i} \cos\beta
\right)
\right)
,
\end{align}
where $i$ in the suffix is again the same suffix of $H_i$ which couples to
down-type quarks.
Sometime the followings are useful:
\begin{align}
 g_{\overline{u} d H^{+}}
&=
 g_{\overline{u} d H^{+}}^L P_L +  g_{\overline{u} d H^{+}}^R P_R \nonumber\\
 &=
+
\sqrt{2}
\left[
\left(
-
\frac{m_u^{\text{diag.}}}{v}
V_{\text{CKM}}
\frac{1}{\tan\beta}
\right)
P_L
+
\left(
V_{\text{CKM}}
\frac{m_d^{\text{diag.}}}{v}
\left(
- \delta_{1 i} \tan\beta
+ \delta_{2 i} \frac{1}{\tan\beta}
\right)
\right)
P_R
\right]
, \nonumber\\
g_{\overline{d} u H^{-}}
&=
g_{\overline{d} u H^{-}}^L P_L + g_{\overline{d} u H^{-}}^R P_R \nonumber\\
&=
-
\sqrt{2}
\left[
\left(
-
\frac{m_d^{\text{diag.}}}{v}
V_{\text{CKM}}^{\dagger}
\left(
- \delta_{1 i} \tan\beta
+ \delta_{2 i} \frac{1}{\tan\beta}
\right)
\right)
P_L
+
\left(
V_{\text{CKM}}^{\dagger}
\frac{m_u^{\text{diag.}}}{v}
\frac{1}{\tan\beta}
\right)
P_R
\right]
.
\end{align}

\subsubsection{${\cal L}_{\mbox{\scriptsize WWW}}$}
These couplings are the same as the SM case, but we show them here to
establish our conventions.
\begin{align}
 {\cal L}
 \supset&
-
\sum_{G=\gamma,Z} i g_{WWG}\biggl\{(\partial^\alpha
W^{+\beta})W^{-\mu}G^\nu(g_{\alpha\mu}g_{\beta\nu}-g_{\alpha\nu}g_{\beta\mu})
\nonumber\\  
 & \hspace{3cm} + W^{+\beta}(\partial^\alpha
 W^{-\mu})G^\nu(g_{\alpha\nu}g_{\beta\mu}-g_{\alpha\beta}g_{\mu\nu})
\nonumber\\
 & \hspace{3cm} + W^{+\beta} W^{-\mu}(\partial^\alpha
  G^\nu)(g_{\alpha\beta}g_{\mu\nu}-g_{\alpha\mu}g_{\beta\nu})\biggr\}
,
\end{align}
where
\begin{align}
 g_{WW\gamma} =& e ,\nonumber\\
 g_{WWZ} =& \frac{e}{s} c.
\end{align}

\subsubsection{$W$-$W$-$h$ couplings}
\label{WWh}

\begin{align}
{\cal L}
\supset &
\sum_h
g_{WWh} W^{+}_{\mu} W^{- \mu} h
+
\frac{1}{2} g_{ZZh} Z_{\mu} Z^{\mu} h
,
\end{align}
where
\begin{align}
g_{WWh}
=&
2 \frac{m_W^2}{v}
\left[
\cos\beta
\omega^{\sigma_1}_h
+
\sin\beta
\omega^{\sigma_2}_h
\right]
, 
\nonumber\\ 
g_{ZZh}
=&
2 \frac{m_Z^2}{v}
\left[
\cos\beta
\omega^{\sigma_1}_h
+
\sin\beta
\omega^{\sigma_2}_h
\right]
.
\end{align}
By using Eq.~(\ref{eq:orthonormal}),
we find that 
\begin{align}
 \sum_h g_{\ell \ell h}^{A} g_{WWh} = 0.
\end{align}

\subsubsection{$V$-$H^{+}$-$H^{-}$ couplings}
\begin{align}
 {\cal L}
\supset&
+
i
\left(
H^{+} \partial_{\mu} H^{-}
-
H^{-} \partial_{\mu} H^{+}
\right)
\left(
g_{\gamma H^{+} H^{-}} 
A^{\mu}
+
g_{Z H^{+} H^{-}} 
Z^{\mu}
\right)
,
\end{align}
where
\begin{align}
g_{\gamma H^{+} H^{-}} 
=&
e
, \nonumber\\
g_{Z H^{+} H^{-}} 
=&
\frac{1}{2}
\frac{e}{sc}
(c^2 - s^2)
.
\end{align}

\subsubsection{$W^{\pm}$-$H^{\mp}$-$h$ couplings}

\begin{align}
 {\cal L}
\supset&
+
i g_{W^{-} H^{+}h }
\left(
h \partial_{\mu} H^{+}
-
 H^{+}  \partial_{\mu} h
\right)
W^{- \mu}
\nonumber\\ 
&+
i g_{W^{+} H^{-} h}
\left(
h   \partial_{\mu} H^{-}
-
 H^{-}  \partial_{\mu} h
\right)
W^{+ \mu}
,
\end{align}
where
\begin{align}
g_{W^{\pm} H^{\mp} h}
=&
\pm
\frac{1}{2}
\frac{e}{s}
\left(
-
\sin\beta \omega^{\sigma_1}_{h}
+
\cos\beta \omega^{\sigma_2}_{h}
\mp
i \omega^{\pi_A}_{h}
\right)
.
\end{align}
By using Eq.~(\ref{eq:orthonormal}),
we find that 
\begin{align}
 \sum_h g_{W^{+} H^{-} h} g_{WWh} = 0.
\end{align}

\subsubsection{$s^{+}$-$s^{-}$-$h$ couplings}
\begin{align}
{\cal L}
\supset &
+
g_{H^{+}H^{-}h}
H^{+}H^{-}h
\nonumber\\
&+
g_{\pi_{W^{+}} \pi_{W^{-}} h}
\pi_{W^{+}} \pi_{W^{-}} h
\nonumber\\
&
+
g_{\pi_{W^{+}} H^{-} h}
\pi_{W^{+}} H^{-} h
+
g_{H^{+} \pi_{W^{-}} h}
H^{+} \pi_{W^{-}} h
\nonumber\\
&
+
\frac{1}{2}
g_{\pi_Z \pi_Z h}
\pi_Z \pi_Z h
,
\end{align}
where
\begin{align}
g_{H^{+}H^{-}h}
=&
+
\frac{v_1}{v^2}
\left(
- v_1^2 \lambda_3
+ v_2^2 (- \lambda_1 + \lambda_4 + \lambda_5 \cos(2\phi) )
\right)
\omega^{\sigma_1}_{h}
\nonumber\\ 
&
+
\frac{v_2}{v^2}
\left(
-v_2^2 \lambda_3
+
v_1^2 (-\lambda_2 + \lambda_4 + \lambda_5 \cos(2\phi))
\right)
\omega^{\sigma_2}_{h}
\nonumber\\ 
&
+
\frac{v_1 v_2}{v}
\lambda_5
\sin(2\phi)
\omega^{\pi_A}_{h}
,\nonumber\\ 
g_{\pi_{W^{+}} \pi_{W^{-}} h}
=&
-
\frac{m_h^2}{2 m_W^2}
g_{WWh}
,
\nonumber\\ 
 g_{\pi_{W^{+}} H^{-} h}
=&
-
\frac{ m_{H^{\pm}}^2  -  m_h^2}{m_W}
g_{W^{+} H^{-} h}
, 
\nonumber\\ 
 g_{H^{+} \pi_{W^{-}} h}
=&
+
\frac{ m_{H^{\pm}}^2  -  m_h^2}{m_W}
g_{W^{-} H^{+} h}
,
\nonumber\\ 
 g_{\pi_Z \pi_Z h}
=&
-
\frac{m_h^2}{2 m_Z^2}
g_{ZZh}
.
\label{eq:hNG}
\end{align}

\subsubsection{$W^{\pm}$-$\pi^{\mp}$-$h$ couplings}

\begin{align}
 {\cal L}
\supset&
+
i g_{ W^{-} \pi^{+} h}
\left(
h   \partial_{\mu} \pi^{+}
-
 \pi^{+}  \partial_{\mu} h
\right)
W^{- \mu}
\nonumber\\ 
&+
i g_{W^{+} \pi^{-} h}
\left(
h   \partial_{\mu} \pi^{-}
-
 \pi^{-}  \partial_{\mu} h
\right)
W^{+ \mu}
,
\end{align}
where
\begin{align}
g_{ W^{\pm} \pi^{\mp} h}
=&
\pm
\frac{1}{2 m_W}
g_{WWh}
.
\end{align}

\subsubsection{$V$-$W^{\pm}$-$\pi^{\mp}$ couplings}

\begin{align}
 {\cal L}
\supset&
+
\sum_{G = \gamma, Z}
\left(
g_{G W^{-} \pi^{+}}
G_{\mu} W^{- \mu} \pi^{+}
+
g_{G W^{+} \pi^{-}}
G_{\mu} W^{+ \mu} \pi^{-}
\right)
,
\end{align}
where
\begin{align}
g_{\gamma W^{\mp} \pi^{\pm}}
&=
+e m_{W},
\nonumber\\
g_{Z W^{\mp} \pi^{\pm}}
&=
-e s m_{Z}
.
\end{align}

\subsubsection{Some four-point couplings}

\begin{align}
 {\cal L}
\supset&
+
g_{H^{-} \pi_{W^{+}} \pi_{W^{-}} \pi_{W^{+}}}
H^{-} \pi_{W^{+}} \pi_{W^{-}} \pi_{W^{+}}
\nonumber\\ 
&
+
\frac{1}{2}
g_{H^{-} \pi_{W^{+}} \pi_{Z} \pi_{Z}}
H^{-} \pi_{W^{+}} \pi_{Z} \pi_{Z}
\nonumber\\ 
&
+
g_{H^{-} \pi_{W^{+}} H^{-} H^{+}}
H^{-} \pi_{W^{+}} H^{-} H^{+},
\end{align}
where
\begin{align}
g_{H^{-} \pi_{W^{+}} \pi_{W^{-}} \pi_{W^{+}}}
=&
\sum_h
\frac{1}{m_W}
g_{W^{+} H^{-} h}
g_{\pi_{W^{+}} \pi_{W^{-}} h}
,\nonumber \\
g_{H^{-} \pi_{W^{+}} \pi_{Z} \pi_{Z}}
=&
\sum_h
\frac{1}{m_W}
g_{ W^{+} H^{-} h}
g_{\pi_Z \pi_Z h}
, \nonumber\\
g_{H^{-} \pi_{W^{+}} H^{-} H^{+}}
=&
\sum_h
\frac{1}{m_W}
g_{W^{+} H^{-}h}
g_{H^{+} H^{-} h}
.
\end{align}

\section{EDM formula details}
\label{sec:EDMdetails}
In this section we present formulae for the Barr-Zee contributions to fermionic EDMs and cEDMs.

\subsection{Fermion loops ($h \gamma \gamma$ and $h Z \gamma$)}
After substituting
Eqs.~(\ref{eq:fermion_sub-diagram1}) and
(\ref{eq:fermion_sub-diagram2}) for Eq.~(\ref{eq:df_formula}), we find
the fermion loop contributions to the EDMs for fermion $\ell$ are
\begin{align}
\left(
 \frac{d_{\ell}}{e}
\right)_{\text{fermion}}
&=
- 
\frac{m_{\ell}}{(4\pi)^4}
\sqrt{2} G_F
\sum_f
\sum_h
\sum_{G = \gamma, Z}
N_c
Q_f
\left(
g_{G \ell \ell}^L
+
g_{G \ell \ell}^R
\right) \nonumber \\
& \quad \quad \quad \times
\Biggl[
\frac{g_{\ell \ell h}^A}{m_{\ell}/v}
\frac{g_{ff h}^V}{m_f/v}
{\cal I}_1^{G}(m_f, m_h)
+
\frac{g_{\ell \ell h}^V}{m_{\ell}/v}
\frac{g_{ff h}^A}{m_f/v}
{\cal I}_2^{G}(m_f, m_h)
\Biggr]
,
\label{eq:edm_fermion}
\end{align}
where
\begin{align}
{\cal I}^G_1(m_f, m_h)
=&
\left(
g_{G ff}^L
+
g_{G ff}^R
\right)
\frac{m_f^2}{m_h^2 - m_G^2}
\left(
I_1 ( m_f, m_G) - I_1 ( m_f, m_h)  
\right)
, \nonumber\\ 
{\cal I}^G_2(m_f, m_h)
=&
\left(
g_{G ff}^L
+
g_{G ff}^R
\right)
\frac{m_f^2}{m_h^2 - m_G^2}
\left(
I_2 ( m_f, m_G) - I_2 ( m_f, m_h)  
\right)
,
\end{align}
and where\footnote{
 The functions $f(z)$ and $g(z)$ in
Refs.~\cite{Barr:1990vd, Leigh:1990kf} 
are related to $I_1$ and $I_2$ as follows:
\begin{align}
 I_1(m_1, m_2)
=
-2 \frac{m_2^2}{m_1^2}
f\left( \frac{m_1^2}{m_2^2} \right)
, \quad
 I_2(m_1, m_2)
=
-2 \frac{m_2^2}{m_1^2}
g\left( \frac{m_1^2}{m_2^2} \right)
.
\end{align}
}
\begin{align}
 I_1(m_1, m_2)
=&
 \int_0^1 dz
\left(
1 - 2z (1-z)
\right)
\frac{m_2^2}{m_1^2 - m_2^2 z (1-z)}
\ln
\frac{m_2^2 z (1-z)}{m_1^2}
,\nonumber \\ 
 I_2(m_1, m_2)
=&
 \int_0^1 dz
\frac{m_2^2}{m_1^2 - m_2^2 z (1-z)}
\ln
\frac{m_2^2 z (1-z)}{m_1^2}
.
\label{eq:def_I2}
\end{align}

\subsection{Charged Higgs loops ($h \gamma \gamma$ and $h Z \gamma$)}

By substituting the result in
Eq.~(\ref{eq:scalar_eff_vertex})
into Eq.~(\ref{eq:df_formula}), 
we find the charged Higgs contribution to the EDMs,
\begin{align}
\left(
 \frac{d_{\ell}}{e}
\right)_{\text{scalar}}
=
+
 \frac{m_{\ell}}{(4\pi)^4}
\sqrt{2} G_F
\sum_h
\sum_{G = \gamma, Z}
&
\left(
g_{G \ell \ell}^L
+
g_{G \ell \ell}^R
\right)
\frac{g_{\ell \ell h}^A}{m_{\ell}/v}
\frac{g_{H^{+} H^{-}h}}{v}
{\cal I}^G_3(m_{H^{\pm}}, m_h),
\label{eq:scalarEDM}
\end{align}
where
\begin{align}
{\cal I}^G_3 (m_{H^{\pm}}, m_h)
=&
-
\frac{1}{2}
g_{G H^{+} H^{-}}
\frac{v^2}{m_h^2 - m_G^2} \nonumber \\
& \quad \quad  \times
\Biggl[
\left(
 I_1(m_{H^{\pm}}, m_G)
-
 I_1(m_{H^{\pm}}, m_h)
\right)
- 
\left(
 I_2(m_{H^{\pm}}, m_G)
-
 I_2( m_{H^{\pm}}, m_h) 
\right)
\Biggr]
.
\end{align}

\subsection{$W$ loops ($h \gamma \gamma$ and $h Z \gamma$)}

The EDM contributions from $W$ boson loops are
\begin{align}
\left(
 \frac{d_{\ell}}{e}
\right)_{W}
=&
+
\frac{m_{\ell}}{(4\pi)^{4}} 
\sqrt{2} G_F
\sum_h
\sum_{G = \gamma, Z}
\left(
g_{G \ell \ell}^L
+
g_{G \ell \ell}^R
\right)
\frac{g_{\ell \ell h}^{A}}{m_{\ell} /v}
\frac{g_{WWh}}{2 m_W^2 /v}
 {\cal I}^G_W(m_h)
,
\label{eq:EDM_W_2ndline}
\end{align}
where
\begin{align}
 {\cal I}^G_W(m_h)
&=
g_{WWG}
\frac{2 m_W^2}{m_h^2 - m_G^2}
\nonumber\\ 
& \times
\Biggl[
-\frac{1}{4}
\left\{
\left(
6
-
\frac{m_{G}^2}{m_W^2}
\right)
+
\left(
1
-
\frac{m_{G}^2}{2 m_W^2}
\right)
\frac{m_h^2}{m_W^2}
\right\}
\bigl[
I_1(m_W, m_h)
-
I_1(m_W, m_G)
\bigr]
\nonumber\\ 
&
+
\left\{
\left(
-4 + \frac{m_{G}^2}{m_W^2}
\right)
+
\frac{1}{4}
\left(
6 - \frac{m_{G}^2}{m_W^2}
+
\left(
1 - \frac{m_{G}^2}{2 m_W^2}
\right)
\frac{m_h^2}{m_W^2}
\right)
\right\}
\bigl[
I_2(m_W, m_h)
-
I_2(m_W, m_G)
\bigr]
\Biggr]
.
\label{eq:edm_gauge}
\end{align}
We note that when one chooses $G = \gamma$ in Eq.~\eqref{eq:EDM_W_2ndline} and drops $m_{h}^2/m_W^2$ terms in Eq.~\eqref{eq:edm_gauge}, 
the EDM contribution from $W$ boson loops becomes consistent with original result of Barr and Zee \cite{Barr:1990vd}, where they ignored diagrams which contain only NG boson in the loop in Fig.~\ref{fig:Barr-Zee}  for simplicity and 
Higgs-NG bosons interaction is proportional to 
$m_h^2 /m_W^2$  (see Eq.~\eqref{eq:hNG}).

\subsection{$H^{\mp} W^{\pm} \gamma$}
In this paper we compute for the first time the EDM contributions from $H^{\mp}
W^{\pm} \gamma$ vertices which are generated by $W$ and charged Higgs
boson loops. 
The detail of this derivation is given in
Appendix~\ref{sec:HWA_coupling_appendix}.  The contributions to the EDMs are
\begin{align}
\frac{d_{\ell}}{e}
=&
-
\frac{m_{\ell}}{(4\pi)^4}
\sqrt{2} G_F  {\cal S}_{\ell}
\sum_h
\left(
\frac{g_{\ell \ell h}^{A}}{m_{\ell}/v}
\frac{g_{WWh}}{2 m_W^2/v}
\frac{e^2}{2 s^2}
 {\cal I}_4 (m_h, m_{H^{\pm}})
+
\frac{g_{\ell \ell h}^{A}}{m_{\ell}/v}
\frac{g_{H^{+}H^{-}h}}{v}
{\cal I}_5 (m_h, m_{H^{\pm}})
\right)
,
\end{align}
where
\begin{align}
 {\cal I}_4 (m_h, m_{H^{\pm}})
=&
\frac{m_W^2}{m_{H^{\pm}}^2 - m_W^2}
\left(
I_4 (m_W, m_h)
-
I_4 (m_{H^{\pm}}, m_h)
\right)
,
\nonumber\\
{\cal I}_5 (m_h, m_{H^{\pm}})
=&
\frac{m_W^2}{m_{H^{\pm}}^2 - m_W^2}
\left(
I_5 (m_W, m_h)
-
I_5 (m_{H^{\pm}}, m_h)
\right)
,
\end{align}
and where
\begin{align}
 I_4 (m_1, m_h)
=&
\int_0^1 dz
\left(
z(1-z)^2 - 4(1-z)^2 + \frac{m_{H^{\pm}}^2 -m_h^2}{m_W^2} z (1-z)^2
\right)
\nonumber\\
& 
\qquad \qquad
\times
\frac{m_1^2 }{m_W^2 (1-z) + m_h^2 z - m_1^2 z(1-z)}
\ln
\left(
\frac{m_W^2 (1-z) + m_h^2 z}{m_1^2 z(1-z)}
\right)
, \nonumber\\
 I_5 (m_1, m_h)
=&
2
\int_0^1 dz
\frac{m_1^2 z(1-z)^2}{m_{H^{\pm}}^2 (1-z) + m_h^2 z - m_1^2 z(1-z)}
\ln
\left(
\frac{m_{H^{\pm}}^2 (1-z) + m_h^2 z}{m_1^2 z(1-z)}
\right).
\end{align}
Here we have used the following relations among the coupling,
\begin{align}
 \text{Im}
\left(
\frac{g_{\bar{\nu} e H^{+}}^R}{\sqrt{2} m_{e}/v} 
\frac{g_{W^{+} H^{-} h}}{e/(2 s)}
\right)
=&
\frac{g_{eeh}^{A}}{m_e/v}
,
\nonumber\\ 
 \text{Im}
\left(
\frac{g_{ \bar{u} d H^{+}}^R}{\sqrt{2} m_{d}/v} 
\frac{g_{W^{+} H^{-} h}}{e/(2 s)}
\right)
=&
\frac{g_{ddh}^{A}}{m_d/v}
,
\nonumber\\ 
 \text{Im}
\left(
\frac{g_{\bar{d} u H^{-} }^R}{\sqrt{2} m_{u}/v} 
\frac{g_{W^{-} H^{+} h}}{e/(2 s)}
\right)
=&
\frac{g_{uuh}^A}{m_u/v}
.
\end{align}

\subsection{CEDMs}
The effective Hamiltonian for the cEDM is  defined as Eq.~(\ref{eq:cEDM}).
We find
\begin{align}
 d_{q}^c
&=
+ 
\frac{m_q}{(4\pi)^4}
\sqrt{2} G_F
\sum_f
\sum_h
2 g_s^2
\frac{m_f^2}{m_h^2}
\Biggl[
\frac{g_{qqh}^A}{m_q/v}
\frac{g_{ff h}^V}{m_f/v}
I_1 ( m_f, m_h)  
+
\frac{g_{qqh}^V}{m_{q}/v}
\frac{g_{ffh}^A}{m_f/v}
I_2 ( m_f, m_h)  
\Biggr]
.
\end{align}

\section{Derivation for effective $H^{-} W^{+} \gamma$ vertex}
\label{sec:HWA_coupling_appendix}
In this appendix, we present explicit derivation of the effective
$H^{-} W^{+} \gamma$ vertex, which is generated from bosonic loop
diagrams, in 2HDMs.

\begin{figure}[tb]
\begin{minipage}{0.24\hsize}
\subfigure[]{\includegraphics[bb=0 0 350 150,
 width=1.3\hsize]{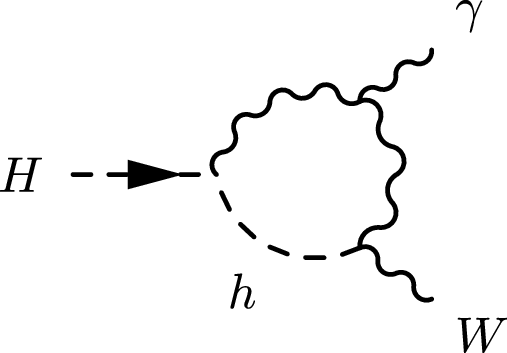} \label{fig:a1}}  
\end{minipage}
\begin{minipage}{0.24\hsize}
\subfigure[]{\includegraphics[bb=0 0 350 150,
 width=1.3\hsize]{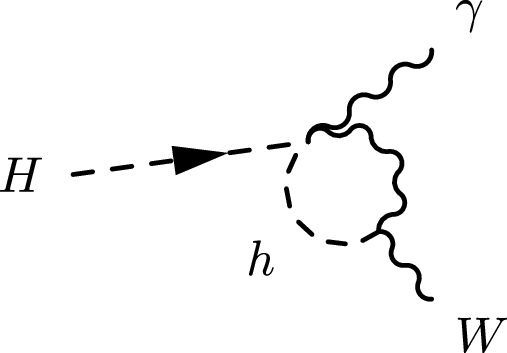} \label{fig:a2}}  
\end{minipage}
\begin{minipage}{0.24\hsize}
\subfigure[]{\includegraphics[bb=0 0 350 150,
 width=1.3\hsize]{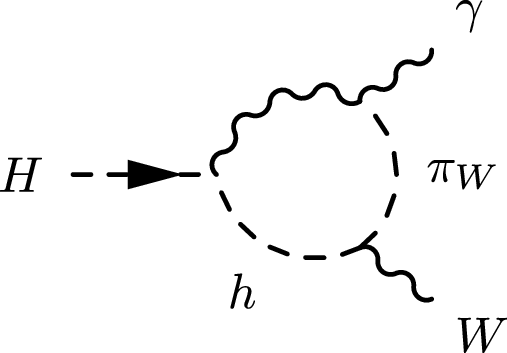} \label{fig:a3}}  
\end{minipage}
\begin{minipage}{0.24\hsize}
\subfigure[]{\includegraphics[bb=0 0 350 200,
 width=1.3\hsize]{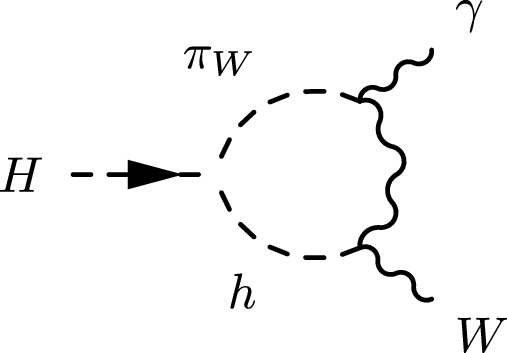} \label{fig:a4}}  
\end{minipage}
\\
\begin{minipage}{0.24\hsize}
\subfigure[]{\includegraphics[bb=0 0 350 200,
 width=1.3\hsize]{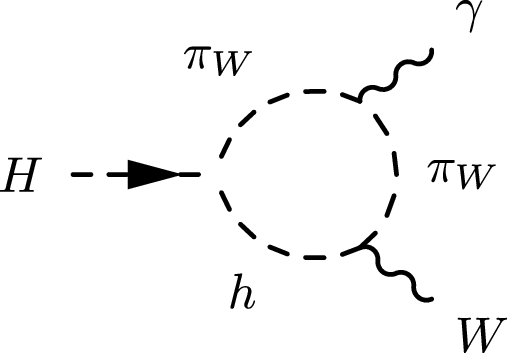} \label{fig:a5}}  
\end{minipage}
\begin{minipage}{0.24\hsize}
\subfigure[]{\includegraphics[bb=0 0 350 200,
 width=1.3\hsize]{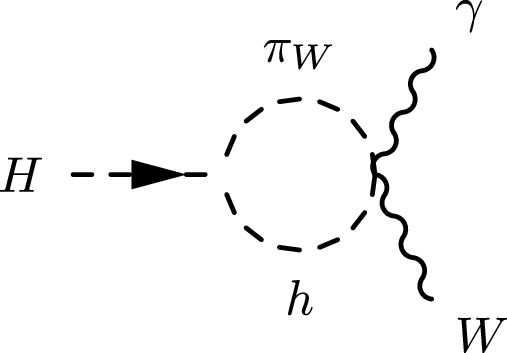} \label{fig:a6}}  
\end{minipage}
\begin{minipage}{0.24\hsize}
\subfigure[]{\includegraphics[bb=0 0 350 200,
 width=1.3\hsize]{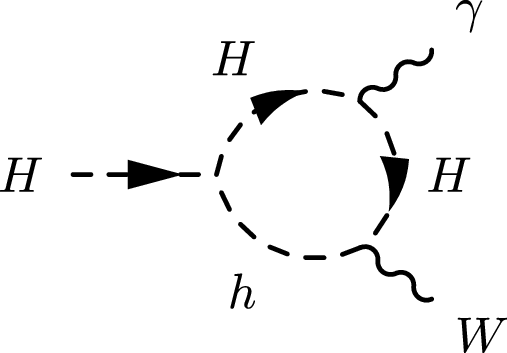} \label{fig:b1}} 
\end{minipage}
\begin{minipage}{0.24\hsize}
\subfigure[]{\includegraphics[bb=0 0 350 200,
 width=1.3\hsize]{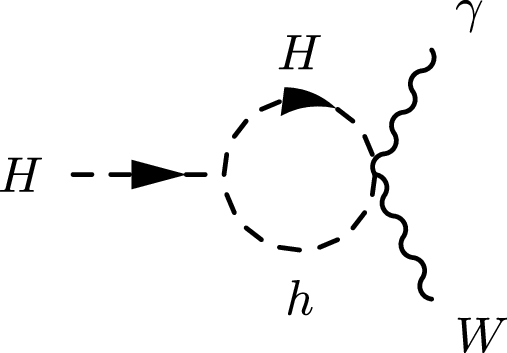} \label{fig:b2}} 
\end{minipage}
\caption{Diagrams for the vertex corrections to $H^{-} W^{+}
 \gamma$. Figs.~\ref{fig:a1}--\ref{fig:a6} depend on the gauge fixing parameter $\xi$, while Figs.~\ref{fig:b1} and \ref{fig:b2} are independent of $\xi$.}
\label{fig:HWA_vertex}
\end{figure}

\begin{figure}[tb]
\begin{minipage}{0.24\hsize}
\subfigure[]{\includegraphics[bb=0 0 450 150,
 width=1.2\hsize]{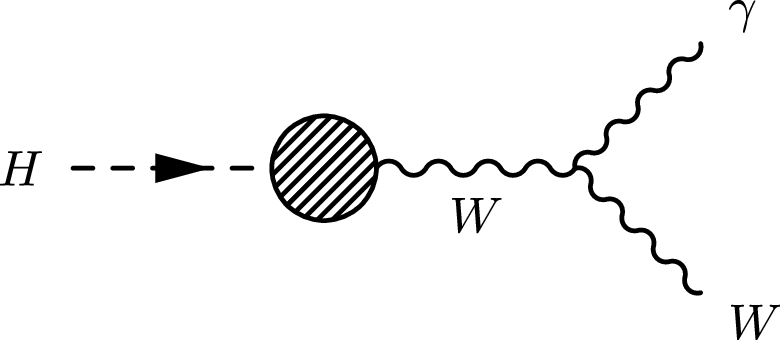} 
 \label{fig:wavefunction_HW}} 
\end{minipage}
\begin{minipage}{0.24\hsize}
\subfigure[]{\includegraphics[bb=0 0 450 150,
width=1.2\hsize]{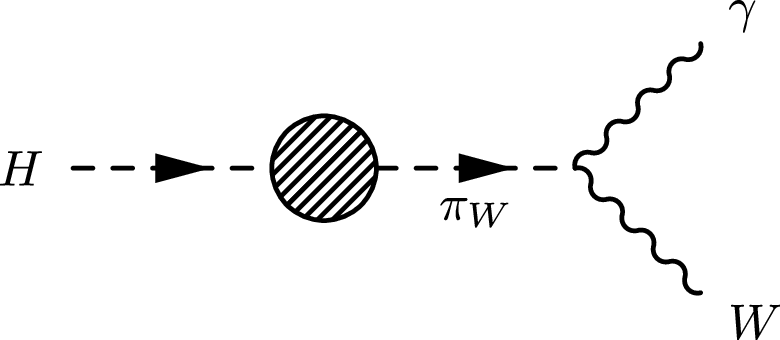}
 \label{fig:wavefunction_Hpi}} 
\end{minipage}
\begin{minipage}{0.24\hsize}
\subfigure[]{\includegraphics[bb=0 0 450 300,
width=1.2\hsize]{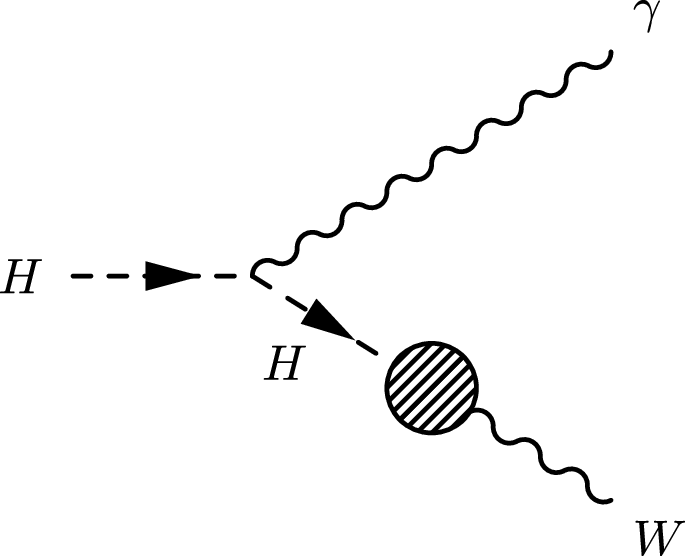}
 \label{fig:wavefunction_W}} 
\end{minipage}
\begin{minipage}{0.24\hsize}
\subfigure[]{\includegraphics[bb=0 0 450 300,
width=1.2\hsize]{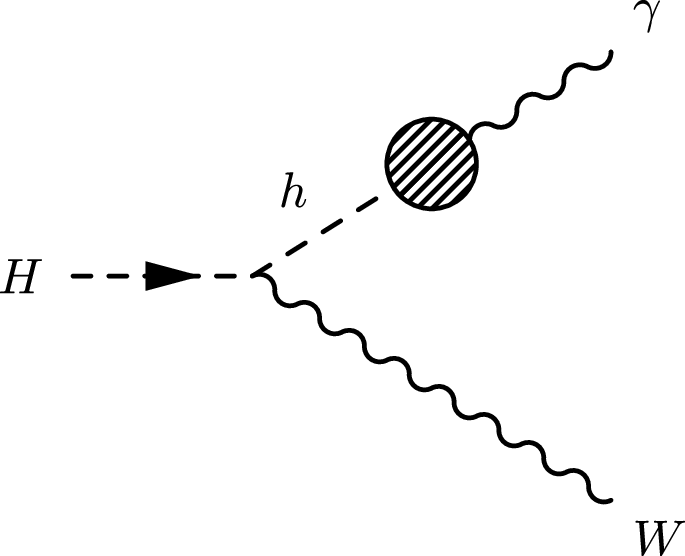}
 \label{fig:wavefunction_gamma}} 
\end{minipage}
\caption{Diagrams of wave function type corrections.}
\label{fig:wavefunctions}
\end{figure}

There are two types of loop diagrams; vertex corrections (Fig.~\ref{fig:HWA_vertex}) and wave function corrections (Fig.~\ref{fig:wavefunctions}).
The diagrams in Fig.~\ref{fig:wavefunction_gamma} give nothing
because of $C$-invariance.
The contributions from Fig.~\ref{fig:wavefunction_W} is always
proportional to $p_2^{\nu}$. Thus they do not contribute
to the on-shell amplitude of $H^{\mp} \to W^{\mp} \gamma$ nor the EDM at
two-loop level by the same discussion in Sec.~\ref{sec:subdiagram}.
Hence what we need to calculate are only the diagrams in 
Figs.~\ref{fig:HWA_vertex}, \ref{fig:wavefunction_HW}, and  
\ref{fig:wavefunction_Hpi}. In this section, we calculate these diagrams
in 't Hooft-Feynman gauge.


First, let us consider the diagrams in Figs.~\ref{fig:a1}--\ref{fig:a6}.
These diagrams depend on the gauge fixing parameter of $W$ boson. We find 
\begin{align}
\sum\text{{\small Figs.~\ref{fig:a1}-\ref{fig:a6}}}
&=
+
\frac{i}{(4\pi)^{D/2}} \Gamma(3 - D/2)
\left(
p_{2 \mu} p_{1 \nu} - p_2 p_1 g_{\mu \nu}
\right)
\sum_h
e g_{W^{+} H^{-} h} g_{WWh}
\nonumber\\ 
&
\quad
\times
\int_{x+y+z=1}
\frac{
-2 yz - 4z + 4 - \frac{m_{H^{\pm}}^2 - m_h^2}{m_W^2} 2 y z
}{
\left[
m_W^2 (1-z) + m_h^2 z - p_2^2 z (1-z) -2 p_1 p_2 yz
\right]^{3-D/2}
}
\nonumber\\
&
+
\frac{i}{(4\pi)^{D/2}} 
g_{\mu \nu}
\sum_h
e g_{W^{+} H^{-} h} g_{WWh}
\nonumber\\ 
&
\quad
\times
\Biggl[
\Gamma(2-D/2)
\int_{0}^{1} dz
\frac{
-(1+z)
}{
\left[
m_W^2 z + m_h^2 (1-z) - p_H^2 z (1-z) 
\right]^{2 - D/2}
}
\nonumber\\
&
\quad 
-
\frac{m_{H^{\pm}}^2 - m_h^2}{m_W^2}
\Gamma(2-D/2)
\int_{0}^{1} dz
\frac{
\frac{1}{2}(-1+2z)
}{
\left[
m_W^2 z + m_h^2 (1-z) - p_H^2 z (1-z) 
\right]^{2 - D/2}
}
\nonumber
\end{align}
\begin{align}
&
\quad 
+
\left(
p_H^2 - m_{H^{\pm}}^2
\right)
\Gamma(3-D/2)
\nonumber\\
& \quad \quad \quad \Biggl.
\times
\int_{x+y+z=1}
\frac{
1
}{
\left[
m_W^2 (1-z) + m_h^2 z - p_2^2 z (1-z)  - 2 p_1 p_2 y z
\right]^{3 - D/2}
}
\Biggr]
,\label{eq:vertex_suicide_3}
\end{align}
where $p_H^2 = (p_1 + p_2)^2$.
We find $g_{\mu \nu}$ terms, which are not gauge invariant.
We will show these terms are canceled with other diagrams, that is, the pinch
contributions. 


The diagrams in Figs.~\ref{fig:b1} and
\ref{fig:b2} are independent from the gauge fixing
parameter. 
\begin{align}
\text{Fig.~\ref{fig:b1} + Fig.~\ref{fig:b2}}
=
&
- \frac{i}{(4\pi)^{D/2}} \Gamma(3 - D/2)
\left(
p_{2 \mu} p_{1 \nu} - p_2 p_1 g_{\mu \nu}
\right)
\sum_h
e g_{W^{+} H^{-} h} g_{H^{+}H^{-}h}
\nonumber\\ 
&
\qquad
\times
\int_{x+y+z=1}
\frac{
4 y z
}{
\left[
m_{H^{\pm}}^2 (1-z) + m_h^2 z - p_2^2 z (1-z) -2 p_1 p_2 yz
\right]^{3-D/2}
}
\nonumber\\
&
- \frac{i}{(4\pi)^{D/2}} \Gamma(2 - D/2)
g_{\mu \nu}
\sum_h
e g_{W^{+} H^{-} h} g_{H^{+}H^{-}h}
\nonumber\\ 
&
\qquad 
\times
\int_{0}^{1} dz
\frac{
-1 + 2z
}{
\left[
m_{H^{\pm}}^2 z + m_h^2 (1-z) - p_H^2 z (1-z) 
\right]^{2-D/2}
}
.
\label{eq:scalar_gauge_variant_terms}
\end{align}

Next we calculate the diagrams in Figs.~\ref{fig:wavefunction_HW} and
\ref{fig:wavefunction_Hpi}.  First we define the following notation
for self-energies:
\begin{align}
\begin{minipage}{0.25\hsize}
\includegraphics[bb=0 0 350 70, 
 width=0.95\hsize]{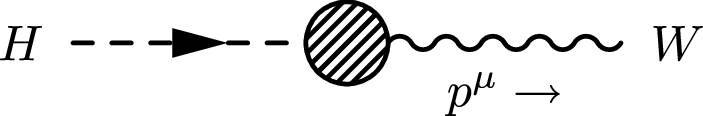}
\end{minipage}
&=
i \Pi^{\mu}_{H^{-}W^{+}}(p)
=
i p^{\mu} \Pi_{H^{-}W^{+}}(p^2)
, \\
\begin{minipage}{0.25\hsize}
\includegraphics[bb=0 0 350 50,
 width=0.95\hsize]{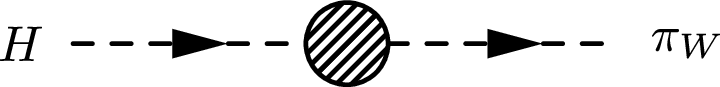}
\end{minipage}
&=
i \Pi_{H^{-}\pi_{W^{+}}}(p^2).
\end{align}
The direction of the momentum of $\Pi_{H^{-}W^{+}}^{\mu}$ is shown in the 
figure.
Using this notation, we find
\begin{align}
 \text{Fig.~\ref{fig:wavefunction_HW} + Fig.~\ref{fig:wavefunction_Hpi}}
=&
\frac{- i g_{\mu \nu}}{ p_H^2 - m_W^2}
\left(
-e m_W i \Pi_{H^{-} \pi_{W^{+}}} ( p_H^2)
-e m_W^2 \Pi_{H^{-} W^{+}} ( p_H^2)
\right)
\nonumber\\ 
&
+
\left( p_2^2 - m_W^2 \right)
\frac{- i g_{\mu \nu}}{ p_H^2 - m_W^2}
\left(
-e \Pi_{H^{-} W^{+}} ( p_H^2)
\right)
.
\label{eq:wave_WHA_1}
\end{align}
Here we ignored $p_2^{\mu} p_2^{\nu}$ terms because they do not
contribute to the EDMs as we discussed in Sec.~\ref{sec:subdiagram}. 
Note that the $\left( p_2^2 - m_W^2 \right)$ term does not also contribute
to the on-shell amplitudes nor the EDMs at two-loop  level. If we
calculate the EDMs with this term, we immediately see that $q^2$
dependence completely canceled out.
Thus, we only need the first term in Eq.~(\ref{eq:wave_WHA_1}).

\begin{figure}[tb]
\begin{center}
\begin{minipage}{0.3\hsize}
\subfigure[]{\includegraphics[bb=0 0 150 100,
 width=0.8\hsize]{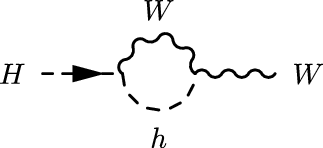} \label{fig:c1}}  
\end{minipage}
\begin{minipage}{0.3\hsize}
\subfigure[]{\includegraphics[bb=0 0 150 100,
 width=0.8\hsize]{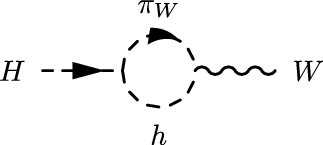} \label{fig:c2}}  
\end{minipage}
\begin{minipage}{0.3\hsize}
\subfigure[]{\includegraphics[bb=0 0 150 100,
 width=0.8\hsize]{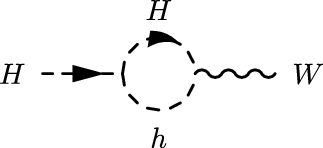} \label{fig:c3}}  
\end{minipage}
\caption{Diagrams for $\Pi_{H^{-} W^{+}}$.}
\label{fig:HW_mixing}
\end{center}
\end{figure}
Fig.~\ref{fig:HW_mixing} shows the diagrams for $\Pi_{H^{-} W^{+}} (p^2)$. We find
\begin{align}
 i \Pi_{H^{-} W^{+}} (p^2)
=&
+
\frac{i}{(4 \pi)^{D/2}} 
\Gamma( 2 - D/2)
g_{WWh} g_{W^{+} H^{-} h}
\int_{0}^{1} dx
\frac{ - (2 - x) + \frac{m_{H^{\pm}}^2 - m_h^2}{m_W^2} ( x - \frac{1}{2})
}{
\left[
m_W^2 ( 1 - x) + m_h^2 x - p^2 x (1-x)
\right]^{2 - D/2}
}
\nonumber \\ 
&
+
\frac{i}{(4 \pi)^{D/2}} 
\Gamma( 2 - D/2)
g_{H^{+}H^{-} h} g_{W^{+} H^{-} h}
\int_{0}^{1} dx
\frac{ 1 - 2x
}{
\left[
m_{H^{\pm}}^2 x + m_h^2 (1-x) - p^2 x (1-x)
\right]^{2 - D/2}
}
.
\label{eq:Pi_{HW}}
\end{align} 

\begin{figure}[tb]
\begin{center}
\begin{minipage}{0.24\hsize}
\subfigure[]{\includegraphics[bb=0 0 200 100,
 width=0.8\hsize]{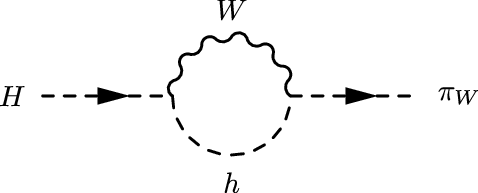} \label{fig:e1}}  
\end{minipage}
\begin{minipage}{0.24\hsize}
\subfigure[]{\includegraphics[bb=0 0 200 100,
 width=0.8\hsize]{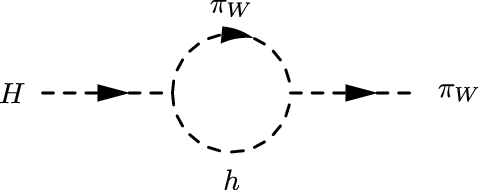} \label{fig:e2}}  
\end{minipage}
\begin{minipage}{0.24\hsize}
\subfigure[]{\includegraphics[bb=0 0 200 100,
 width=0.8\hsize]{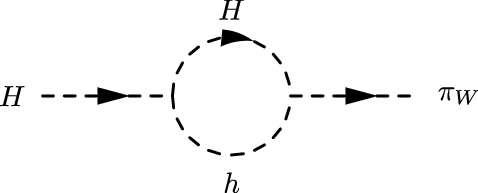} \label{fig:f1}}  
\end{minipage}
\begin{minipage}{0.24\hsize}
\subfigure[]{\includegraphics[bb=0 0 200 100,
 width=0.8\hsize]{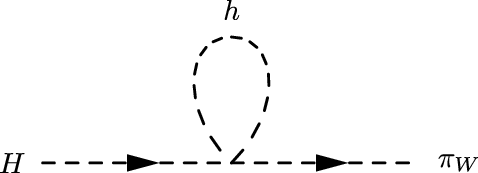} \label{fig:g1}}  
\end{minipage}
\\
\begin{minipage}{0.24\hsize}
\subfigure[]{\includegraphics[bb=0 0 200 100,
 width=0.8\hsize]{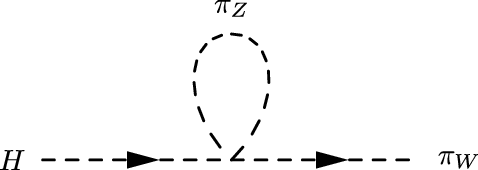} \label{fig:g2}}  
\end{minipage}
\begin{minipage}{0.24\hsize}
\subfigure[]{\includegraphics[bb=0 0 200 100,
 width=0.8\hsize]{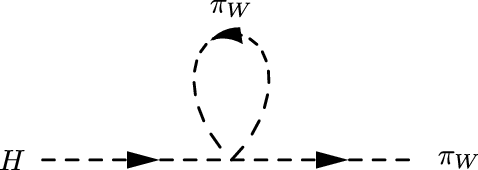} \label{fig:g3}}  
\end{minipage}
\begin{minipage}{0.24\hsize}
\subfigure[]{\includegraphics[bb=0 0 200 100,
 width=0.8\hsize]{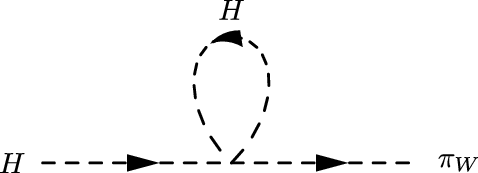} \label{fig:g4}}  
\end{minipage}
\begin{minipage}{0.24\hsize}
\subfigure[]{\includegraphics[bb=0 0 231 84,
 width=0.8\hsize]{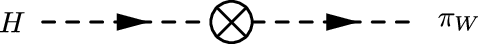} \label{fig:ct_Hpi}}  
\end{minipage}
\caption{Diagrams for $\Pi_{H^{-} \pi_{W^{+}}}$. The last one is for the
 counter term.}
\label{fig:Hpi_mixing}
\end{center}
\end{figure}
\begin{figure}[tb]
\begin{center}
\begin{minipage}{0.19\hsize}
\subfigure[]{\includegraphics[bb=0 0 100 50,
 width=0.8\hsize]{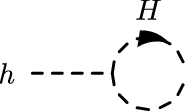} \label{fig:tad_H}}  
\end{minipage}
\begin{minipage}{0.19\hsize}
\subfigure[]{\includegraphics[bb=0 0 100 50,
 width=0.8\hsize]{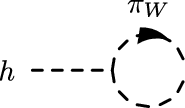} \label{fig:tad_piw}}  
\end{minipage}
\begin{minipage}{0.19\hsize}
\subfigure[]{\includegraphics[bb=0 0 100 50,
 width=0.8\hsize]{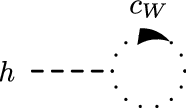} \label{fig:tad_cw}}  
\end{minipage}
\begin{minipage}{0.19\hsize}
\subfigure[]{\includegraphics[bb=0 0 91 53,
 width=0.8\hsize]{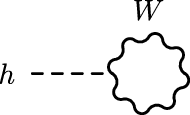} \label{fig:tad_W}}  
\end{minipage}
\begin{minipage}{0.19\hsize}
\quad
\end{minipage}
\\
\begin{minipage}{0.19\hsize}
\subfigure[]{\includegraphics[bb=0 0 100 50,
 width=0.8\hsize]{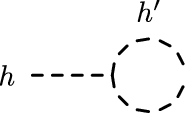} \label{fig:tad_h0}}  
\end{minipage}
\begin{minipage}{0.19\hsize}
\subfigure[]{\includegraphics[bb=0 0 100 50,
 width=0.8\hsize]{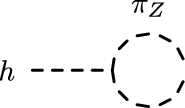} \label{fig:tad_piz}}  
\end{minipage}
\begin{minipage}{0.19\hsize}
\subfigure[]{\includegraphics[bb=0 0 100 50,
 width=0.8\hsize]{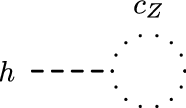} \label{fig:tad_cz}}  
\end{minipage}
\begin{minipage}{0.19\hsize}
\subfigure[]{\includegraphics[bb=0 0 100 50,
 width=0.8\hsize]{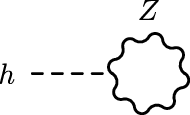} \label{fig:tad_Z}}  
\end{minipage}
\begin{minipage}{0.19\hsize}
\subfigure[]{\includegraphics[bb=0 0 118 56,
 width=0.8\hsize]{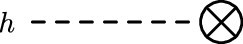} \label{fig:ct_tadpole}}  
\end{minipage}
\caption{Tadpoles diagrams.}
\label{fig:tadpoles}
\end{center}
\end{figure}

Fig.~\ref{fig:Hpi_mixing} shows the diagrams for $\Pi_{H^{-} \pi_{W^{+}}} (p^2)$.
We find Figs.~\ref{fig:g1}--\ref{fig:g4} are canceled by
Fig.~\ref{fig:ct_Hpi}, so we do not calculate
them. Fig.~\ref{fig:ct_Hpi} is the counter term for $H$--$\pi_W$ mixing, and it
is also related with the counter terms for the Higgs tadpoles
(Fig.~\ref{fig:ct_tadpole}), 
\begin{align}
 \delta_{H^{-} \pi_{W^{+}}}
=&
\sum_h
\frac{1}{m_W}
g_{W^{+} H^{-} h}
\delta_h
,
\label{eq:relation_between_CT}
\end{align}
where $\delta$'s are defined through
\begin{align}
 {\cal L}
\supset&
-
\delta_{H^{-} \pi_{W^{+}}}
H^{-} \pi_{W^{+}}
+
\sum_h
\delta_h 
h
.
\label{eq:ct_relation}
\end{align}
It is easy to find this relation by analyzing the Higgs potential. 
We take renormalization conditions in which all tadpole diagrams are
completely canceled by their counter terms. Then $\delta_{H^{-}
\pi_{W^{+}}}$ is not arbitrary but should be calculated from the tadpole
diagrams and Eq.~(\ref{eq:ct_relation}). We show the tadpole diagrams in
Fig.~\ref{fig:tadpoles}. 
After calculating tadpole diagrams, using
Eq.~(\ref{eq:relation_between_CT}), we find
\begin{align}
\text{Fig.~\ref{fig:ct_Hpi}}
=&
-
\left(
\text{Fig.~\ref{fig:g1}}
+
\text{Fig.~\ref{fig:g2}}
+
\text{Fig.~\ref{fig:g3}}
+
\text{Fig.~\ref{fig:g4}}
\right)
\nonumber\\ 
&
+
i
\sum_h
\left(
\frac{m_{H^{\pm}}^2 - m_h^2}{2 m_W^3}
g_{WWh} g_{W^{+} H^{-} h}
+
\frac{g_{H^{+}H^{-} h} g_{W^{+} H^{-} h}}{m_W}
\right)
\int_{\ell}
\frac{1}{\ell^2 - m_h^2}
\nonumber\\ 
&
-
i
\sum_h
\frac{g_{H^{+}H^{-}h} g_{W^{+} H^{-} h}}{m_W}
\int_{\ell}
\frac{1}{\ell^2 - m_{H^{\pm}}^2}
\nonumber\\ 
&
+
i
\sum_h
\frac{m_h^2}{2 m_W^3}
g_{H^{+}H^{-}h} g_{W^{+} H^{-}h}
\int_{\ell}
\frac{1}{\ell^2 - m_W^2}
.
\end{align}
Now we have calculated all the diagrams shown in
Fig.~\ref{fig:Hpi_mixing}, and we find
\begin{align}
 i \Pi_{H^{-} \pi_{W^{+}}}
=&
+
\frac{\Gamma( 2 - D/2)}{(4 \pi)^{D/2}}
\frac{1}{2 m_W}
g_{WWh} g_{W^{+} H^{-} h}
\nonumber \\ 
&
\quad 
\times
\int_{0}^{1} dx
\Biggl[
\frac{p^2 ( 1 + 2x) + m_h^2
}{
\left[
m_W^2 x + m_h^2 (1-x) - p^2 x (1-x)
\right]^{2 - D/2}
} \Biggr. \nonumber \\
& \qquad \qquad  \Biggl.
+
\frac{m_{H^{\pm}}^2 - m_h^2}{m_W^2}
\frac{m_W^2 - p^2 ( 1 - 2x) 
}{
\left[
m_W^2 x + m_h^2 (1-x) - p^2 x (1-x)
\right]^{2 - D/2}
}
\Biggr]
\nonumber \\ 
&
-
\frac{\Gamma( 2 - D/2)}{(4\pi)^{D/2}}
\frac{1}{m_W}
g_{H^{+}H^{-}h} g_{W^{+} H^{-} h}
p^2
\int_{0}^{1} dx
\frac{1-2x
}{
\left[
m_{H^{\pm}}^2 x + m_h^2 (1-x) -p^2 x (1-x)
\right]^{2-D/2}
}
.
\label{eq:Pi_{Hpi}}
\end{align}

We have finished preparing to calculate 
Fig.~\ref{fig:wavefunction_HW} + Fig.~\ref{fig:wavefunction_Hpi}.
Substituting Eqs.~(\ref{eq:Pi_{HW}}) and (\ref{eq:Pi_{Hpi}}) into
the first term in Eq.~(\ref{eq:wave_WHA_1}), then we find
\begin{align}
\text{Fig.~\ref{fig:wavefunction_HW} + Fig.~\ref{fig:wavefunction_Hpi}}
=&
-i g_{\mu \nu}
\frac{1}{(4 \pi)^{D/2}}
\Gamma(2 - D/2) e g_{W^{+} H^{-} h}
\nonumber \\
&
 \times
\Biggl[
 g_{WWh}
\int_{0}^{1} dx
\frac{
-
(1+x)
+
\frac{m_{H^{\pm}}^2 - m_h^2}{2 m_W^2}
( 1 - 2 x)
}{
\left[
m_W^2 x + m_h^2 (1-x) - p_H^2 x (1-x)
\right]^{2 - D/2}
}
\nonumber
\\ 
&
\quad  
+
 g_{WWh} 
\frac{p_H^2 - m_{H^{\pm}}^2}{2 (p_H^2 - m_W^2)}
\int_{0}^{1} dx
\frac{
1
}{
\left[
m_W^2 x + m_h^2 (1-x) - p_H^2 x (1-x)
\right]^{2 - D/2}
}
\nonumber
\\ 
&
\quad 
+
 g_{H^{+}H^{-}h} 
\int_{0}^{1} dx
\frac{
 1 - 2 x 
}{
\left[
m_{H^{\pm}}^2 x + m_h^2 (1-x) - p_H^2 x (1-x)
\right]^{2 - D/2}
}
\Biggr]
\label{eq:self_HHh_suicide}
.
\end{align} 
Here we dropped the $(p_2^2 - m_W^2) g_{\mu \nu}$ term because it does not
contribute to what we are interested in. 
Note that the first term in the bracket in
Eq.~(\ref{eq:self_HHh_suicide}) is canceled with
Eq.~(\ref{eq:vertex_suicide_3}), and
the second term is canceled with
Eq.~(\ref{eq:scalar_gauge_variant_terms}).

So far we have calculated many diagrams, vertex corrections and
wave function corrections. The corrections are not so simple and some of
them canceled out, so we give a short summary so far here. After summing
up all the correction we have calculated so far, we find
\begin{align}
&
+
\frac{i}{(4\pi)^{D/2}} \Gamma(3 - D/2)
\left(
p_{2 \mu} p_{1 \nu} - p_2 p_1 g_{\mu \nu}
\right)
\nonumber\\ 
&
\qquad \qquad
\times
\Biggl(
+
\sum_h
e g_{W^{+} H^{-} h} g_{WWh}
\int_{x+y+z=1}
\frac{
-2 yz - 4z + 4 - \frac{m_{H^{\pm}}^2 - m_h^2}{m_W^2} 2 y z
}{
\left[
m_W^2 (1-z) + m_h^2 z - p_2^2 z (1-z) -2 p_1 p_2 yz
\right]^{3-D/2}
}
\nonumber\\ 
&
\qquad \qquad \qquad
-
\sum_h
e g_{W^{+} H^{-} h} g_{H^{+}H^{-} h}
\int_{x+y+z=1}
\frac{
4 y z
}{
\left[
m_{H^{\pm}}^2 (1-z) + m_h^2 z - p_2^2 z (1-z) -2 p_1 p_2 yz
\right]^{3-D/2}
}
\Biggr)
\nonumber\\
&
+
\frac{i}{(4\pi)^{D/2}} 
g_{\mu \nu}
\left(
p_H^2 - m_{H^{\pm}}^2
\right)
\sum_h
e g_{W^{+} H^{-} h} g_{WWh}
\nonumber\\ 
&
\qquad \qquad
\times
\Biggl[
+
\Gamma(3-D/2)
\int_{x+y+z=1}
\frac{
1
}{
\left[
m_W^2 (1-z) + m_h^2 z - p_2^2 z (1-z)  - 2 p_1 p_2 y z
\right]^{3 - D/2}
}
\nonumber\\
&
\qquad \qquad \qquad
-
\Gamma(2 - D/2)
\frac{1}{2 (p_H^2 - m_W^2)}
\int_{0}^{1} dx
\frac{
1
}{
\left[
m_W^2 x + m_h^2 (1-x) - p_H^2 x (1-x)
\right]^{2 - D/2}
}
\Biggr]
.
\label{eq:pinch_de_byebye2}
\end{align}
Note that the last two terms are not gauge invariant in the sense that we
discussed in Sec.~\ref{sec:subdiagram}. Since they are proportional to 
$p_H^2 - m_{H^{\pm}}^2$, if we take the charged Higgs boson on-shell, they are
dropped and the result becomes gauge invariant. However, now we need to
take the charged Higgs boson off-shell, so we still need some other
terms to cancel them. 

\begin{figure}[tb]
\begin{center}
\begin{minipage}{0.19\hsize}
\subfigure[]{\includegraphics[bb=0 0 150 100,
 width=0.7\hsize]{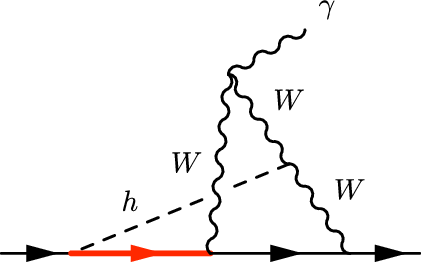} \label{fig:pinch_a2}}  
\end{minipage}
\begin{minipage}{0.19\hsize}
\subfigure[]{\includegraphics[bb=0 0 150 100,
 width=0.7\hsize]{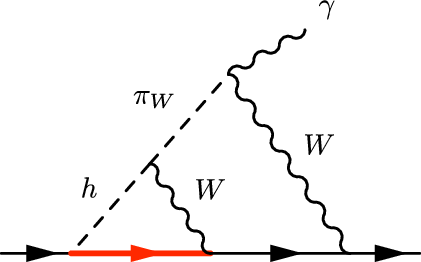} \label{fig:pinch_a3}}  
\end{minipage}\quad \quad
\begin{minipage}{0.09\hsize}
$\longrightarrow$
\end{minipage}
\begin{minipage}{0.19\hsize}
\subfigure[]{\includegraphics[bb=0 0 150 100,
 width=0.7\hsize]{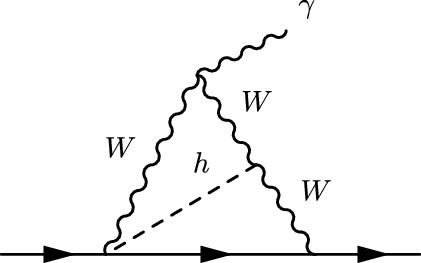} \label{fig:pinch_b2}}  
\end{minipage}
\begin{minipage}{0.19\hsize}
\subfigure[]{\includegraphics[bb=0 0 150 100,
 width=0.7\hsize]{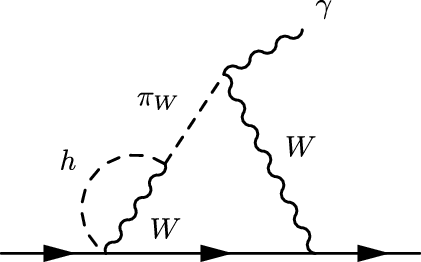} \label{fig:pinch_b3}}  
\end{minipage}
\caption{Pinch contributions.
}
\label{fig:pinch}
\end{center}
\end{figure}
To find a gauge invariant set for the Barr-Zee diagrams, we need to take into
account for the pinch contributions shown in Fig.~\ref{fig:pinch}.
After pinching the fermion propagators with red color in
Figs.~\ref{fig:pinch_a2} and \ref{fig:pinch_a3}, the pinch contributions
for $H^{-} W^{+} \gamma$ effective vertex for the Barr-Zee diagrams
arise. They are schematically shown in Figs.~\ref{fig:pinch_b2} and
\ref{fig:pinch_b3}. We denote their contributions as $\Gamma^{\mu
\nu}_{\text{P}}$ and $i \Pi_{\text{P}}$, respectively. Then we find
\begin{align}
 i \Gamma_{\text{P}}^{\mu \nu} (p_1, p_2)
=&
-i
\frac{\Gamma( 3 - D/2)}{(4 \pi)^{D/2}}
 e g_{W^{+} H^{-} h} g_{WWh} g^{\mu \nu}
\nonumber \\ 
&
\quad \times
\left(
p_H^2 - m_{H^{\pm}}^2
\right)
\int_{x+y+z=1}
\frac{1}{
\left[
m_W^2 (1-z) + m_h^2 z - p_2^2 z (1-z) - 2 p_1 p_2 yz)
\right]^{3-D/2}
}
, 
\end{align}
\begin{align}
 i \Pi_{\text{P}} (p_H^2)
=&
+
\frac{\Gamma(2 - D/2)}{(4 \pi)^{2-D/2}}
\frac{1}{2 m_W}
g_{W^{+} H^{-} h} g_{WWh} 
\left(
p_H^2 - m_{H^{\pm}}^2
\right) \nonumber \\
&\quad \times
\int_{0}^{1} dx
\frac{1}{
\left[
m_W^2 x + m_h^2 (1-x) - p_H^2 x (1-x)
\right]^{2-D/2}
}
.
\end{align}
Using Eq.~(\ref{eq:wave_WHA_1}), we find that
$\Gamma^{\mu\nu}_{\text{P}}$ and $i \Pi_{\text{P}}$ completely cancel
the second term in Eq.~(\ref{eq:pinch_de_byebye2}), namely these pinch
contributions really make the effective vertex correction gauge
invariant.

\providecommand{\href}[2]{#2}
\begingroup\raggedright

\endgroup
\end{document}